\documentclass[11pt,reqno,twoside]{amsart}
\usepackage{amsfonts,amsmath,amssymb}
\usepackage{mathrsfs,mathtools}
\usepackage{enumerate}
\usepackage{hyperref}
\usepackage{esint}
\usepackage{graphicx}
\usepackage{bm}
\usepackage{commath}
\usepackage{esint}
\usepackage{placeins}
\usepackage{flafter}
\usepackage{tikz}
\usepackage[dvips,bottom=1.4in,right=1in,top=1in, left=1in]{geometry}

\usepackage{fancyvrb}
\fvset{fontsize=\footnotesize,xleftmargin=2em}
\usepackage{lettrine}
\usepackage{graphicx}

\font\code=cmtt10

\def\pmb#1{\setbox0=\hbox{$#1$}%
             \kern-.027em\copy0\kern-\wd0
             \kern+.009em\copy0\kern-\wd0
             \kern+.009em\copy0\kern-\wd0
             \kern+.009em\copy0\kern-\wd0
             \kern+.009em\copy0\kern-\wd0
             \kern+.009em\copy0\kern-\wd0
             \kern+.009em\copy0\kern-\wd0
             \kern-.045em\raise+.012em\copy0\kern-\wd0
             \kern+.009em\raise+.012em\copy0\kern-\wd0
             \kern+.009em\raise+.012em\copy0\kern-\wd0
             \kern+.009em\raise-.012em\copy0\kern-\wd0
             \kern+.009em\raise-.012em\copy0\kern-\wd0
             \kern-.018em\copy0\kern-\wd0\raise-.012em\box0}
\def\Pmb#1{\setbox0=\hbox{$#1$}%
             \kern-.033em\copy0\kern-\wd0
             \kern+.011em\copy0\kern-\wd0
             \kern+.011em\copy0\kern-\wd0
             \kern+.011em\copy0\kern-\wd0
             \kern+.011em\copy0\kern-\wd0
             \kern+.011em\copy0\kern-\wd0
             \kern+.011em\copy0\kern-\wd0
             \kern-.055em\raise+.015em\copy0\kern-\wd0
             \kern+.011em\raise+.015em\copy0\kern-\wd0
             \kern+.011em\raise+.015em\copy0\kern-\wd0
             \kern+.011em\raise-.015em\copy0\kern-\wd0
             \kern+.011em\raise-.015em\copy0\kern-\wd0
             \kern-.022em\copy0\kern-\wd0\raise-.015em\box0}
\def\ms{\medskip}

\def\ub{\underbar}
\def\noi{\noindent}

\def\fvec{{\bf f}}
\def\gvec{{\bf g}}

\def\rvec{{\bf r}}
\def\svec{{\bf s}}

\def\uvec{{\bf u}}
\def\vvec{{\bf v}}

\def\xvec{{\bf x}}

\def\Dvec{{\bf D}}

\def\Mmat{{\bf M}}

\def\dt{{\Delta t}}

\def\Grd{\nabla}
\def\Div{\nabla \cdot}

\def\duvec{\Delta {\bf u}}

\vfill

\DeclareMathAlphabet{\mathpzc}{OT1}{pzc}{m}{it}
\numberwithin{equation}{section}
\makeatletter
\def\eqnarray{\stepcounter{equation}\let\@currentlabel=\theequation
\global\@eqnswtrue
\tabskip\@centering\let\\=\@eqncr
$$\halign to \displaywidth\bgroup\hfil\global\@eqcnt\z@
  $\displaystyle\tabskip\z@{##}$&\global\@eqcnt\@ne
  \hfil$\displaystyle{{}##{}}$\hfil
  &\global\@eqcnt\tw@ $\displaystyle{##}$\hfil
  \tabskip\@centering&\llap{##}\tabskip\z@\cr}
\def\endeqnarray{\@@eqncr\egroup
      \global\advance\c@equation\m@ne$$\global\@ignoretrue}

\numberwithin{equation}{section}

\thanks{H. Antil is partially supported by NSF grants DMS-1818772, DMS-1913004,
the Air Force Office of Scientific Research (AFOSR) under
Award NO: FA9550-19-1-0036, and Department of Navy, Naval PostGraduate
School under Award NO: N00244-20-1-0005.
}

\begin{document}

\title[AEROSOL PROPAGATION WITH MOVING PEDESTRIANS]{HIGH FIDELITY MODELING OF AEROSOL PATHOGEN PROPAGATION
       IN BUILT ENVIRONMENTS WITH MOVING PEDESTRIANS}

\author{Rainald L\"ohner}
\address{Rainald L\"ohner, Center for Computational Fluid Dynamics,
               College of Science, George Mason University, \\
               Fairfax, VA 22030-4444, USA,
}
\email{rlohner@gmu.edu}
\author{Harbir Antil}
\address{Harbir Antil, Center for Mathematics and Artificial Intelligence (CMAI),
               College of Science, \\
               George Mason University,
               Fairfax, VA 22030-4444, USA
}
\email{hantil@gmu.edu}

\begin{abstract}
A high fidelity model for the propagation of pathogens via
aerosols in the presence of moving pedestrians is proposed.
The key idea is the tight coupling of computational fluid
dynamics and computational crowd dynamics in order to
capture the emission, transport and inhalation of pathogen
loads in space and time. \\
An example simulating pathogen propagation in a narrow corridor 
with moving pedestrians clearly shows the considerable effect 
that pedestrian motion has on airflow,
and hence on pathogen propagation and potential infectivity.
\end{abstract}

\maketitle

\section{Introduction}
Advances in computational fluid and crowd dynamics (CFD, CCD), as well 
as computer hardware and software, have enabled fast and reliable 
simulations in both disciplines. A natural next step is the coupling of 
both disciplines. This would be of high importance for evacuation 
studies where fire, smoke, visibility and inhalation of toxic
materials influence the motion of people, and where a large crowd can 
block or influence the flow in turn. The same capability could
also be used to simulate with high fidelity the transmission of 
pathogens in the presence of moving pedestrians, enabling a much
needed extension of current simulation technologies \cite{Loh20} . \\
The present work considers a tight, bi-directional coupling, whereby 
the flow (and any pathogens in it) and the motion of the crowd are 
computed concurrently and with mutual influences. 
Enabling technologies that made this tight coupling feasible include: 
\begin{itemize}
\item[a)] Development of immersed boundary methods; 
\item[b)] Implementation of fast search techniques for information
transfer between codes; and
\item[c)] Strong scaling to tens of thousands of cores for CFD codes.
\end{itemize}
Before describing the numerical methodologies, a quick overview of
pathogen, and in particular virus transmission is given. 
This defines the relevant physical
phenomena, which in turn define the ordinary and partial differential
equations that describe the flow and the particles. Thereafter, the
models used for pedestrian motion are outlined. The sections on
numerical methodologies conclude with a description of the 
coupling methodology employed. Several examples illustrate the
influence of pedestrian motion on air motion, and hence aerosol 
and pathogen transport, and show the potential of the 
proposed methodology.

\section{Virus Infection}
Before describing the proposed model for
virus transmission a brief description of virus propagation and
lifetime is given. Viruses are usually present
in the air or some surface, and make their way into the body either
via inhalation (nose, mouth), ingestion (mouth) or attachment
(eyes, hands, clothes). In many cases the victim inadvertently
touches an infected surface or viruses are deposited on its hands,
and then the hands or clothes touch either the nose, the eyes or 
the mouth, thus allowing the virus to enter the body. \\
An open question of great importance is
how many viruses it takes to overwhelm the body's natural defense
mechanism and trigger an infection. This number, which is sometimes
called the {\sl viral load} or the {\sl infectious dose} will depend on
numerous factors, among them the state of immune defenses of the
individual, the timing of viral entry (all at once, piece by piece),
and the amount of hair and mucous in the nasal vessels.
In principle, a single organism in a favourable environment
may replicate sufficiently to cause disease \cite{Tan06}.
Data from research performed on biological
warfare agents \cite{Fra97} suggests that both bacteria and
viruses can produce disease with as few as 1-100
organisms (e.g. brucellosis 10-100, Q fever 1-10,
tularaemia 10-50, smallpox 10-100, viral haemorrhagic fevers
1-10 organisms, tuberculosis 1). Compare these numbers and consider
that as many as 3,000 organisms can be produced by talking for
5~minutes or a single cough, with sneezing producing many more
\cite{Lou67,Lin10,Teu10,Mil13,Wei16}.
Figure~1, reproduced from \cite{Teu10}, shows a typical number and
size distribution.

\begin{figure}
	\centering
	\includegraphics[width=10.0cm]{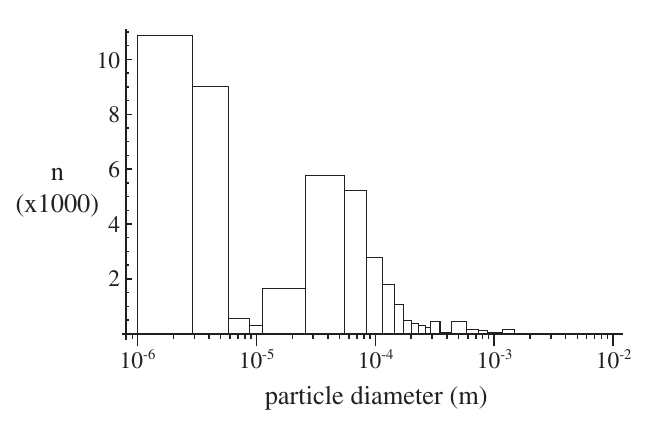}
	\caption{Counts of Particles of Various Diameters in Air Expelled 
	by 90 Coughs \cite{Lou67}}
\end{figure}

\section{Virus Lifetime Outside the Body}
Current evidence for Covid-19 points to lifetimes outside the 
body that can range
from 1-2 hours in air to several days on particular surfaces
(so-called fomite transmission mode) \cite{Kam20,vDo20}.
There has also been some documentation of lifetime variation depending
on humidity.

\section{Virus Transmission}

\subsection{Sneezing and Coughing}
In the sequel, we consider sneezing and coughing as the main
conduits of virus transmission. Clearly, breathing and talking
will lead to the exhalation of air, and, consequently the exhalation
of viruses for infected victims \cite{Asa20a,Asa20b}.
However, it stands to reason that
the size and amount of particles released -~and hence the amount of
viruses in them~- is much higher and much more concentrated when
sneezing or coughing \cite{Fab08,Teu10,Joh11,Lin12,Asa20a,Asa20b}. \\
The velocity of air at a person's mouth during sneezing and coughing
has been a source of heated debate, particularly in the media.
The experimental evidence points to exit velocities of the order
of 2-14~m/sec \cite{Gup09,Gup10,Tan12,Tan13}.
A typical amount and size of particles can be seen in Figure~1.

\subsection{Sink Velocities}
Table~1 lists the terminal sink velocities for water droplets
in air based on the diameter \cite{Sch79}. One can see that below
diameters of $O(0.1~mm)$ the sink velocity is very low,
implying that these particles remain in and move with the air for
considerable time (and possibly distances).

\begin{table}[htbp]
\begin{center}
\caption{Sink Velocities and Reynolds Number For Water Particles in Air}
\label{tab:Sinkvelo}
\begin{tabular}{c|c|c}
\hline
Diameter [mm] & sink velocity [m/sec] & Re \\
\hline
1.00E-01      & 3.01E-01      & 1.99E+00 \\
1.00E-02      & 3.01E-03      & 1.99E-03 \\
1.00E-03      & 3.01E-05      & 1.99E-06 \\
1.00E-04      & 3.01E-07      & 1.99E-09 \\
\hline
\end{tabular}
\end{center}
\end{table}

\subsection{Evaporation}
Depending on the relative humidity and the temperature of the
ambient air, the smaller particles can evaporate in milliseconds.
However, as the mucous and saliva evaporate, they build a gel-like
structure that surrounds the virus, allowing it to survive. This implies
that extremely small particles with possible viruses will remain
infectious for extended periods of times - up to an hour according to
some studies \cite{Kam20,vDo20}. \\
An important question is whether a particle/droplet will
first reach the ground or evaporate. Figure~2, taken from \cite{Xie07},
shows that below 0.12~$mm$ the particles evaporate before
falling 2~m (i.e. reaching the ground).

\begin{figure}
	\centering
	\includegraphics[width=10.0cm]{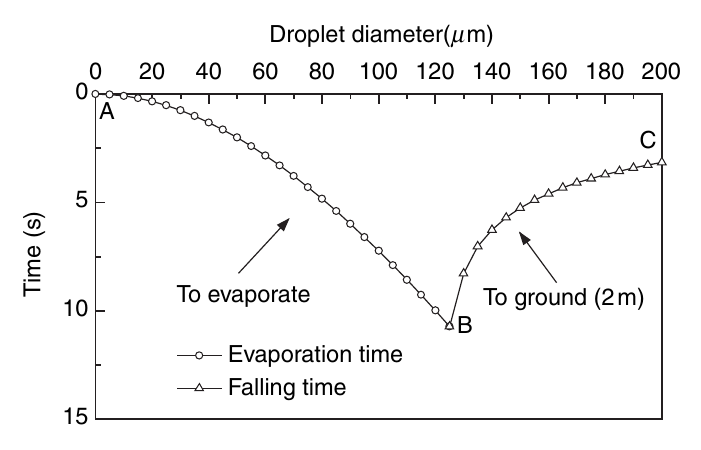}
	\caption{Evaporation Time and Falling Time of Droplets of Varying Diameter}
\end{figure}

\section{Physical Modeling of Aerosol Propagation}
\label{physmodelaerosols}
When solving the two-phase equations, the air, as a continuum,
is best represented by a set of partial differential equations 
(the Navier-Stokes equations) that are numerically solved on a mesh. 
Thus, the gas characteristics are calculated 
at the mesh points within the flowfield. 
The droplets/particles, which are relatively sparse in the flowfield, 
are modeled using a Lagrangian description, where individual 
particles (or groups of particles) are monitored and tracked in the 
flow, allowing for an exchange of mass, momentum and energy between
the air and the particles.

\subsection{Equations Describing the Motion of the Air}
As seen from the experimental evidence, the velocities of air
encountered during coughing and sneezing never exceed a Mach-number
of $Ma=0.1$. Therefore, the air may be assumed as a Newtonian,
incompressible liquid, where buoyancy effects are modeled via
the Boussinesq approximation. The equations describing the
conservation of momentum, mass and energy for incompressible, 
Newtonian flows may be written as

$$ \rho \vvec_{,t} + \rho \vvec \cdot \Grd \vvec + \Grd p = 
 \nabla \cdot \mu \nabla \vvec + \rho \gvec + \beta \rho \gvec(T - T_0) 
      + \svec_v   ~~, 
                                                         \eqno(5.1.1) $$
$$                             \Div \vvec = 0      ~~,   \eqno(5.1.2) $$
$$ \rho c_p T_{,t} + \rho c_p \vvec \cdot \Grd T = 
       \nabla \cdot k \nabla T + s_e ~~.
                                                         \eqno(5.1.3) $$

\noi
Here $\rho, \vvec, p, \mu, \gvec, \beta, T, T_0, c_p, k$ denote 
the density, velocity vector, pressure, viscosity, gravity vector, 
coefficient of thermal expansion, temperature, reference temperature,
specific heat coefficient and conductivity respectively, and
$\svec_v, s_e$ momentum and energy source terms (e.g. due to particles
or external forces/heat sources).
For turbulent flows both the viscosity and the conductivity are
obtained either from additional equations or directly via a
large eddy simulation (LES) assumption through monotonicity induced
LES (MILES) \cite{Bor92,Fur99,Gri02,Ide19}.

\subsection{Equations Describing the Motion of Particles/Droplets}
\label{partmotion}
In order to describe the interaction of particles/droplets with the 
flow, the mass, forces and energy/work exchanged between the 
flowfield and the particles must be defined.
As before, we denote for {\bf fluid (air)}
by $\rho, p, T, k, v_i, \mu$ and $c_p$ the density, pressure,
temperature, conductivity, velocity in direction $x_i$,
viscosity, and the specific heat at constant pressure.
For the {\bf particles}, we denote by 
$\rho_p, T_p, v_{pi}, d, c_{pp}$ and 
$Q$ the density, temperature, velocity in direction $x_i$, 
equivalent diameter, specific heat coefficient and heat transferred 
per unit volume. In what follows we will refer to droplet and 
particles collectively as particles. \\
Making the classical assumptions that the particles may be represented
by an equivalent sphere of diameter $d$, the drag forces $\Dvec$
acting on the particles will be due to the difference of fluid and 
particle velocity:

$$ \Dvec = {{\pi d^2} \over 4} \cdot c_d \cdot
         { 1 \over 2} \rho | \vvec - \vvec_p | ( \vvec - \vvec_p ) 
                                                  ~~.  \eqno(5.2.1) $$

\noi
The {\bf drag coefficient} $c_d$ is obtained empirically from the
Reynolds-number $Re$:

$$ Re = {{\rho | \vvec - \vvec_p | d } \over { \mu }}  \eqno(5.2.2) $$

\noi
as (see, e.g. \cite{Sch79}):

$$ c_d = max\left(0.1 ,
{24 \over Re} \left( 1 + 0.15 Re^{0.687} \right) \right) \eqno(5.2.3) $$

\noi
The lower bound of $c_d=0.1$ is required to obtain the proper limit for
the Euler equations, when $Re \rightarrow \infty$.
\noi
The heat transferred between the particles and the fluid is given by

$$ Q = {{\pi d^2} \over 4} \cdot 
      \left[ h_f      \cdot ( T   - T_p   )
           + \sigma^* \cdot ( T^4 - T_p^4 ) \right]
                                                 ~~,  \eqno(5.2.4) $$
\noi
where $h_f$ is the film coefficient and $\sigma^*$ the radiation 
coefficient. For the class of problems considered here, the particle 
temperature and kinetic energy are such that the radiation 
coefficient $\sigma^*$ may be ignored. The film coefficient $h_f$ is 
obtained from the Nusselt-Number $Nu$:

$$ Nu = 2 + 0.459 Pr^{0.333} Re^{0.55} ~~,  \eqno(5.2.5)        $$

\noi
where $Pr$ is the Prandtl-number of the gas

$$ Pr = {k \over \mu } ~~,  \eqno(5.2.6)         $$

\noi
as

$$ h_f = {{ Nu \cdot k }\over d} ~~.      \eqno(5.2.7)    $$

\par \noi
Having established the forces and heat flux, the particle motion 
and temperature are obtained from Newton's law and the first law 
of thermodynamics. For the particle velocities, we have:

$$ \rho_p {{\pi d^3} \over 6 } \cdot {{ d\vvec_p} \over {dt}} = \Dvec 
   + \rho_p {{\pi d^3} \over 6 } \gvec ~~.
                                                      \eqno(5.2.8)    $$

\noi
This implies that:

$$ {{ d\vvec_p} \over {dt}} = {{3 \rho} \over {4 \rho_p d}} \cdot c_d
                               | \vvec - \vvec_p | ( \vvec - \vvec_p ) 
                    + \gvec
                    = \alpha_v | \vvec - \vvec_p | ( \vvec - \vvec_p )
                    + \gvec  ~~,
                                                      \eqno(5.2.9)   $$

\noi
where $\alpha_v=3\rho c_d / (4 \rho_p d)$.
The particle positions are obtained from:

$$ {{ d\xvec_p} \over {dt}} = \vvec_p ~~.     \eqno(5.2.10)    $$

\noi
The temperature change in a particle is given by:

$$ \rho_p c_{pp} {{\pi d^3} \over 6 } \cdot {{ dT_p} \over {dt}} = Q ~~,
                                                      \eqno(5.2.11) $$

\noi
which may be expressed as:

$$ {{ dT_p} \over {dt}} = {{3 k}\over{2 c_{pp} \rho_p d^2}} \cdot Nu \cdot
                          ( T - T_p )
                        = \alpha_T ( T - T_p ) ~~,    \eqno(5.2.12) $$

\noi
with $\alpha_T=3 k/(2 c_{pp} \rho_p d^2)$.
Equations (5.2.9, 5.2.10, 5.2.12) may be formulated as a system 
of Ordinary Differential Equations (ODEs) of the form:

$$ {{d\uvec_p} \over {dt}} = \rvec(\uvec_p, \xvec, \uvec_f) ~~,
                                                     \eqno(5.2.13)  $$

\noi
where $\uvec_p, \xvec, \uvec_f$ denote the particle unknowns, the
position of the particle and the fluid unknowns at the position of
the particle.

\subsection{Equations Describing the Motion of Diluted Viral Loads}
\label{viralmotion}
Viral loads may be obtained directly from the particles in the
flowfield. An alternative for small, diluted particles that are
floating in air is the use of a transport equation of the form:

$$ c_{,t} + \vvec \cdot \Grd c = 
       \nabla \cdot d_c \nabla c + s_c ~~, 
                                                     \eqno(5.3.1) $$

\noi
Here $c, d_c, s_c$ denote pathogen concentration, the diffusivity
and the source terms (due to exhalation or inhalation).

\subsection{Numerical Integration of the Motion of the Air}
\label{numintnavto}
The last six decades have seen a large number of schemes that may be
used to solve numerically the incompressible Navier-Stokes
equations given by Eqns.(5.1.1-5.1.3). In the present case, the 
following design criteria were implemented:
\begin{itemize}
\item[-] Spatial discretization using {\bf unstructured grids} 
(in order to allow for arbitrary geometries and adaptive refinement);
\item[-] Spatial approximation of unknowns with 
{\bf simple linear finite elements} (in order to have a simple 
input/output and code structure);
\item[-] Edge-based data structures (for reduced access to memory and
indirect addressing);
\item[-] Temporal approximation using {\bf implicit integration of viscous
terms and pressure} (the interesting scales are the ones associated with
advection);
\item[-] Temporal approximation using {\bf explicit, high-order 
integration of advective terms};
\item[-] {\bf Low-storage, iterative solvers} for the resulting systems of
equations (in order to solve large 3-D problems); and
\item[-] Steady results that are {\bf independent from the timestep} chosen
(in order to have confidence in convergence studies).
\end{itemize} 
\noindent
The resulting discretization in time is given by the following projection
scheme \cite{Loh06,Loh08}:
\begin{itemize}
\item[-] \ub{Advective-Diffusive Prediction}:
$\vvec^n, p^n \rightarrow \vvec^{*}$

$$ \svec' = - \Grd p^n + \rho \gvec 
          + \beta \rho \gvec (T^n - T_0) + \svec_v ~~,
\eqno(5.4.1)
$$

$$
\vvec^i = \vvec^n + \alpha^i \gamma \dt \left(
 - \vvec^{i-1} \cdot \Grd \vvec^{i-1} 
 + \nabla \cdot \mu \nabla \vvec^{i-1} + \svec' \right)  ~~; ~~i=1,k-1~~;
\eqno(5.4.2)
$$

$$
 \left[ { 1 \over \dt} - \theta \nabla \cdot \mu \nabla \right]
   \left( \vvec^{k} - \vvec^n \right)
 + \vvec^{k-1} \cdot \Grd \vvec^{k-1} = 
   \nabla \cdot \mu \nabla \vvec^{k-1} + \svec' ~~.  \eqno(5.4.3)
$$

\ms \noi
\item[-] \ub{Pressure Correction}: $p^n \rightarrow p^{n+1}$

$$
 \Div \vvec^{n+1} = 0                       ~~; \eqno(5.4.4)
$$
$$
 {{ \vvec^{n+1} - \vvec^{*} }\over \dt} + \Grd ( p^{n+1} - p^n )
   = 0                                      ~~; \eqno(5.4.5)
$$

\noi
\item[ ] which results in

$$
 \nabla^2 ( p^{n+1} - p^n ) = {{\Div \vvec^{*} }\over \dt} ~~;
\eqno(5.4.6)
$$

\ms \noi
\item[-] \ub{Velocity Correction}:
$\vvec^{*} \rightarrow \vvec^{n+1}$

$$
 \vvec^{n+1} = \vvec^{*} - \dt \Grd ( p^{n+1} - p^n ) ~~.
\eqno(5.4.7)
$$
\end{itemize}

\noi
$\theta$ denotes the implicitness-factor for the viscous
terms ($\theta=1$: 1st order, fully implicit, $\theta=0.5$: 2nd order,
Crank-Nicholson). 
$\alpha^i$ are the standard low-storage Runge-Kutta coefficients
$\alpha^i=1/(k+1-i)$. The $k-1$ stages of Eqn.(5.4.2) may be seen as a 
predictor (or replacement)
of $\vvec^n$ by $\vvec^{k-1}$. The original right-hand side has not been
modified, so that at steady-state $\vvec^n=\vvec^{k-1}$, preserving the
requirement that the steady-state be independent of the timestep $\dt$.
The factor $\gamma$ denotes the local ratio of the stability limit for
explicit timestepping for the viscous terms versus the timestep chosen.
Given that the advective and viscous timestep limits are proportional to:

$$ \dt_a \approx {h \over {|\vvec|}} ~~;~~
   \dt_v \approx {{\rho h^2} \over \mu} ~~, \eqno(5.4.8)
$$

\noi
we immediately obtain

$$ \gamma = {{\dt_v} \over {\dt_a}}
    \approx {{\rho |\vvec| h }\over{\mu}} \approx Re_h  ~~,
\eqno(5.4.9)
$$

\noi
or, in its final form:

$$ \gamma = min(1,Re_h) ~~. \eqno(5.4.10) $$

\noi
In regions away from boundary layers, this factor is $O(1)$, implying 
that a high-order Runge-Kutta scheme is recovered. Conversely, for 
regions where $Re_h=O(0)$, the scheme reverts back to the usual
1-stage Crank-Nicholson scheme. 
Besides higher accuracy, an important benefit of explicit multistage 
advection schemes is the larger timestep one can employ. The increase in 
allowable timestep is roughly proportional to the number of stages used 
(and has been exploited extensively for compressible flow simulations 
\cite{Jam81}). 
Given that for an incompressible solver of the projection type 
given by Eqns.(5.4.1-5.4.7) most of the CPU time is spent solving the 
pressure-Poisson system Eqn.(5.4.6), the speedup
achieved is also roughly proportional to the number of stages used. \\
At steady state, $\vvec^{*}=\vvec^n=\vvec^{n+1}$ and the residuals of 
the pressure correction vanish,
implying that the result does not depend on the timestep $\dt$. \\
The spatial discretization of these equations is carried out via 
linear finite elements. The
resulting matrix system is re-written as an edge-based solver, allowing
the use of consistent numerical fluxes to stabilize the advection and
divergence operators \cite{Loh08}. \\
The energy (temperature) equation (Eqn.(5.1.3)) is integrated in a 
manner similar to the advective-diffusive prediction (Eqn(5.4.2)), 
i.e. with an explicit, high order Runge-Kutta scheme for the advective 
parts and an implicit, 2nd order Crank-Nicholson scheme for the 
conductivity.

\subsection{Numerical Integration of the Motion of Particles/Droplets}
\label{numintpast}
The equations describing the position, velocity and temperature of a
particle (Eqns.\ 5.2.9, 5.2.10, 5.2.12) may be formulated as a 
system of nonlinear Ordinary Differential Equations of the form:

$$ {{d\uvec_p} \over {dt}} = \rvec(\uvec_p, \xvec, \uvec_f) ~~.
                                                       \eqno(5.5.1) $$

\par \noi
They can be integrated numerically in a variety of ways. Due to its
speed, low memory requirements and simplicity, we have chosen
the following k-step low-storage Runge-Kutta procedure to integrate them:

$$ \uvec^{n+i}_p = \uvec^n_p + \alpha^i \Delta t \cdot
   \rvec(\uvec^{n+i-1}_p, \xvec^{n+i-1}, \uvec^{n+i-1}_f) ~~,
~~ i=1,k  ~~. \eqno(5.5.2) $$

\noi
For linear ODEs the choice

$$ \alpha^i= {1 \over {k+1-i}} ~~,~~ i=1,k  \eqno(5.5.3) $$

\noi
leads to a scheme that is $k$-th order accurate in time.
Note that in each step the location of the particle with respect to the
fluid mesh needs to be updated in order to obtain the proper values for
the fluid unknowns. The default number of stages used is $k=4$. This
would seem unnecessarily high, given that the flow solver is of
second-order accuracy, and that the particles are integrated separately
from the flow solver before the next (flow) timestep, i.e. in a staggered
manner. However, it was found that the 4-stage particle integration
preserves very well the motion in vortical structures and leads to less
`wall sliding' close to the boundaries of the domain \cite{Loh14}.
The stability/ accuracy of the particle integrator should not be a problem
as the particle motion will always be slower than the maximum wave speed
of the fluid (fluid velocity). \\
The transfer of forces and heat flux between the fluid and the particles
must be accomplished in a conservative way, i.e. whatever is added to the
fluid must be subtracted from the particles and vice-versa. The finite
element discretization of the fluid equations will lead to
a system of ODE's of the form:

$$ \Mmat \duvec = \rvec ~~,    \eqno(5.5.4)       $$

\noi
where $\Mmat, \duvec$ and $\rvec$ denote, respectively, the consistent 
mass matrix, increment of the unknowns vector and right-hand side vector. 
Given the `host element' of each particle, i.e. the fluid mesh element 
that contains the particle, the forces and heat transferred 
to $\rvec$ are added as follows:

$$ \rvec^i_D = \sum_{el~surr~i} N^i(\xvec_p) \Dvec_p ~~.  \eqno(5.5.5) $$

\noi
Here $N^i(\xvec_p)$ denotes the shape-function values of the host 
element for the point coordinates $\xvec_p$, and the sum extends
over all elements that surround node $i$. As the sum of all 
shape-function values is unity at every point:

$$ \sum N^i(\xvec) = 1 ~~\forall \xvec ~~,    \eqno(5.5.6)       $$

\noi
this procedure is strictly conservative. \\
From Eqns.\ 5.2.9, 5.2.10, 5.2.12) and their equivalent numerical 
integration via Eqn.(5.5.2), 
the change in momentum and energy for one particle is given by:

$$ \fvec_p =  \rho_p {{\pi d^3}\over 6} 
             {{\left( \vvec^{n+1}_p - \vvec^n_p \right)} \over {\Delta t}}
                                           ~~,    \eqno(5.5.7)       $$

$$ q_p =  \rho_p c_{pp} {{\pi d^3}\over 6} 
             {{\left( T^{n+1}_p - T^n_p \right)} \over {\Delta t}}
                                           ~~.    \eqno(5.5.8)       $$

\noi
These quantities are multiplied by the number of particles in
a packet in order to obtain the final values transmitted to the fluid.
Before going on, we summarize the basic steps required in order to update
the particles one timestep:
\begin{itemize}
\item[-] Initialize Fluid Source-Terms: $\rvec=0$
\item[-] {\code DO}: For Each Particle:
\item[ ] - {\code DO}: For Each Runge-Kutta Stage:
\item[ ] ~~~- Find Host Element of Particle: {\code IELEM}, $N^i(\xvec)$
\item[ ] ~~~- Obtain Fluid Variables Required
\item[ ] ~~~- Update Particle: Velocities, Position, Temperature, ...
\item[-] - {\code ENDDO}
\item[ ] - Transfer Loads to Element Nodes
\item[-] {\code ENDDO}
\end{itemize}

\subsubsection{Particle Parcels}
\label{partparc}
For a large number of very small particles, it becomes impossible to
carry every individual particle in a simulation. The solution is to:
\begin{itemize}
\item[a)] Agglomerate the particles into so-called packets of $N_p$ 
particles;
\item[b)] Integrate the governing equations for one individual particle; 
and
\item[c)] Transfer back to the fluid $N_p$ times the effect of one 
particle.
\end{itemize}
Beyond a reasonable number of particles per element (typically $> 8$),
this procedure produces accurate results without any deterioration in
physical fidelity.

\subsubsection{Other Particle Numerics}
In order to achieve a robust particle integrator, a number of additional
precautions and algorithms need to be implemented. The most
important of these are:
\begin{itemize}
\item[-] Agglomeration/Subdivision of Particle Parcels:
As the fluid mesh may be adaptively refined and coarsened in time,
or the particle traverses elements of different sizes,
it may be important to adapt the parcel concentrations as well. 
This is necessary to ensure that there is sufficient parcel 
representation in each element and yet, that there are not too many 
parcels as to constitute an inefficient use of CPU and memory. 
\item[-] Limiting During Particle Updates:
As the particles are integrated independently from the flow solver, it is
not difficult to envision situations where for the extreme cases of
very light or very heavy particles physically meaningless or unstable
results may be obtained.
In order to prevent this, the changes in 
particle velocities and temperatures are limited in order not to exceed 
the differences in velocities and temperature between the particles and 
the fluid \cite{Loh14}. 
\item[-] Particle Contact/Merging:
In some situations, particles may collide or merge in a certain region
of space.
\item[- ] Particle Tracking:
A common feature of all particle-grid applications is that the particles
do not move far between timesteps. This makes physical sense:
if a particle jumped ten gridpoints during one timestep, it would have
no chance to exchange information with the points along the way, leading
to serious errors. Therefore, the assumption that the new host elements
of the particles are in the vicinity of the current ones is a valid one.
For this reason, the most efficient way to search for the new host elements
is via the vectorized neighbour-to-neighbour algorithm described in
\cite{Loh90,Loh08}. 
\end{itemize}

\subsection{Immersed Body Techniques}
The information required from CCD codes consists of
the pedestrians in the flowfield, i.e. their position, velocity,
temperature, as well inhalation and exhalation. As the CCD 
codes describe the pedestrians as points,
circles or ellipses, a way has to be found to transform this data
into 3-D objects. Two possibilities have been pursued here:
\begin{itemize}
\item{a)} Transform each pedestrian into a set of (overlapping) spheres that
approximate the body with maximum fidelity with the minimum amount of
spheres;
\item{b)} Transform each pedestrian into a set of tetrahedra that
approximate the body with maximum fidelity with the minimum amount of
tetrahedra.
\end{itemize}
The reason for choosing spheres or tetrahedra is that one can perform
the required interpolation/ information transfer much faster than with
other methods. \\
In order to `impose' on the flow the presence of a pedestrian the
immersed boundary methodology is used. The key idea is to prescribe at
every CFD point covered by a pedestrian the velocity and temperature
of the pedestrian. For the CFD code, this translates into an extra
set of boundary conditions that vary in time and space as the
pedestrians move. This is by now a mature technology. Fast search
techniques as well as extensions to higher order boundary conditions
may be found in \cite{Loh08,Loh08b}. Nevertheless, as the pedestrians
potentially change location every timestep, the search for and the
imposition of new boundary conditions can add a considerable amount
of CPU as compared to `flow-only' runs.

\section{Modeling of Pedestrian Motion}
\label{modpedmotion}
The modeling of pedestrian motion has been the focus of research and
development for more than two decades. If one is only interested in
average quantities (average density, velocity), continuum models
\cite{Hug03} are an option. For problems requiring more realism,
approaches that model each individual are required \cite{Tha07}.
Among these, discrete space models (such as cellular automata
\cite{Blu98,Blu02,Tek00,Dij02,Scha02,Kes02,Klu03,Cou05,Lan06}), 
force-based models (such as
the social force model \cite{Hel95,Hel02,Qui03,Lak05,Loh10})
and agent-based techniques 
\cite{Pel06,Sud07,Guy09,Guy10,Vig10,Tor12,Cur12} have been 
explored extensively.
Together with insights from psychology and neuroscience (e.g.
\cite{Vis95,Tor12}) it has become clear that any
pedestrian motion algorithm that attempts to model reality should be
able to mirror the following empirically known facts and behaviours:
\begin{itemize}
\item[-] Newton's laws of motion apply to humans as well:
from one instant to another, we can only move within certain bounds
of acceleration, velocity and space;
\item[-] Contact between individuals occurs for high densities; these
forces have to be taken into account;
\item[-] Humans have a mental map and plan on how they desire to move
globally (e.g. first go here, then there, etc.); 
\item[-] Human motion is therefore governed by strategic 
(long term, long distance), tactical (medium
term, medium distance) and operational (immediate) decisions;
\item[-] In even moderately crowded situations ($O(1~p/m^2)$),
humans have a visual horizon of $O(2.5-5.0 m)$, and a perception range
of 120 degrees; thus, the influence of other humans beyond these
thresholds is minimal;
\item[-] Humans have a `personal comfort zone'; it is dependent on
culture and varies from individual to individual, but it cannot
be ignored;
\item[-] Humans walk comfortably at roughly 2 paces per second
(frequency: $\nu=2~Hz$); they are able to change the frequency for short
periods of time, but will return to $2~Hz$ whenever possible.
\end{itemize}
We remark that many of the important and groundbreaking work cited
previously took place within the gaming/visualization
community, where the emphasis is on `looking right'. Here, the aim
is to answer civil engineering or safety questions such as maximum
capacity, egress times under emergency, or comfort. Therefore,
comparisons with experiments and actual data are seen as essential
\cite{Loh10,Ise14a,Ise14b}.

\subsection{The PEDFLOW Model}
The PEDFLOW model \cite{Loh10} incorporates these requirements
as follows: individuals move according to Newton's laws of motion;
they follow (via will forces) `global movement targets'; at the local
movement level, the motion also considers the presence of other
individuals or obstacles via avoidance forces (also a type of will
force) and, if applicable, contact forces.
Newton's laws:

$$ m {{d \vvec}\over {dt}} = \fvec ~~,~~ 
     {{d \xvec}\over {dt}} = \vvec ~~, \eqno(6.1.1) $$

\noi
where $m, \vvec, \xvec, \fvec, t$ denote, respectively, mass,
velocity, position, force and time, are integrated in time
using a 2nd order explicit timestepping technique.
The main modeling effort is centered on $\fvec$.
In the present case the forces are separated into internal
(or will) forces [I would like to move here or there] and
external forces [I have been hit by another pedestrian or an
obstacle]. For the sake of completeness, we briefly
review the main forces used. For more
information, as well as verification and validation studies, see
\cite{Loh10,Ise14a,Ise14b,Zha14,Ise16,Ise16a,Ise16b}.

\subsubsection{Will Force}
Given a desired velocity $\vvec_d$ and
the current velocity $\vvec$, this force will be of the form

$$ \fvec_{will} = g_w \left( \vvec_d - \vvec \right) ~~. 
                                                    \eqno(6.1.1.1) $$

\noi
The modelling aspect is included in the function $g_w$, which, in the
non-linear case, may itself be a function of $\vvec_d - \vvec$.
Suppose $g_w$ is constant, and that only the will force is acting.
Furthermore, consider a pedestrian at rest. In this case, we have:

$$  m {{d \vvec} \over {dt}} = g_w \left( \vvec_d - \vvec \right) 
                        ~~,~~ \vvec(0)=0 ~~,        \eqno(6.1.1.2) $$

\noi
which implies:

$$ \vvec = \vvec_d \left( 1 - e^{-\alpha t} \right) ~~,~~
  \alpha = {{g_w}\over{m}} = {1 \over t_r} ~~, \eqno(6.1.1.3) $$

\noi
and

$$ {{d \vvec} \over {dt}}(t=0) = \alpha \vvec_d 
                               = {\vvec_d \over t_r} ~~. 
                                                \eqno(6.1.1.4) $$

\noi
One can see that the crucial parameter here is the `relaxation time'
$t_r$ which governs the initial acceleration and `time to desired
velocity'. Typical values are $v_d=1.35~m/sec$ and $t_r=O(0.5~sec)$.
The `relaxation time' $t_r$ is clearly dependent on the fitness of the
individual, the current state of stress, desire to reach a
goal, climate, signals, noise, etc. Slim, strong individuals will have
low values for $t_r$, whereas fat or weak individuals will have high
values for $t_r$.
Furthermore, dividing by the mass of the individual allows all other
forces (obstacle and pedestrian collision avoidance, contact, etc.) to
be scaled by the `relaxation time' as well, simplifying the modeling
effort considerably.
\par \noi
The direction of the desired velocity

$$ \svec = {{\vvec_d} \over {|\vvec_d|}}         \eqno(6.1.1.5) $$

\noi
will depend on the type of pedestrian and the cases considered. A
single individual will have as its goal a desired position
$\xvec_d(t_d)$ that he would like to reach at a certain time $t_d$.
If there are no time constraints, $t_d$ is simply set to a large number.
Given the current position $\xvec$, the direction of the velocity is
given by

$$ \svec = {{ \xvec_d(t_d) - \xvec } \over {|\xvec_d(t_d) - \xvec|} } ~~,
                                                    \eqno(6.1.1.6) $$

\noi
where $\xvec_d(t_d)$ denotes the desired position (location, goal) of
the pedestrian at the desired time of arrival $t_d$. 
For members of groups, the goal is always to stay close to the leader.
Thus, $\xvec_g(t_g)$ becomes the position of the leader. In the case of
an evacuation simulation, the direction is given by the gradient of the
perceived time to exit $\tau_e$ to the closest perceived exit:

$$ \svec = {{ \nabla \tau_e } \over {|\nabla \tau_e}| } ~~. 
                                                    \eqno(6.1.1.7) $$

\noi
The magnitude of the desired velocity $|\vvec_d|$ depends on the fitness
of the individual, and the motivation/urgency to reach a certain place
at a certain time. Pedestrians typically stroll leisurely at
$0.6-0.8~m/sec$, walk at $0.8-1.0~m/sec$, jog at $1.0-3.0~m/sec$,
and run at $3.0-10.0~m/sec$.

\subsubsection{Pedestrian Avoidance Forces}
The desire to avoid collisions with other individuals is modeled by
first checking if a collision will occur. If so, forces are applied
in the direction normal and tangential to the intended motion. The
forces are of the form:

$$ f_i = f_{max}/( 1 + \rho^p) ~~;~~ 
   \rho  = | x_i - x_j |  /  r_i ~~, \eqno(6.1.2.7) $$

\noi
where $x_i$, $x_j$ denote the positions of individuals $i,j$,
$r_i$ the radius of individual $i$, and $f_{max}=O(4)f_{max}(will)$.
Note that the forces weaken with increasing non-dimensional distance
$\rho$. For years we have used $p=2$, but, obviously, this can (and
probably will) be a matter of debate and speculation (perhaps a
future experimental campaign will settle this issue).
In the far range, the forces are mainly orthogonal to the direction of
intended motion: humans tend to move slightly sideways without decelerating.
In the close range, the forces are also in the direction of intended
motion, in order to model the slowdown required to avoid a collision.

\subsubsection{Wall Avoidance Forces}
Any pedestrian modeling software requires a way to input geographical
information such as walls, entrances, stairs, escalators, etc. In
the present case, this is accomplished via a triangulation (the so-called
background mesh). A distance to walls map (i.e. a function $d_w(x)$
is constructed using fast marching techniques on unstructured grids),
and this allows to define a wall avoidance force as follows:

$$ \fvec = - f_{max}{ 1 \over { 1 + ({d_w \over r})^2 }} 
         \cdot \nabla d_w ~~,~~p=2             \eqno(6.1.3.1) $$

\noi
Note that $|d_w|=1$. The default for the maximum wall avoidance force
is $f_{max}=O(8)f_{max}(will)$. The desire to be far/close to a wall
also depends on cultural background.

\subsubsection{Contact Forces}
When contact occurs, the forces can increase markedly. Unlike will
forces, contact forces are symmetric. Defining

$$ \rho_{ij} = |x_i - x_j|/(r_i + r_j) ~~,   \eqno(6.1.3.2) $$

\noi
these forces are modeled as follows:

$$ \rho_{ij} < 1:  f = - [ f_{max} /( 1 +  \rho_{ij}^p)]~~; ~~  p=2
                                                \eqno(6.1.3.3a)  $$
$$ \rho_{ij} > 1:  f = - [2f_{max} /( 1 +  \rho_{ij}^p)]~~; ~~  p=2 
                                                \eqno(6.1.3.3b) $$

\noi
and $f_{max}=O(8)f_{max}(will)$.

\subsubsection{Motion Inhibition}
A key requirement for humans to move is the ability to put one foot
in front of the other. This requires space. Given the comfortable
walking frequency of $\nu=2~Hz$, one is able to limit the comfortable
walking velocity by computing the distance to nearest neighbors and
seeing which one of these is the most `inhibiting'.

\subsubsection{Psychological Factors}
The present pedestrian motion model also incorporates a number of 
psychological factors that,
among the many tried over the years, have emerged as important for
realistic simulations. Among these, we mention:
\begin{itemize}
\item[-] Determination/Pushiness: it is an everyday experience that in
crowds, some people exhibit a more polite behavior than others. This
is modeled in PEDFLOW by reducing the collision avoidance forces of
more determined or `pushier' individuals. Defining a determination or
pushiness parameter $p$, the avoidance forces are reduced by $(1-p)$.
\item[-] Comfort zone: in some cultures (northern Europeans are a
good example) pedestrians want to remain at some minimum distance from
contacting others. This comfort zone is an input parameter in PEDFLOW,
and is added to the radii of the pedestrians when computing collisions
avoidance and pre-contact forces.
\item[-] Right/Left Avoidance and Overtaking: in many western countries
pedestrians tend to avoid incoming pedestrians by stepping towards
their right, and overtake others on the left. However, this is not the
norm everywhere, and one has to account for it.
\end{itemize}

\subsection{Numerical Integration of the Motion of Pedestrians}
\label{numintpeds}
The equations describing the position and velocity of a pedestrian
may be formulated as a system of nonlinear Ordinary Differential 
Equations of the form:

$$ {{d\uvec_p} \over {dt}} = \rvec(\uvec_p, \xvec, \uvec_f) ~~.
                                                   \eqno(6.2.1)    $$

\noi
These ODEs are integrated with explicit Runge-Kutta schemes,
typically of order 2. \\
The geographic information required, such as terrain data
(inclination, soil/water, escalators, obstacles, etc.), climate data
(temperature, humidity, sun/rain, visibility), signs,
the location and accessibility of guidance personnel, as well as doors,
entrances and emergency exits is stored
in a so-called background grid consisting of triangular elements. This
background grid is used to define the geometry of the problem.
At every instance, a pedestrian will be located in one of the elements
of the background grid. Given this `host element' the geographic
data, stored at the nodes of the background grid, is interpolated
linearly to the pedestrian.
The closest distance to a wall $\delta_w$ or exit(s) for any given
point of the background grid evaluated via a fast ($O(N\ln(N))$)
nearest neighbour/heap list technique (\cite{Loh08,Loh10}).
For cases with visual or smoke impediments, the closest distance to
exit(s) is recomputed every few seconds of simulation time.

\subsection{Linkage to CFD Codes}
The information required from CFD codes such as FEFLO consists of
the spatial distribution of temperature, smoke, other toxic or
movement impairing substances in space, as well as pathogen 
distribution. This information is
interpolated to the (topologically 2-D) background mesh at every
timestep in order to calculate properly the visibility/ reachability
of exits, routing possibilities, smoke, toxic substance or
pathogen inhalation, and any other flowfield variable required 
by the pedestrians. As the tetrahedral grid used for the CFD code 
and the triangular
background grid of the CCD code do not change in time, the interpolation
coefficients need to be computed just once at the beginning of the
coupled run. While the transfer of information from CFD to CCD is
voluminous, it is very fast, adding an insignificant amount
to the total run-times.

\section{Coupling Methodology}
The coupling methodology used is shown in Figure~3. The CFD code computes
the flowfield, providing such information as temperature, smoke, toxic
substance and pathogen concentration, and any other flow quantity 
that may affect the movement of pedestrians. These variables are then 
interpolated to the
position where the pedestrians are, and are used with all other
pertinent information (e.g. will-forces, targets, exits, signs, etc.)
to update the position, velocity, inhalation of smoke, toxic
substances or pathogens, state of exhaustion or intoxication, and any other
pertinent quantity that is evaluated for the pedestrians. The position,
velocity and temperature of the pedestrians, together with information
such as sneezing or exhaling air, is then transferred to the
CFD code and used to modify and update the boundary conditions of the
flowfield in the regions where pedestrians are present.

\begin{figure}
	\centering
	\includegraphics[width=10.0cm]{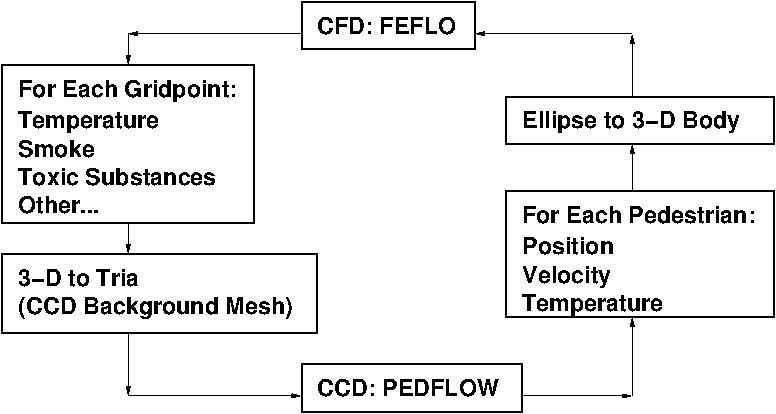}
	\caption{Coupling CFD and CCD Codes}
\end{figure}

\noi
Of the many possible coupling options
(see e.g. \cite{Bun06,Sch08,Ceb05}),
we have implemented the simplest one: loose coupling with
sequential timestepping (\cite{Loh95,Loh11}). This is justified, as
the timesteps of both the flow and pedestrian solvers are very small,
so that possible coupling errors are negligible.

\section{Examples}

\subsection{Corridor With Pedestrians}
This example considers the corridor of 10.0~m x 2.0~m x 2.5~m
shown in Figure~4.
Both entry and exit sides have two doors each of size 0.8~m x 2.0~m.
For climatisation, 4 entry vanes and 1 exit vane are placed
in the ceiling. The vertical air velocity for the entry vanes
as set to $v_z=0.2~m/sec$, while the horizontal velocity was set
as increasing proportional to the distance of the center of the
vane to a maximum of $v_r=0.4~m/sec$.
Two streams of pedestrians enter and exit 
through the doors over time. As stated before, the pedestrian
dynamics code, which only `sees' the
floorplan of the problem at hand, computes position, velocity and
orientation of the pedestrians, and then produces a tetrahedral mesh
for each pedestrian and the sends this information to the transfer
library. This information is then passed on to the flow solver, 
which treats the pedestrians via the immersed body approach 
in the flowfield. Should there be smoke, pollutants or pathogens
in the flowfield, this information is passed
back to the pedestrian dynamics code, which interpolates it at the 
height of pedestrians in order to update inhalation, intoxication 
and infection information.

\begin{figure}
	\centering
	\includegraphics[width=10.0cm]{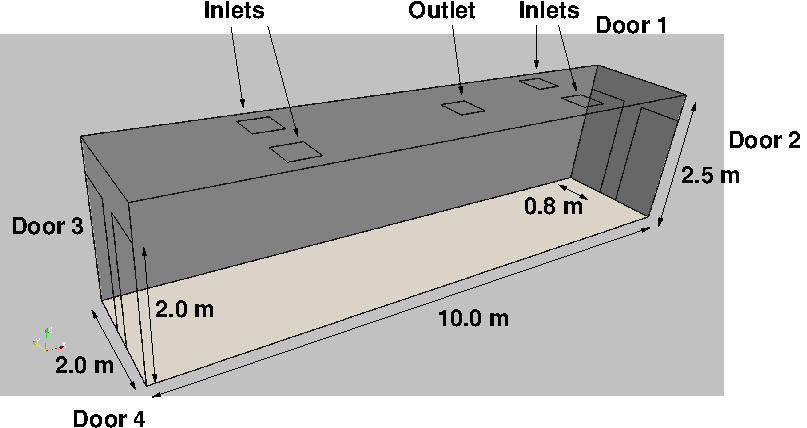}
	\caption{Simple Channel: Geometry and Boundary Conditions}
\end{figure}

This simulation was run in 3 phases:
\begin{itemize}
\item[Phase 1:] Every code is run independently until `things settle down',
i.e until the flow reaches a quasi-steady state and the pedestrian
streams have formed; for the present case this took 20~sec of physical
time;
\item[Phase 2:] The restart files from Phase~1 are taken, and the
run continues in fully coupled mode, until `things settle down';
for the present case this took 20~sec of physical
time;
\item[Phase 3:] The restart files from Phase~2 are taken, and the
run continues in fully coupled mode imposing the boundary conditions
for a sneezing event.
\end{itemize}
In order to see the effects of pedestrians, three simulations were
carried out: a)~Two pedestrians streams in counterflow mode;
b)~Two pedestrians streams in parallel flow mode; and
c)~No pedestrians. \\
Figures~5-11, 12-18 and 19-25 show the solutions obtained for 
these different modes. \\
In the cases shown different temporal scales appear:
\begin{itemize}
\item[-] The fast, ballistic drop of the larger ($d=1~mm$) particles,
occurring in the range of $O(1)~sec$;
\item[-] The slower drop of particles of diameter $d=O(0.1)~mm$,
occurring in the range of $O(10)~sec$; and
\item[-] The transport of the even smaller particles through the air,
occurring in the range of $O(100)~sec$.
\end{itemize}
We have attempted to show these phases in the results, and for this
reason the results are not displayed at equal time intervals.
Unless otherwise noted, the particles have been colored according to
the {\bf logarithm} of the diameter, with red colors representing the
largest and blue the smallest particles. \\
Note the very large differences in the flowfield with and without
pedestrians. The main reason for this difference is the discrepancy
in velocities: humans walk at approximately $v=1.2~m/sec$, while
the perception of discomfort due to air motion being at 
around $v-0.3~m/sec$,
implying that in most of the volume of any built environment where
humans reside, these lower velocities will be encountered. The
different velocities between walking pedestrians, and in particular
counterflows as the one shown, lead to large-scale turbulent mixing,
enhancing the spread of pathogens emanating from infected victims.

\begin{figure}
	\centering
	\includegraphics[width=10.0cm]{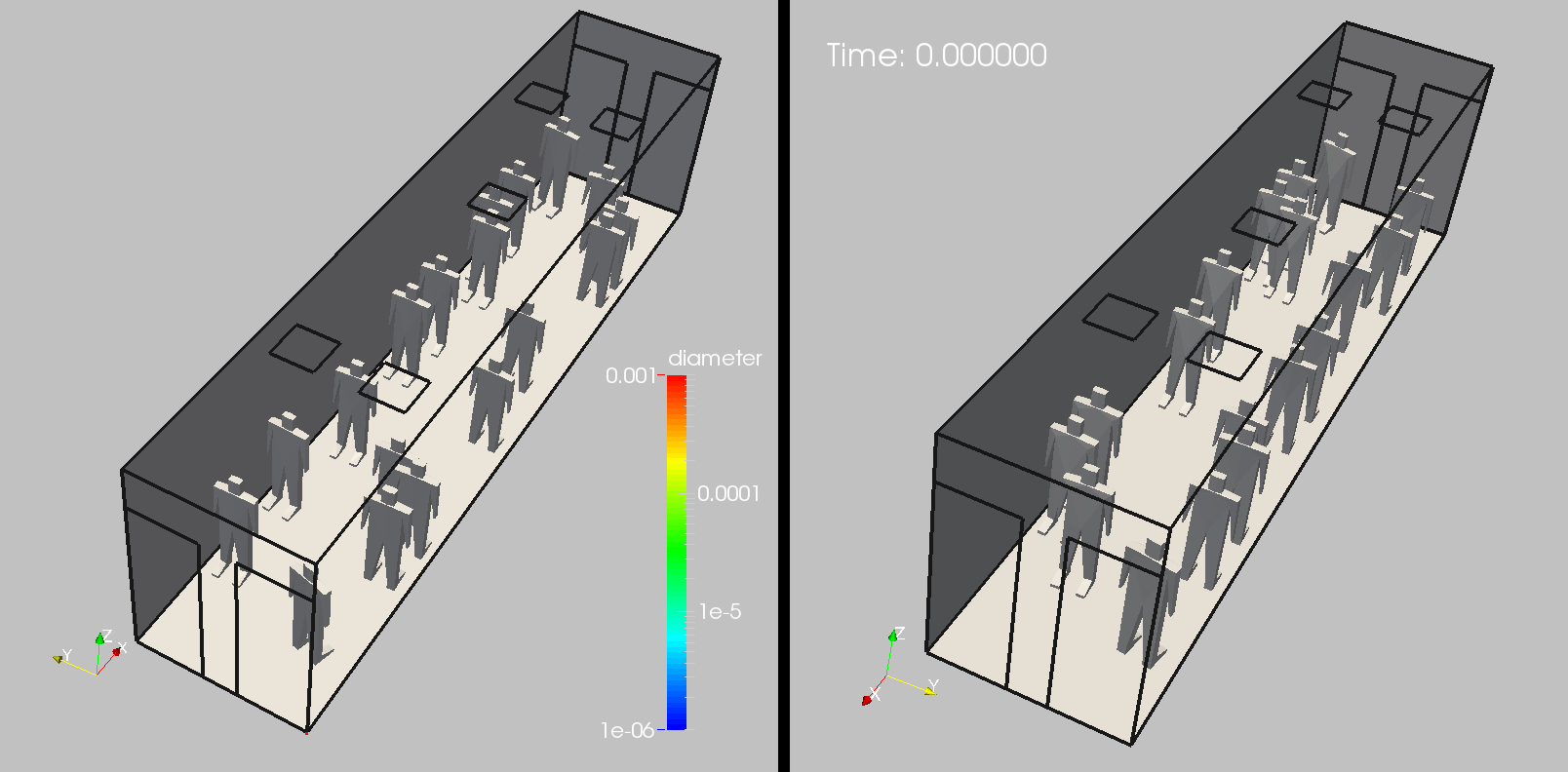}
	\vskip 10pt	
	\includegraphics[width=10.0cm]{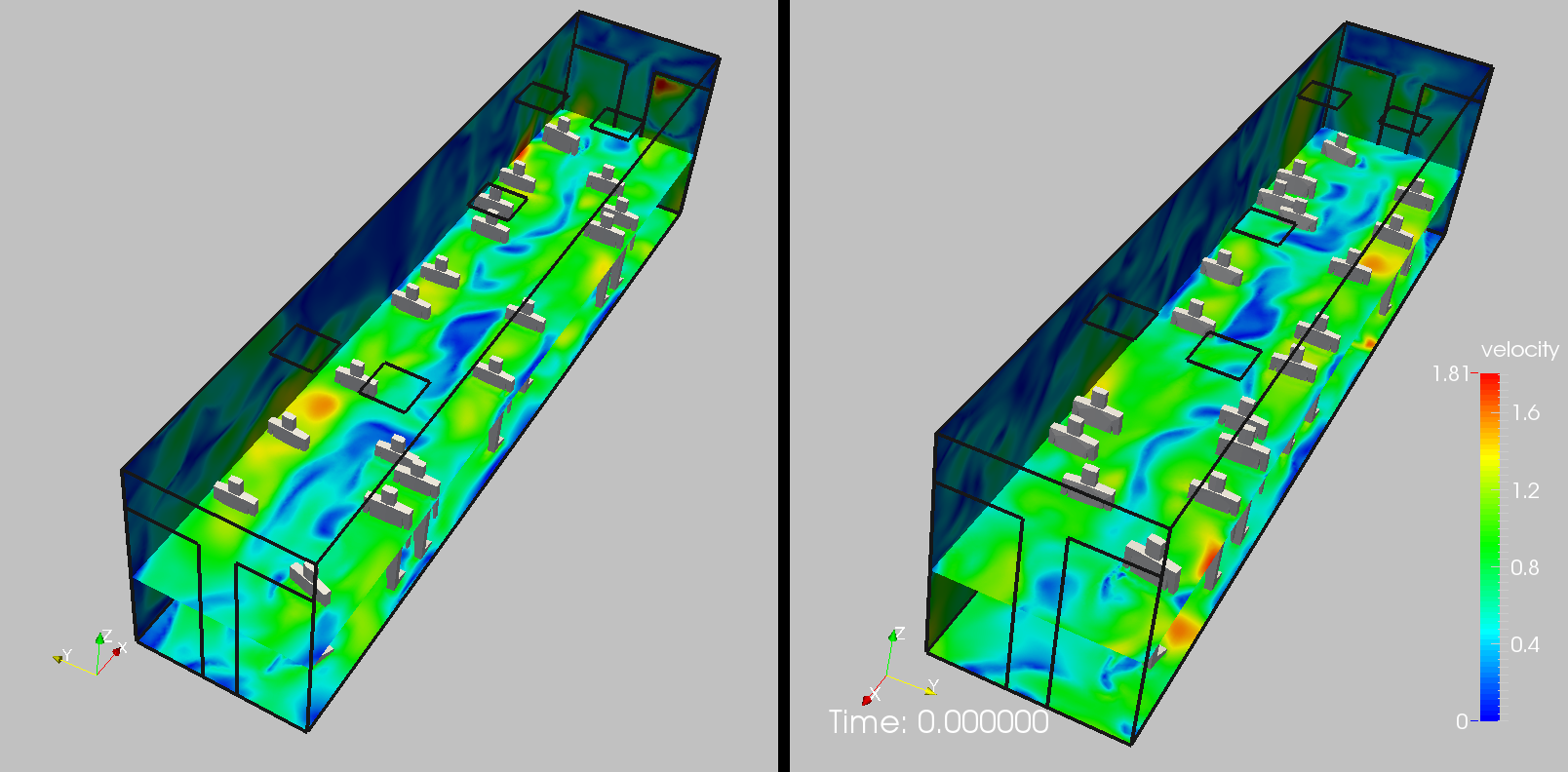}
	\vskip 10pt	
	\includegraphics[width=10.0cm]{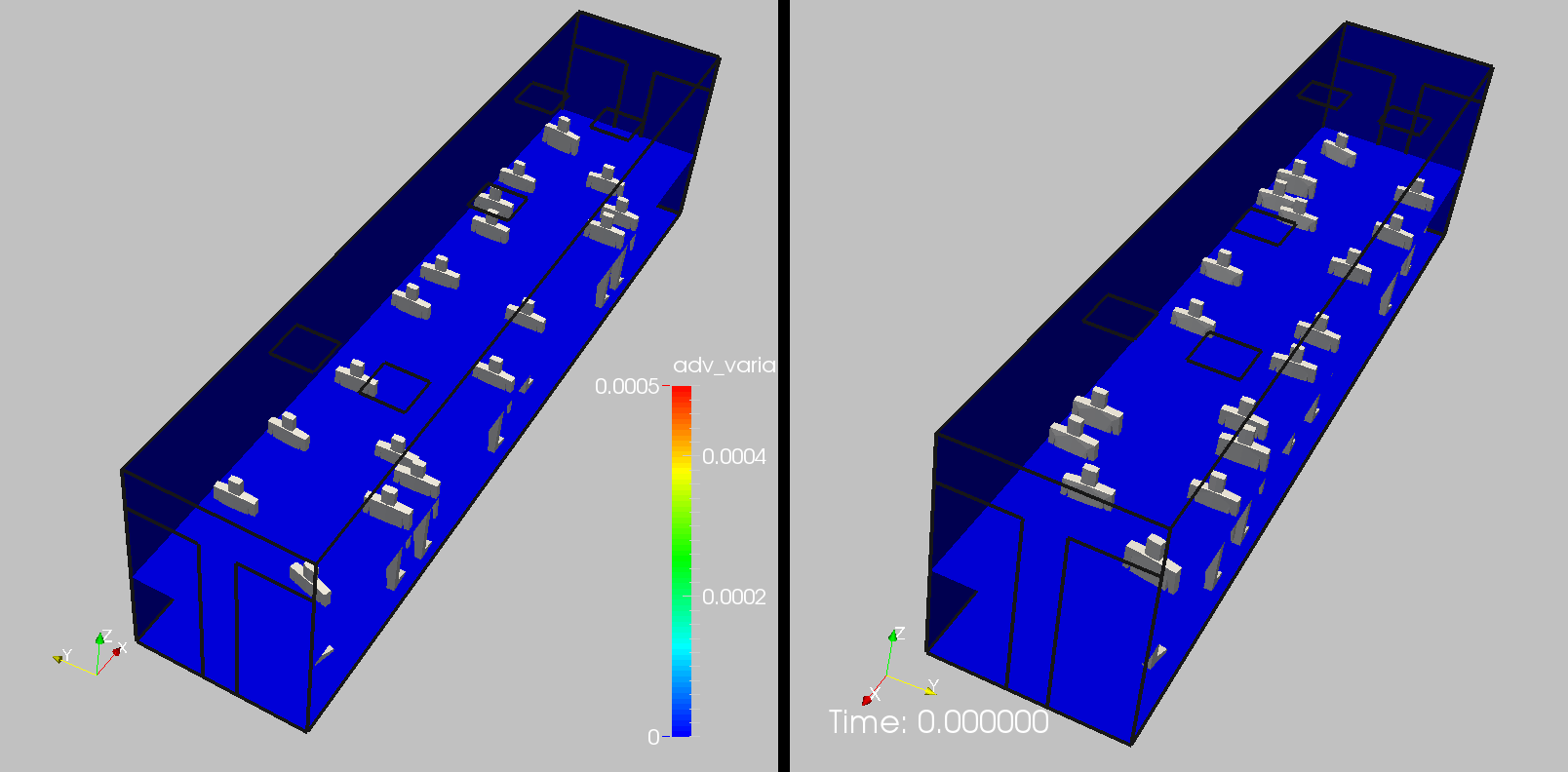}
	\caption{Counterflow Movement: Solution at $t=0.00~sec$}
\end{figure}

\begin{figure}
	\centering
	\includegraphics[width=10.0cm]{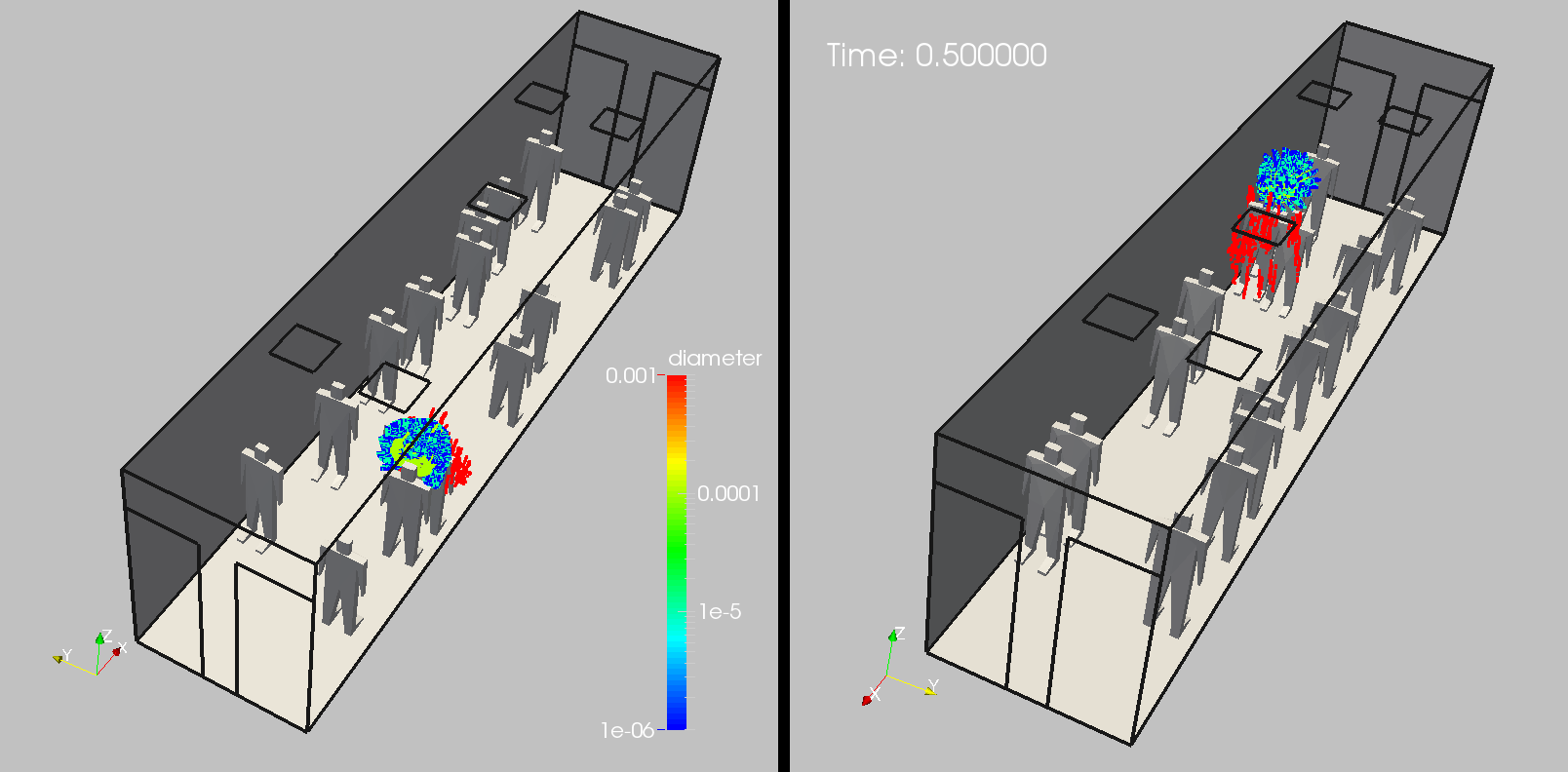}
	\vskip 10pt	
	\includegraphics[width=10.0cm]{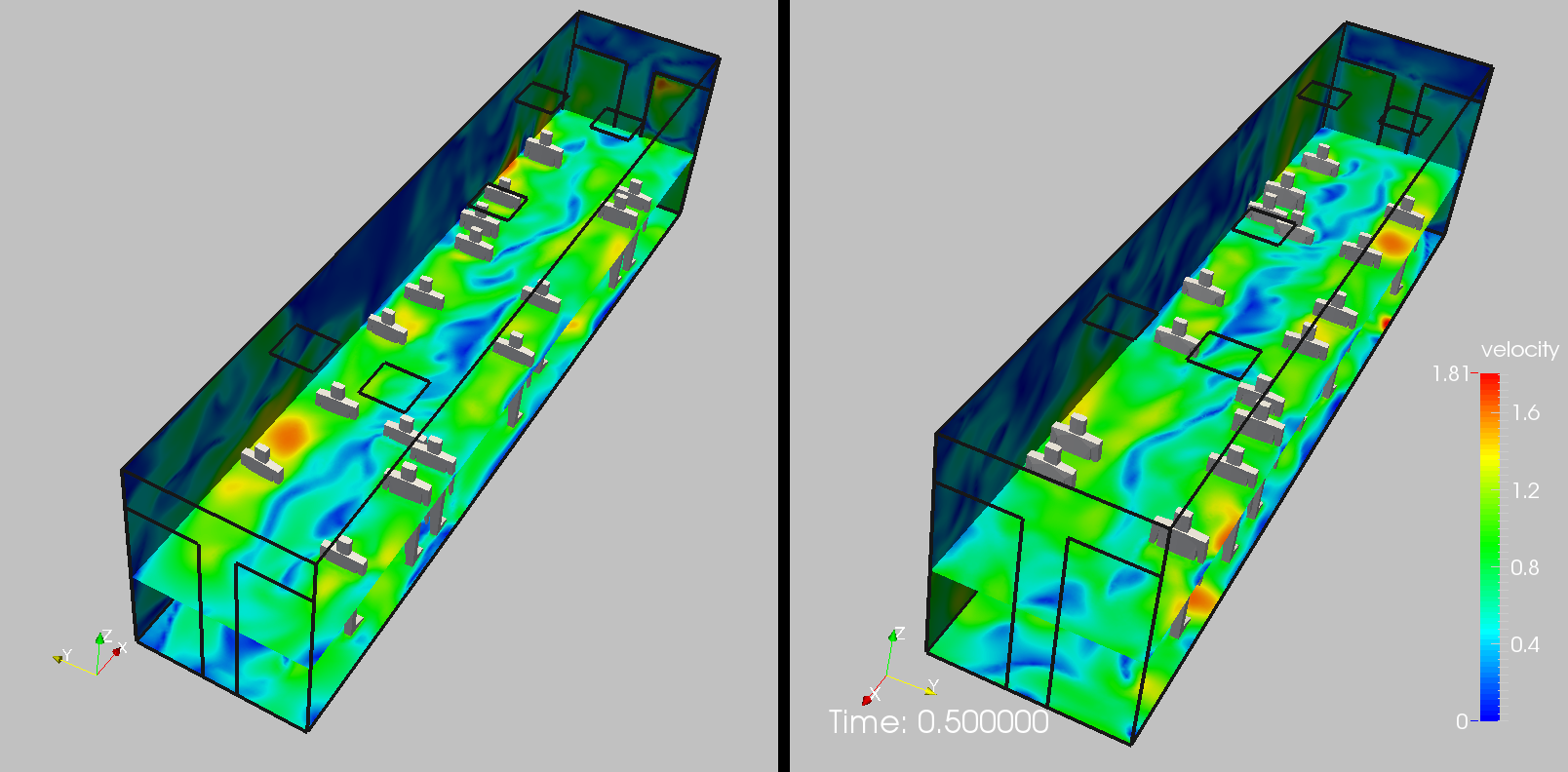}
	\vskip 10pt	
	\includegraphics[width=10.0cm]{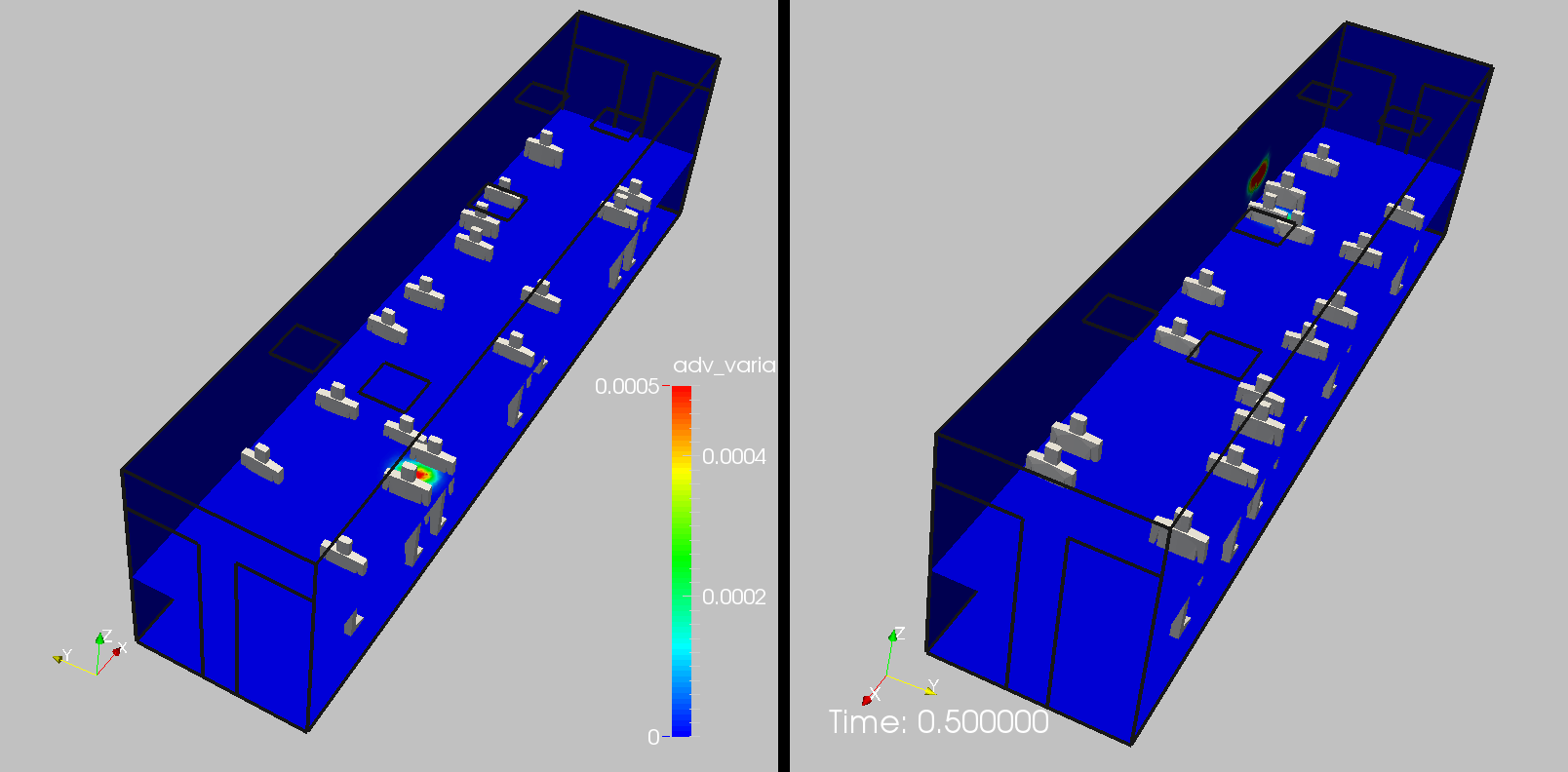}
	\caption{Counterflow Movement: Solution at $t=0.50~sec$}
\end{figure}

\begin{figure}
	\centering
	\includegraphics[width=10.0cm]{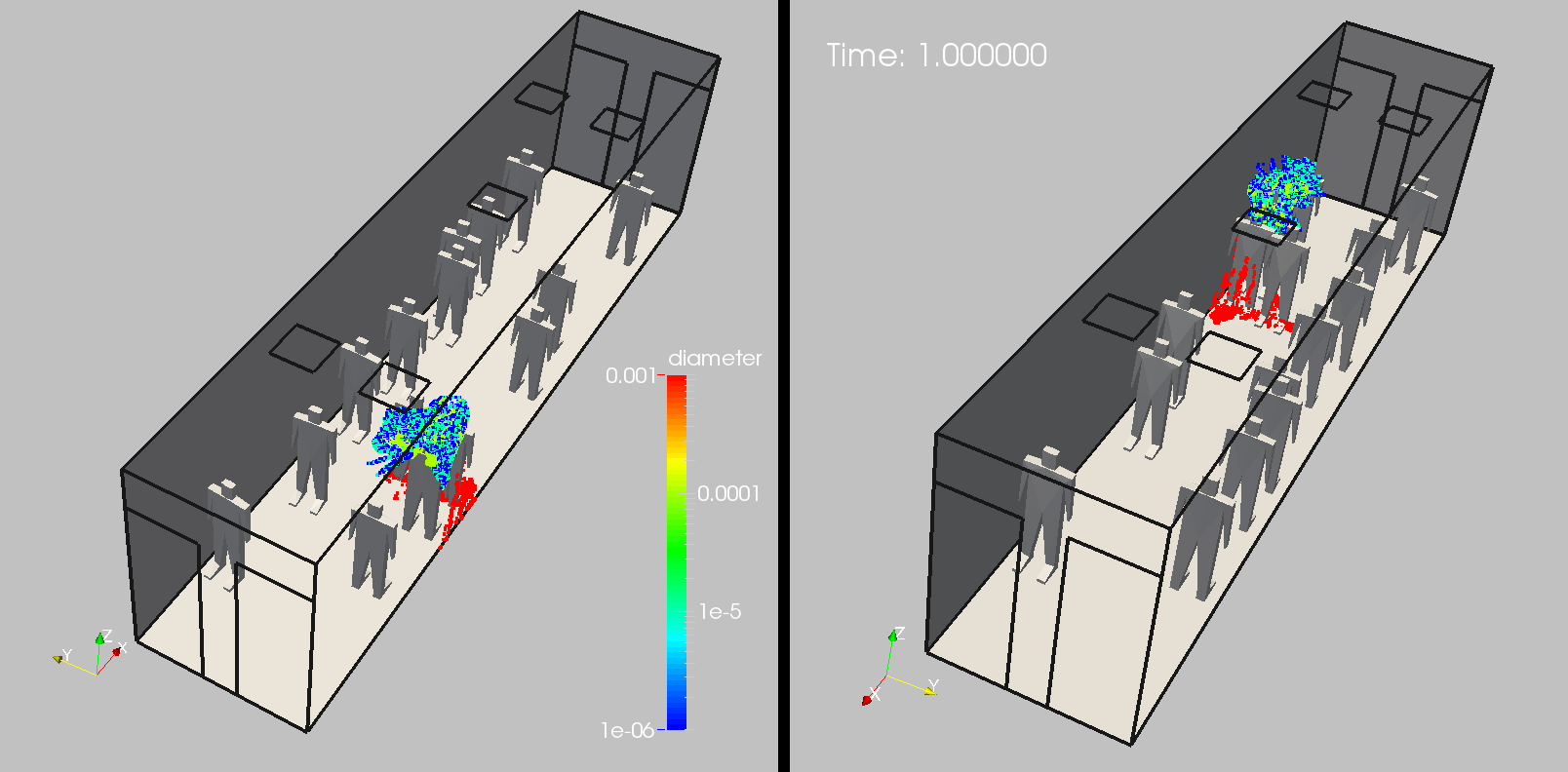}
	\vskip 10pt	
	\includegraphics[width=10.0cm]{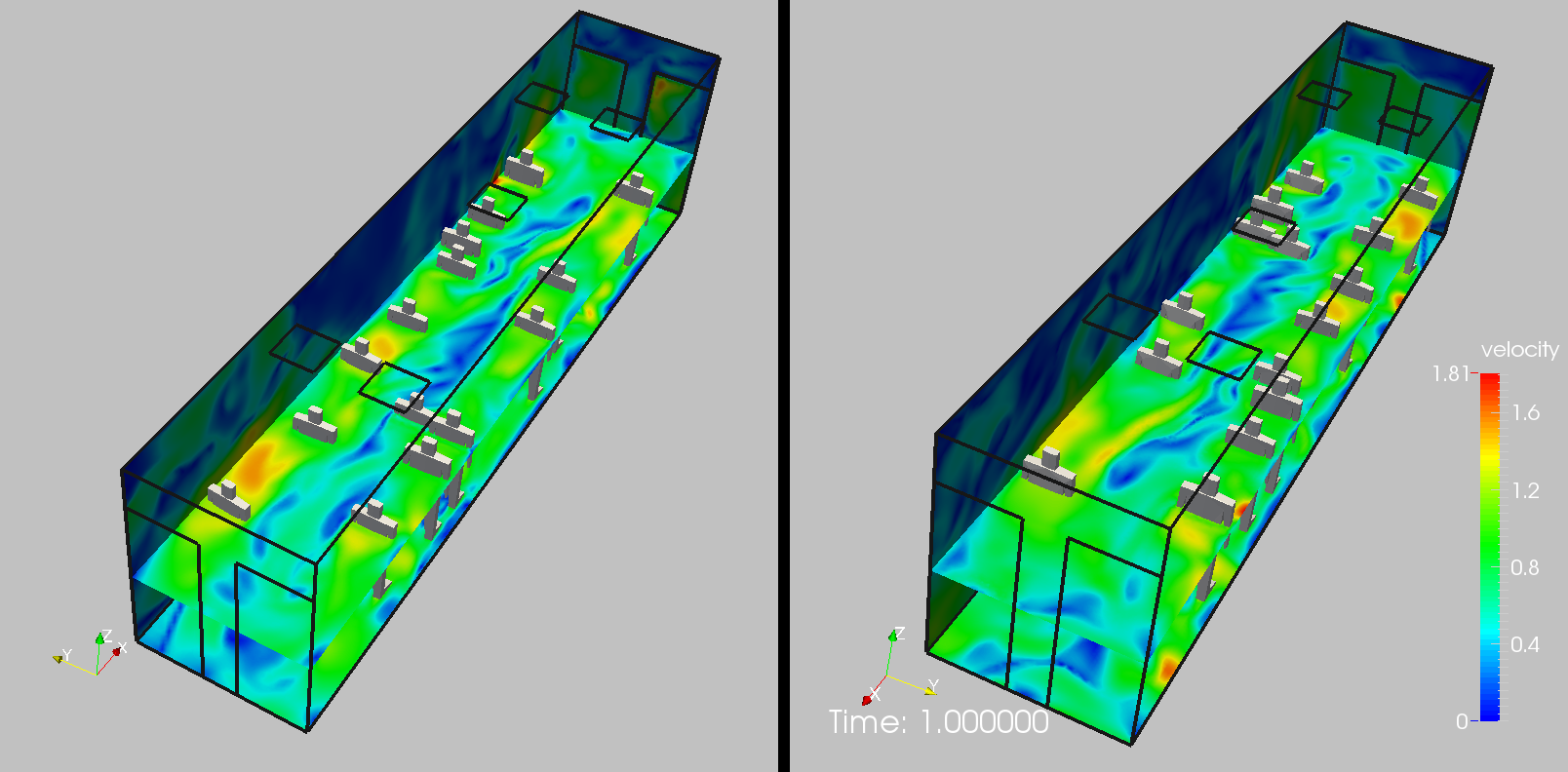}
	\vskip 10pt	
	\includegraphics[width=10.0cm]{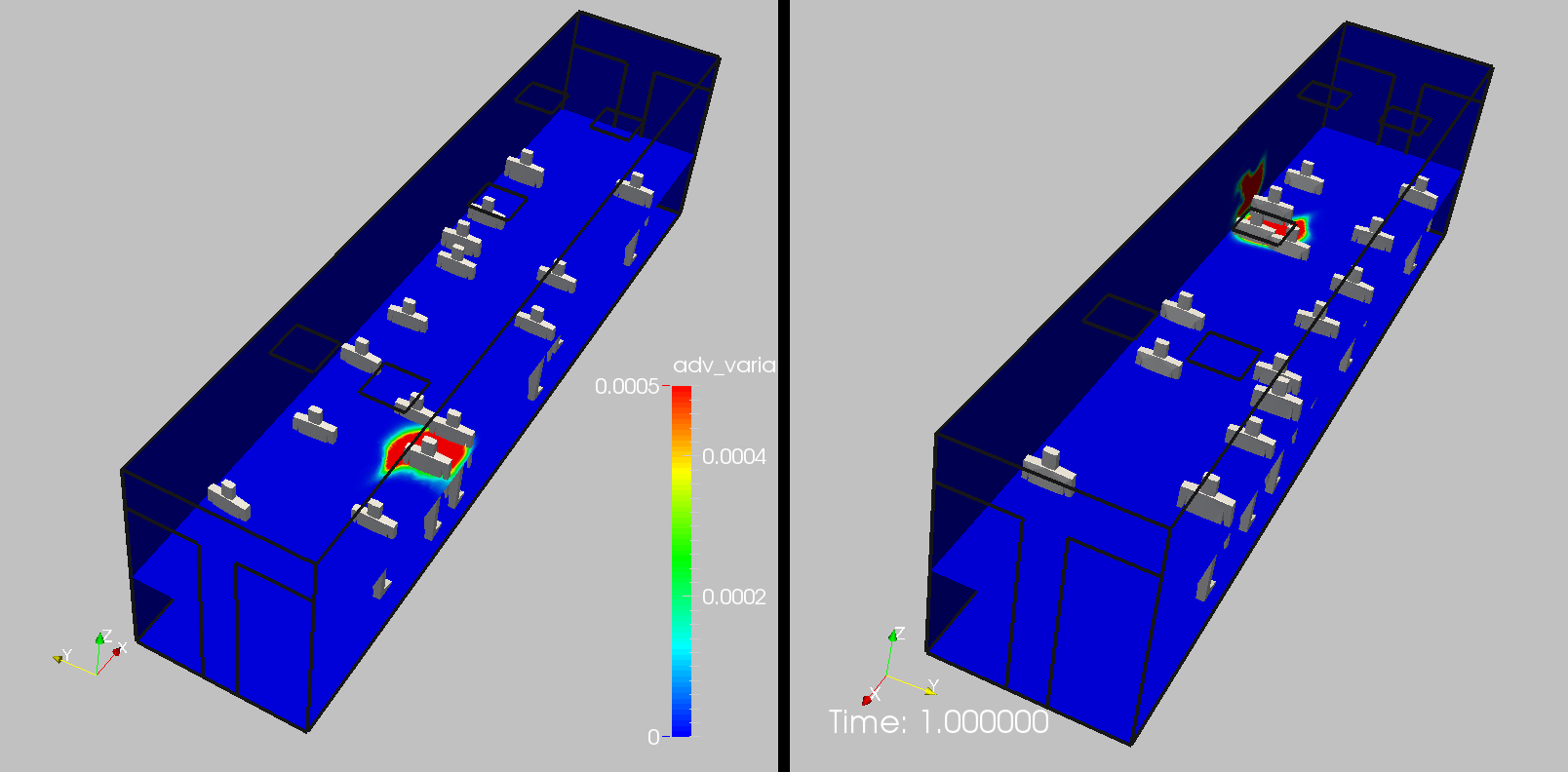}
	\caption{Counterflow Movement: Solution at $t=1.00~sec$}
\end{figure}

\begin{figure}
	\centering
	\includegraphics[width=10.0cm]{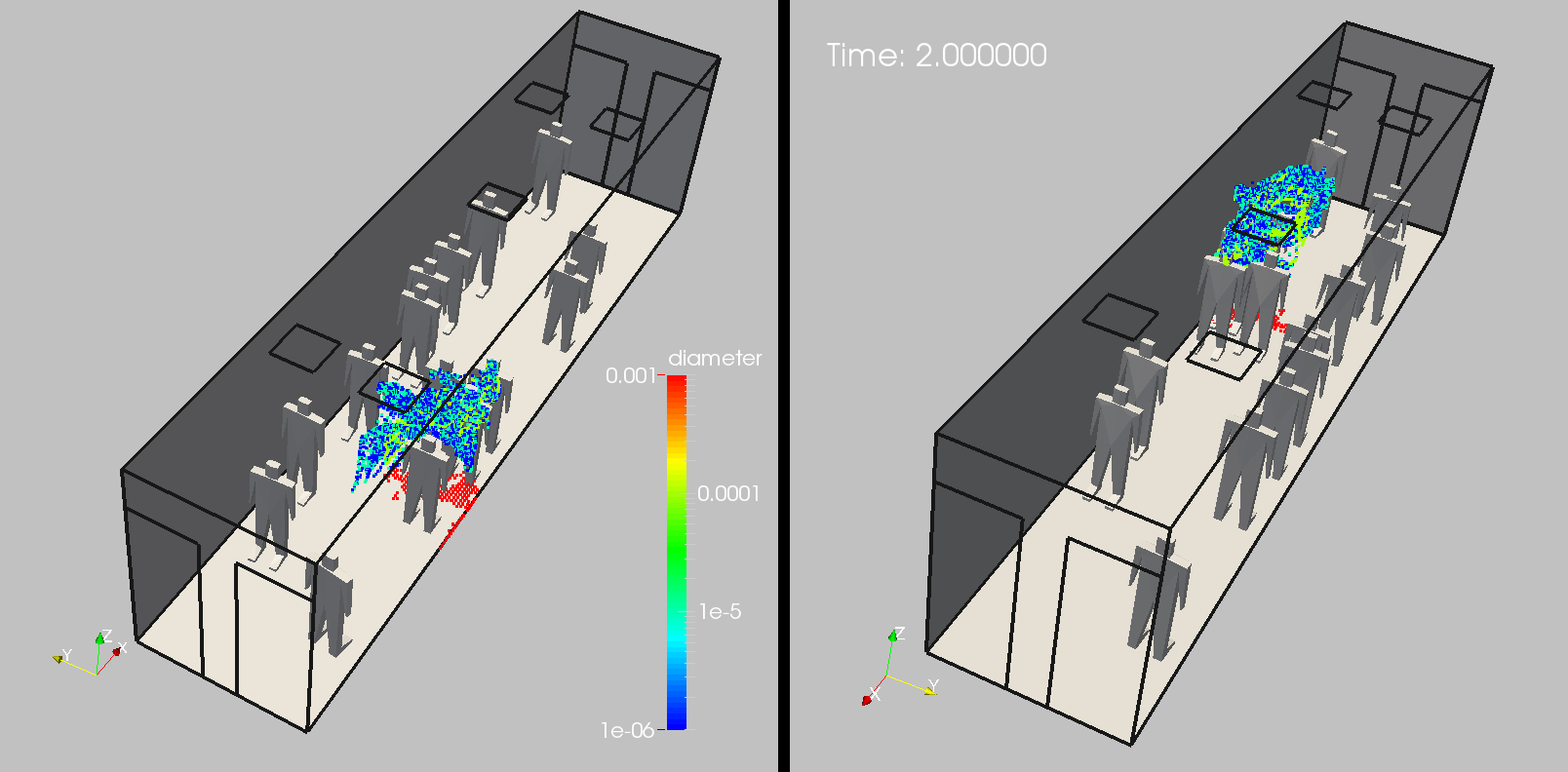}
	\vskip 10pt	
	\includegraphics[width=10.0cm]{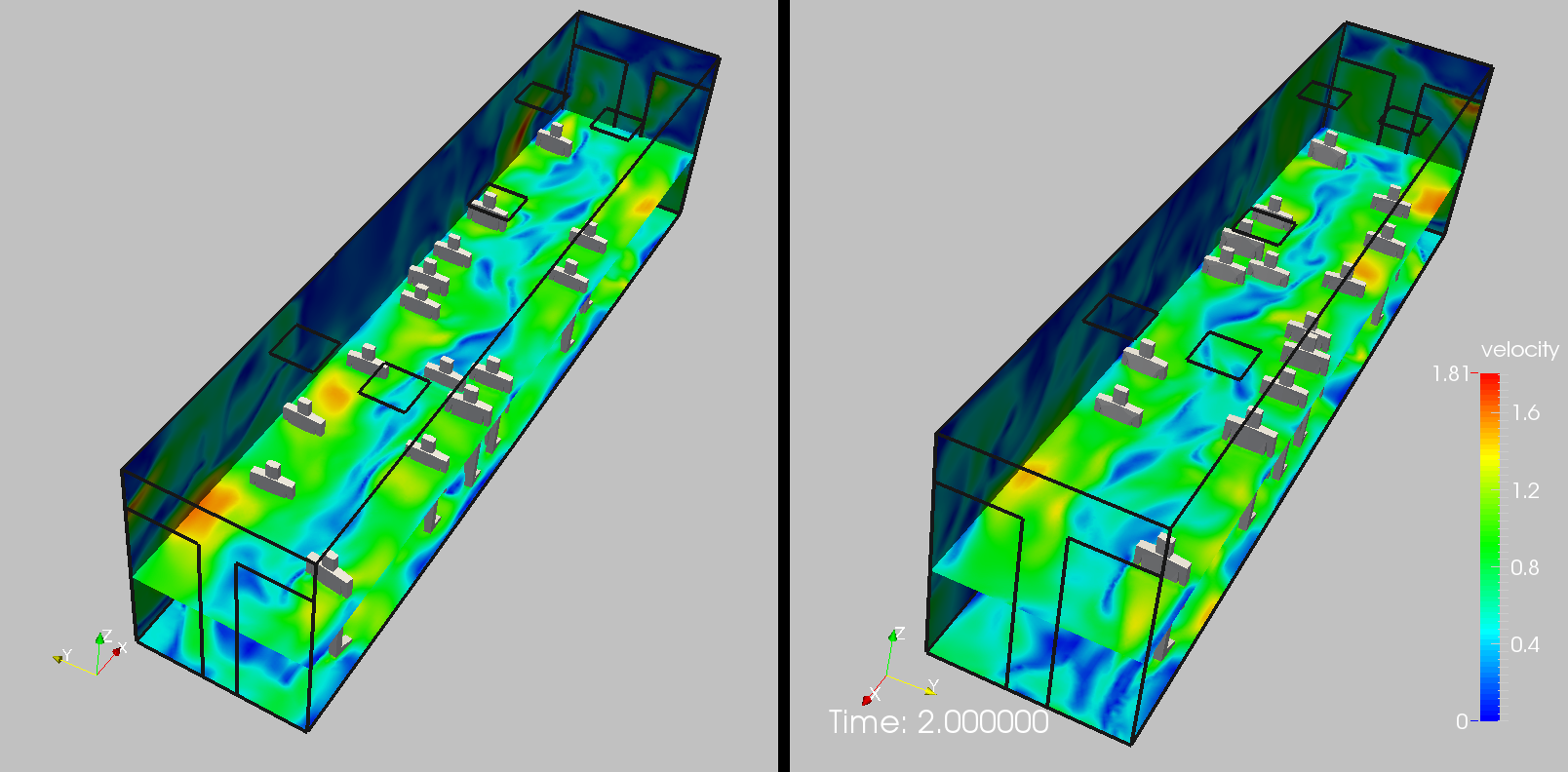}
	\vskip 10pt	
	\includegraphics[width=10.0cm]{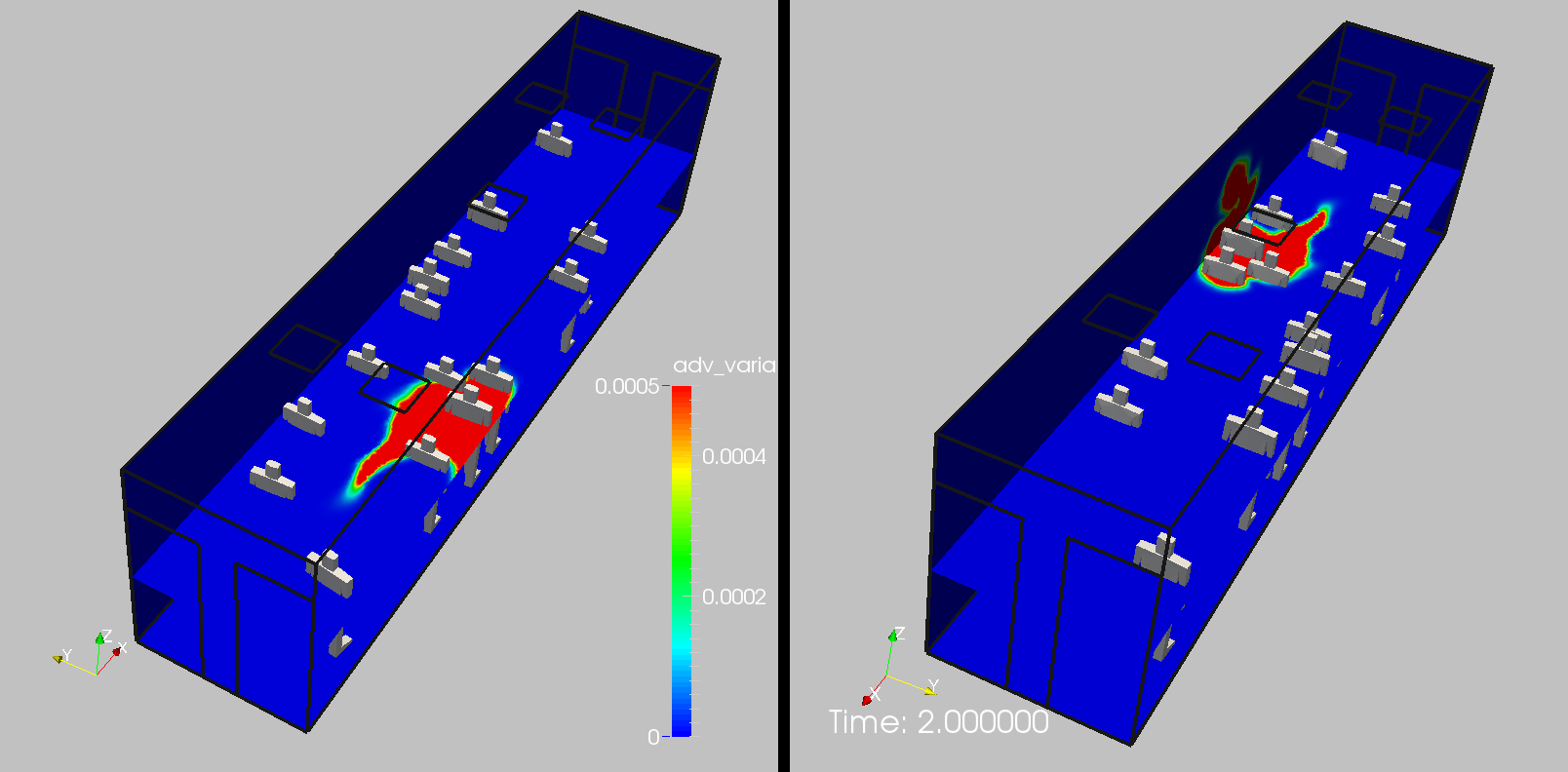}
	\caption{Counterflow Movement: Solution at $t=2.00~sec$}
\end{figure}

\begin{figure}
	\centering
	\includegraphics[width=10.0cm]{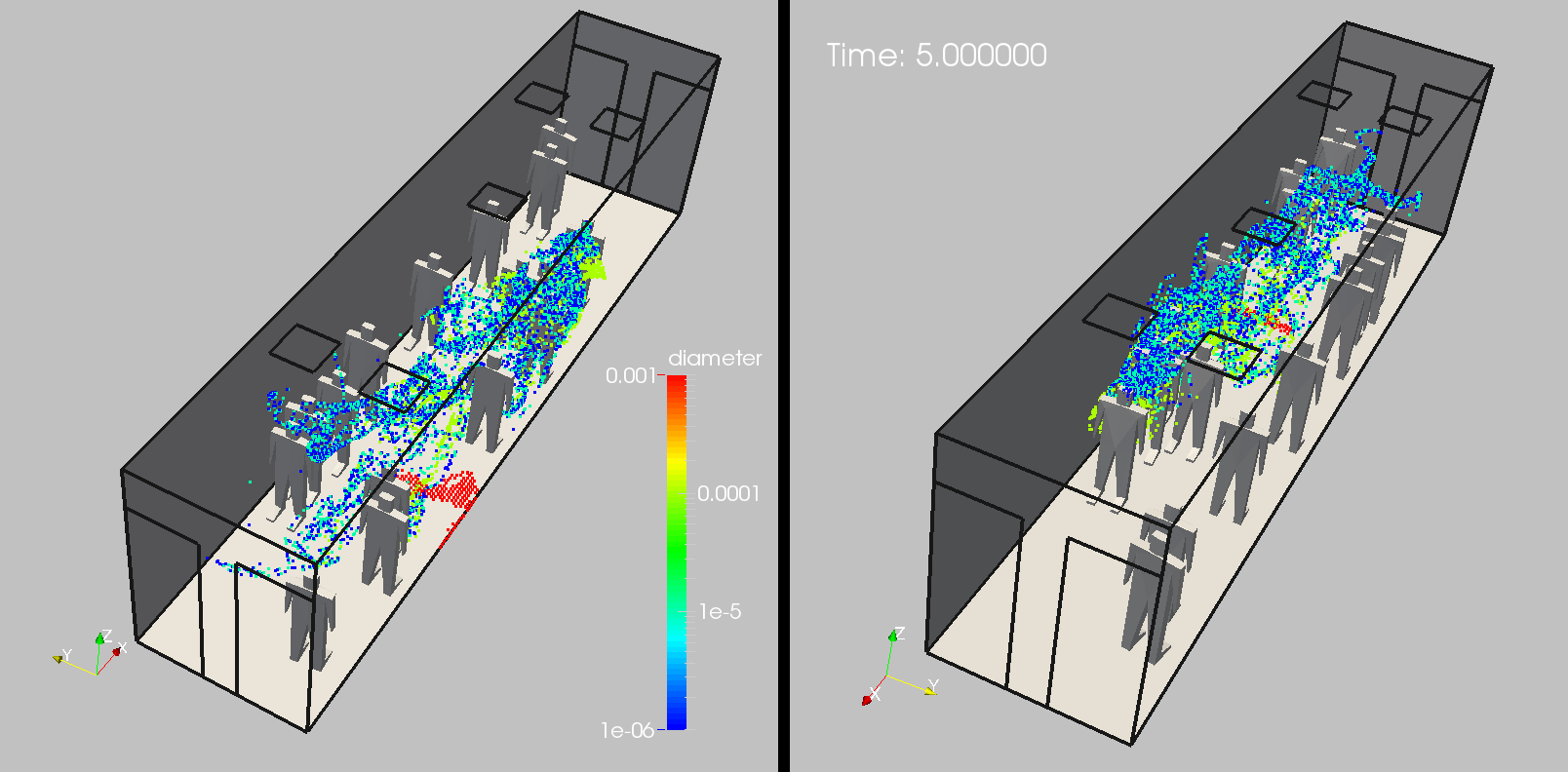}
	\vskip 10pt	
	\includegraphics[width=10.0cm]{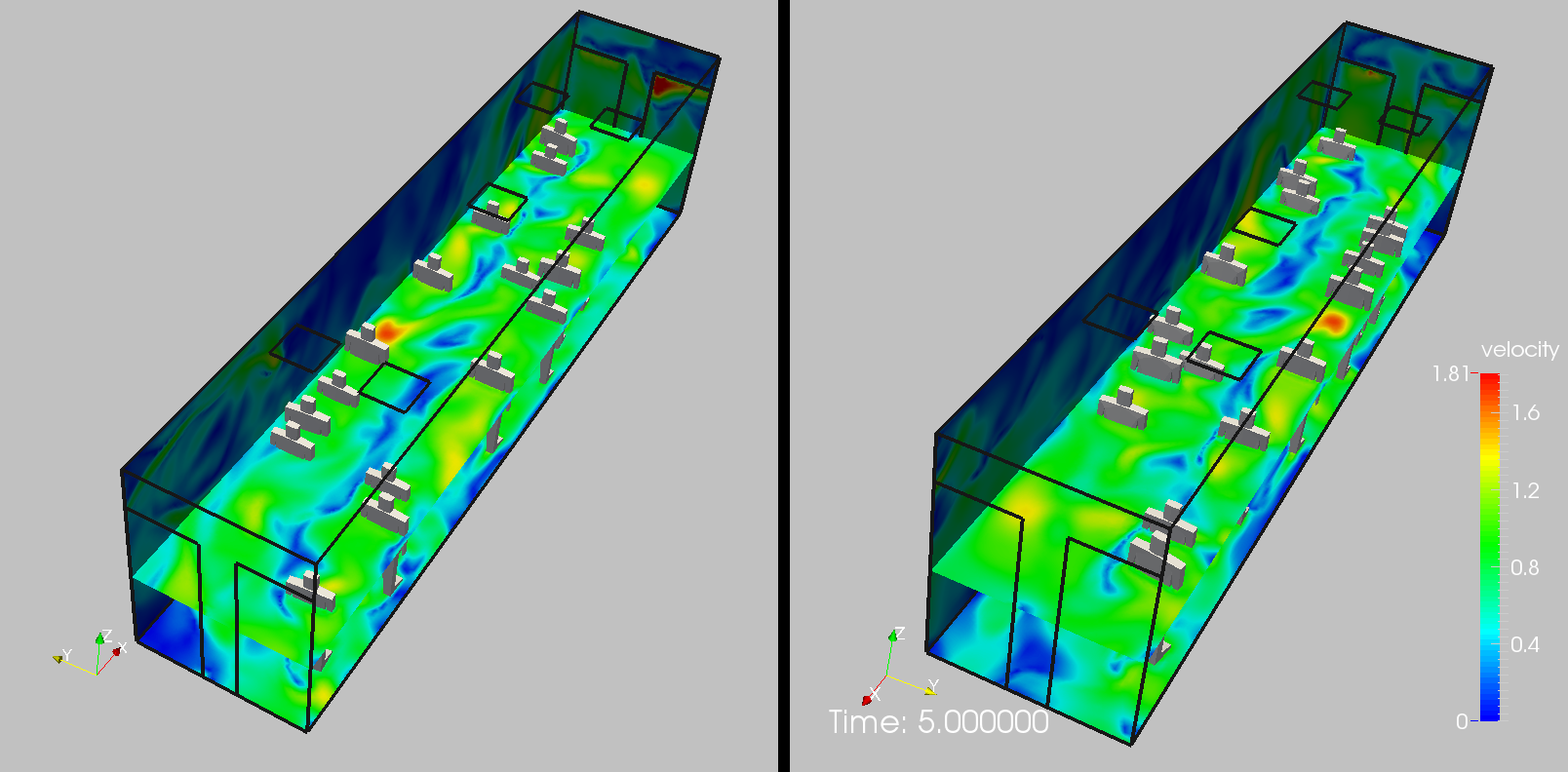}
	\vskip 10pt	
	\includegraphics[width=10.0cm]{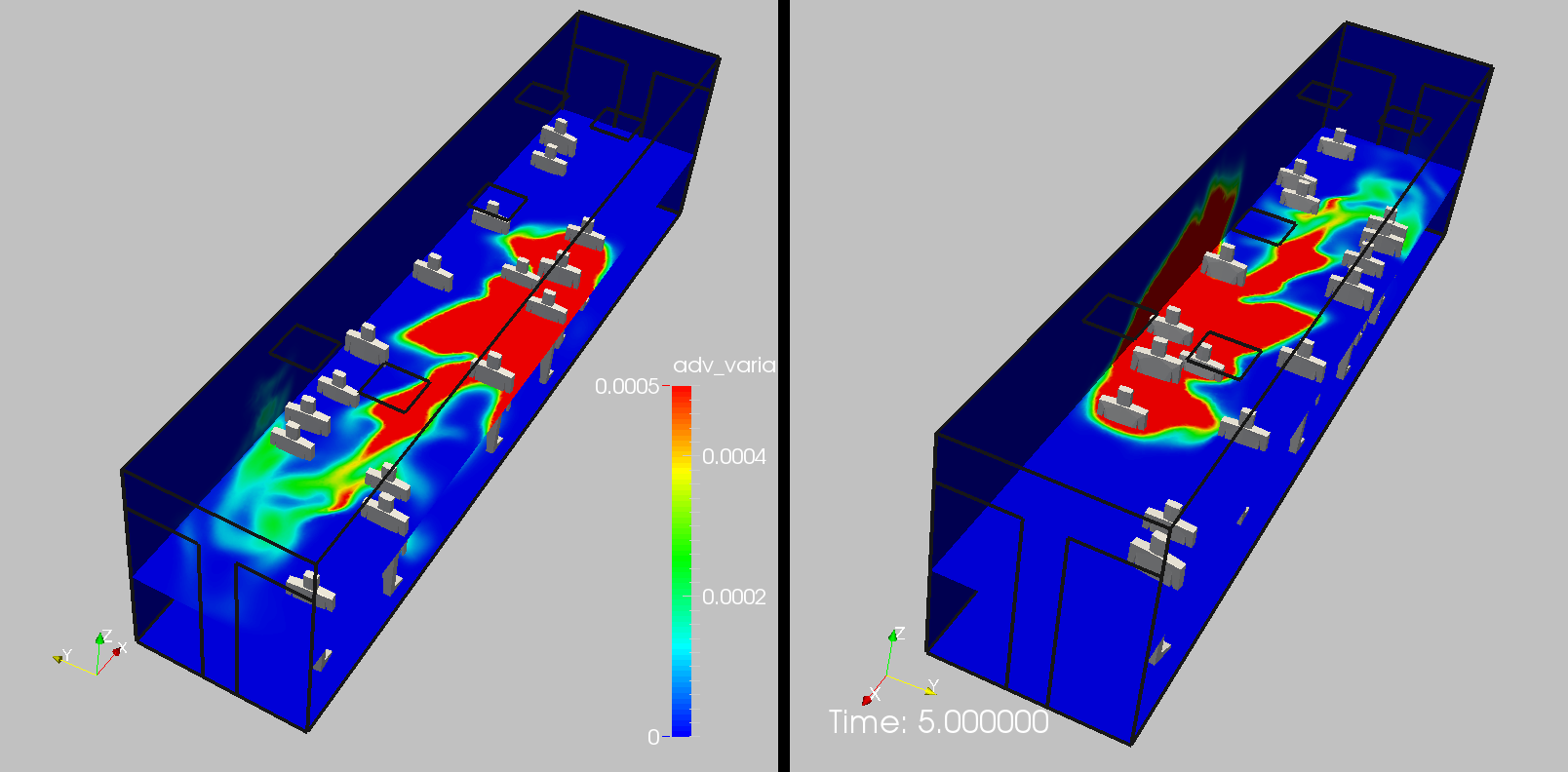}
	\caption{Counterflow Movement: Solution at $t=5.00~sec$}
\end{figure}

\begin{figure}
	\centering
	\includegraphics[width=10.0cm]{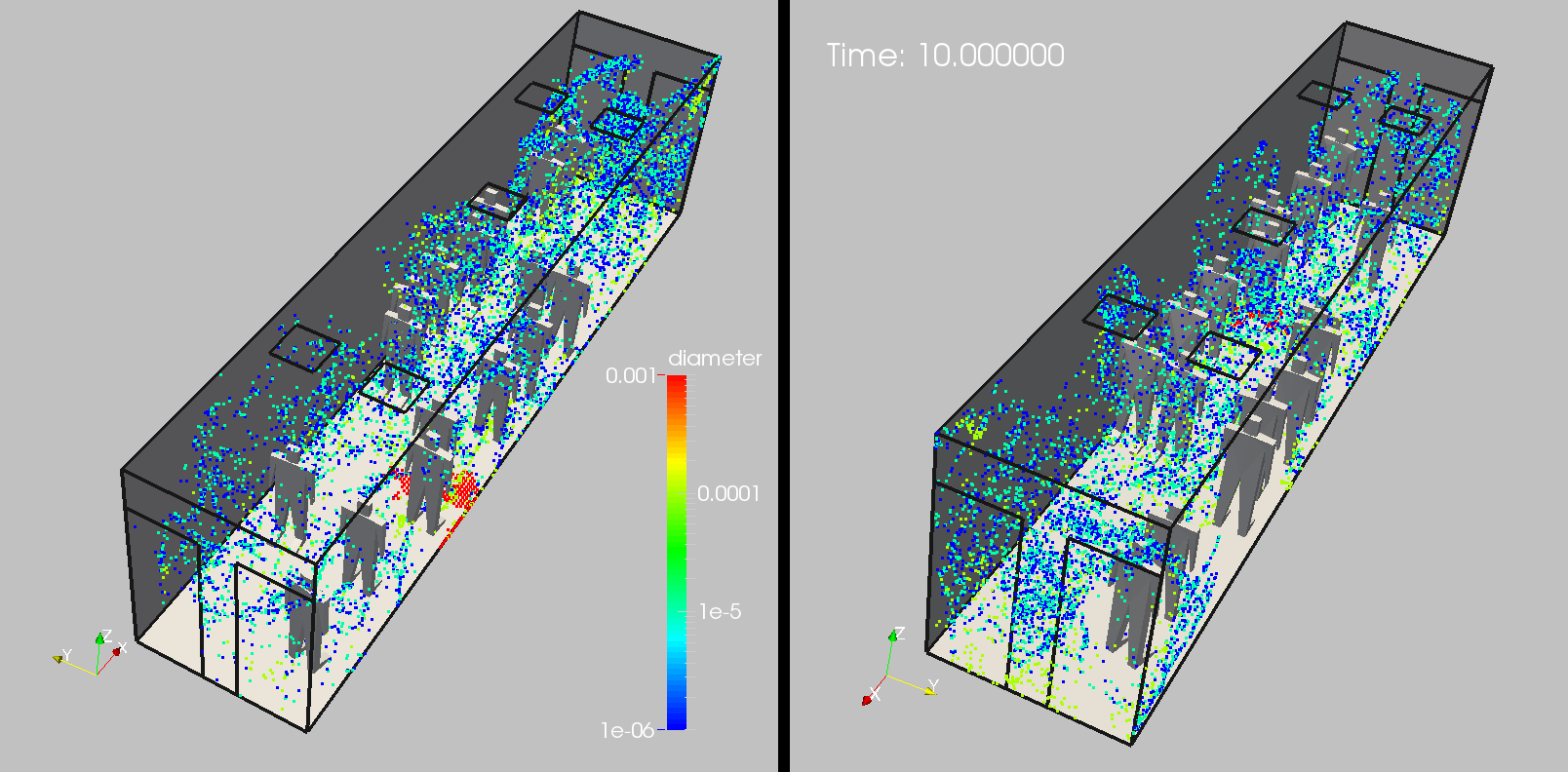}
	\vskip 10pt	
	\includegraphics[width=10.0cm]{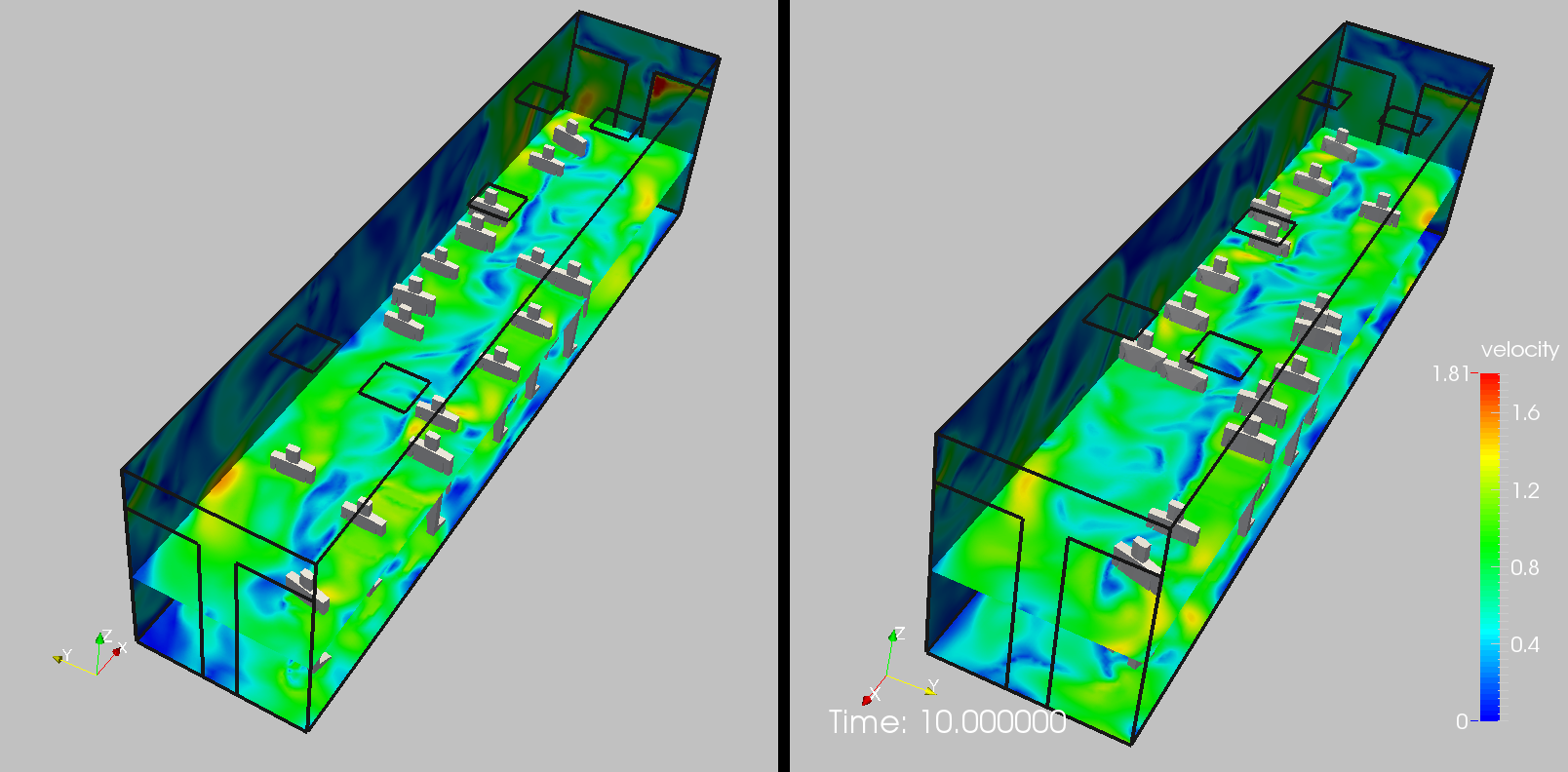}
	\vskip 10pt	
	\includegraphics[width=10.0cm]{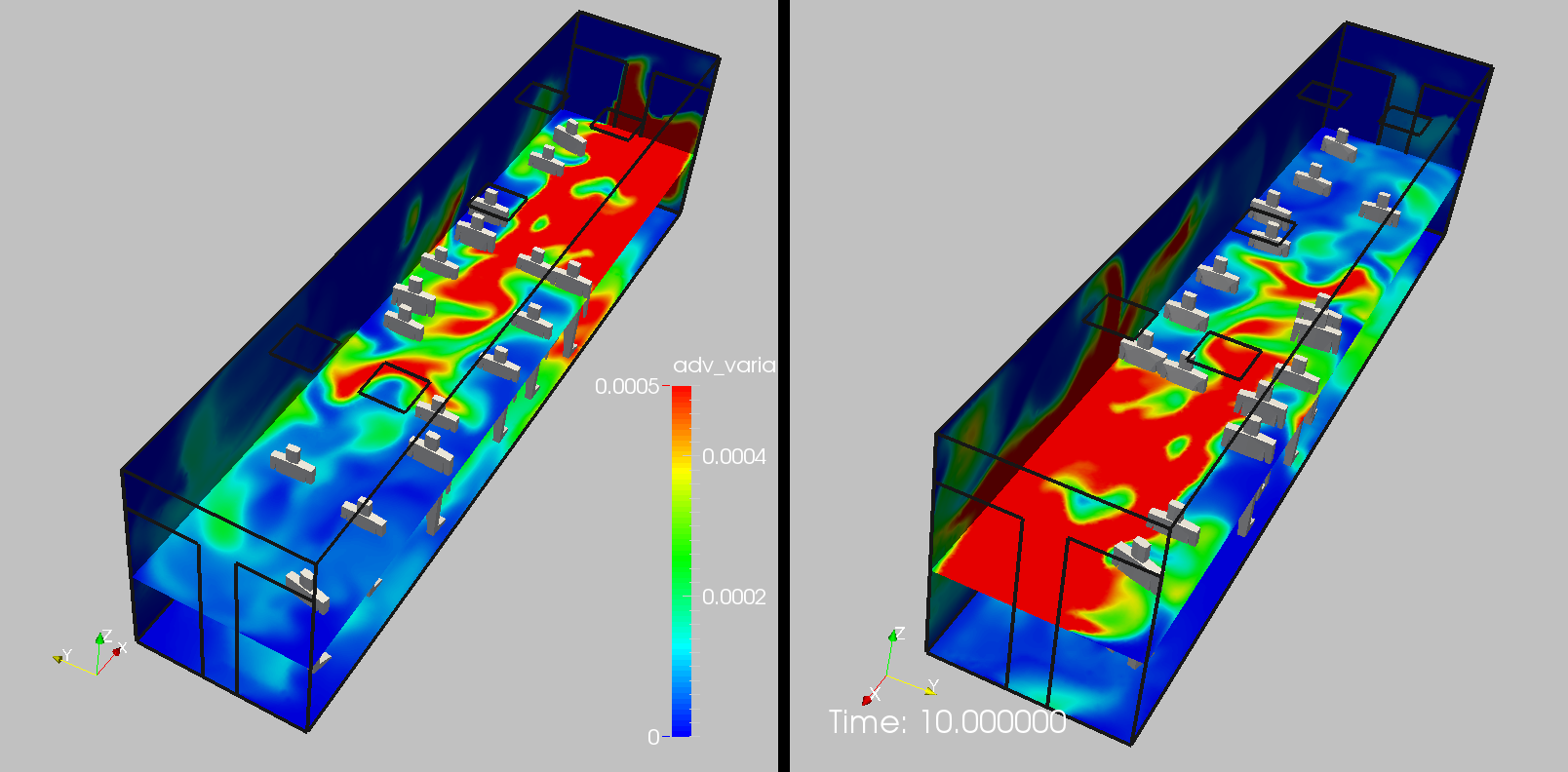}
	\caption{Counterflow Movement: Solution at $t=10.00~sec$}
\end{figure}

\begin{figure}
	\centering
	\includegraphics[width=10.0cm]{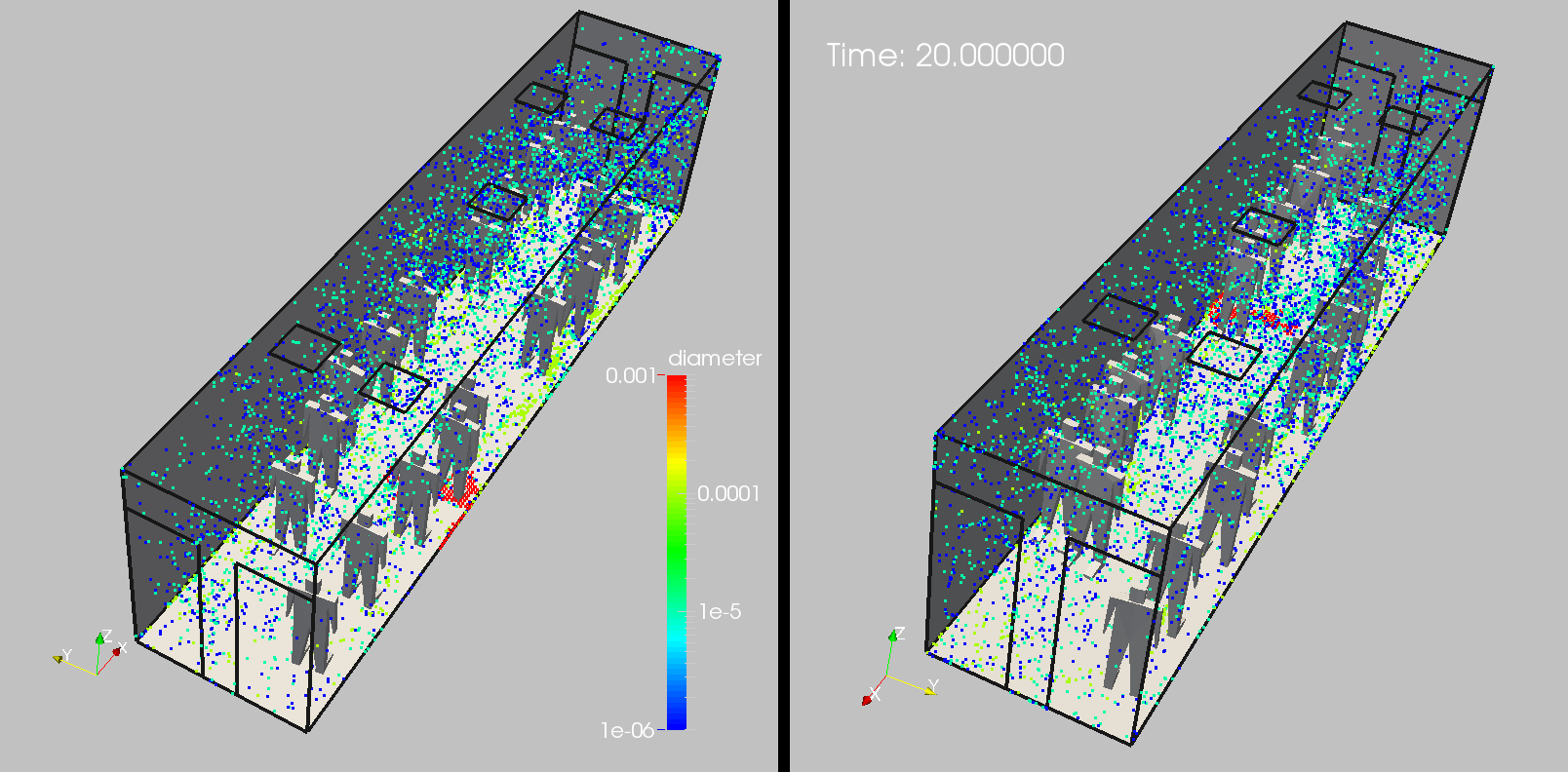}
	\vskip 10pt	
	\includegraphics[width=10.0cm]{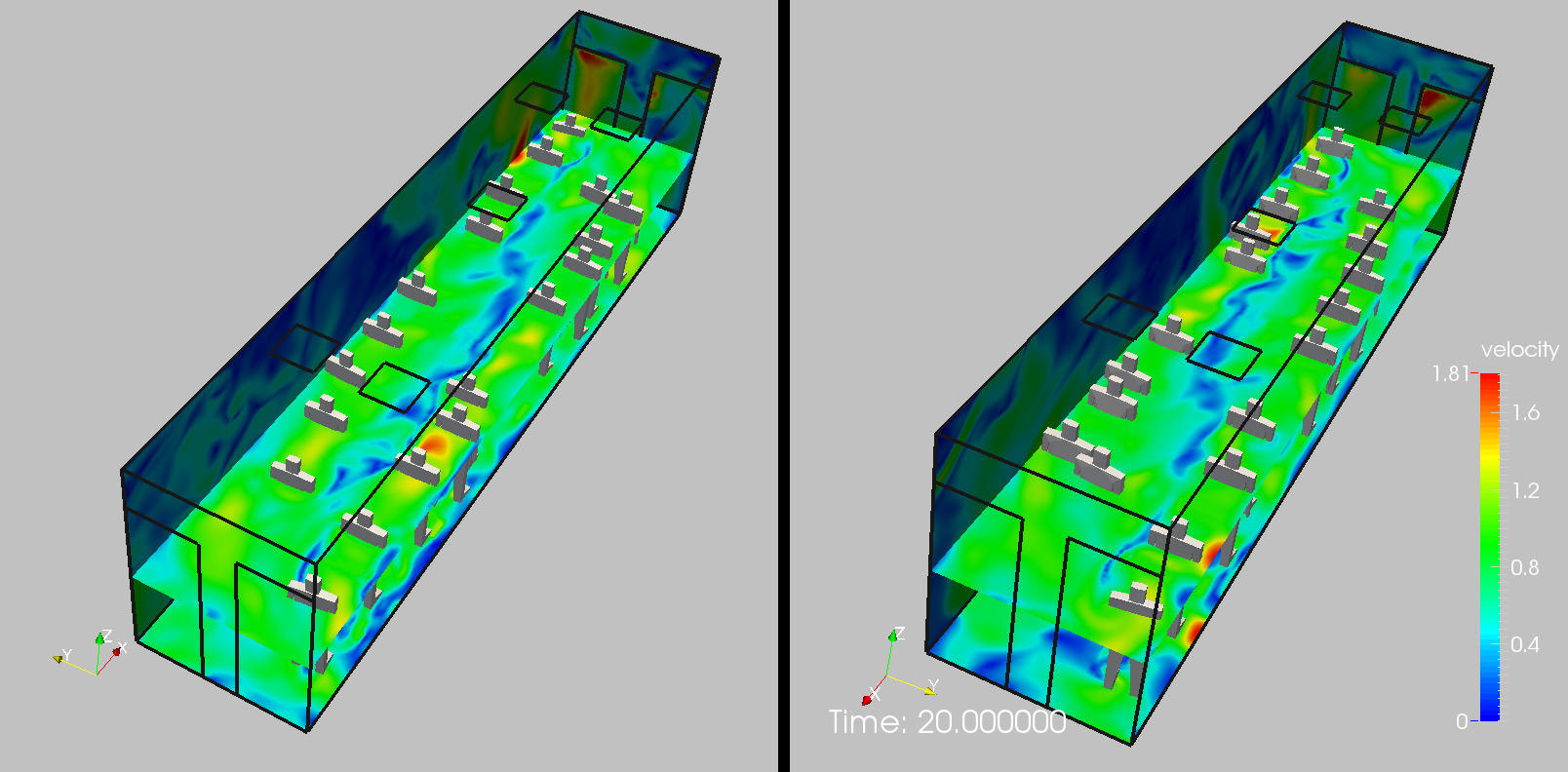}
	\vskip 10pt	
	\includegraphics[width=10.0cm]{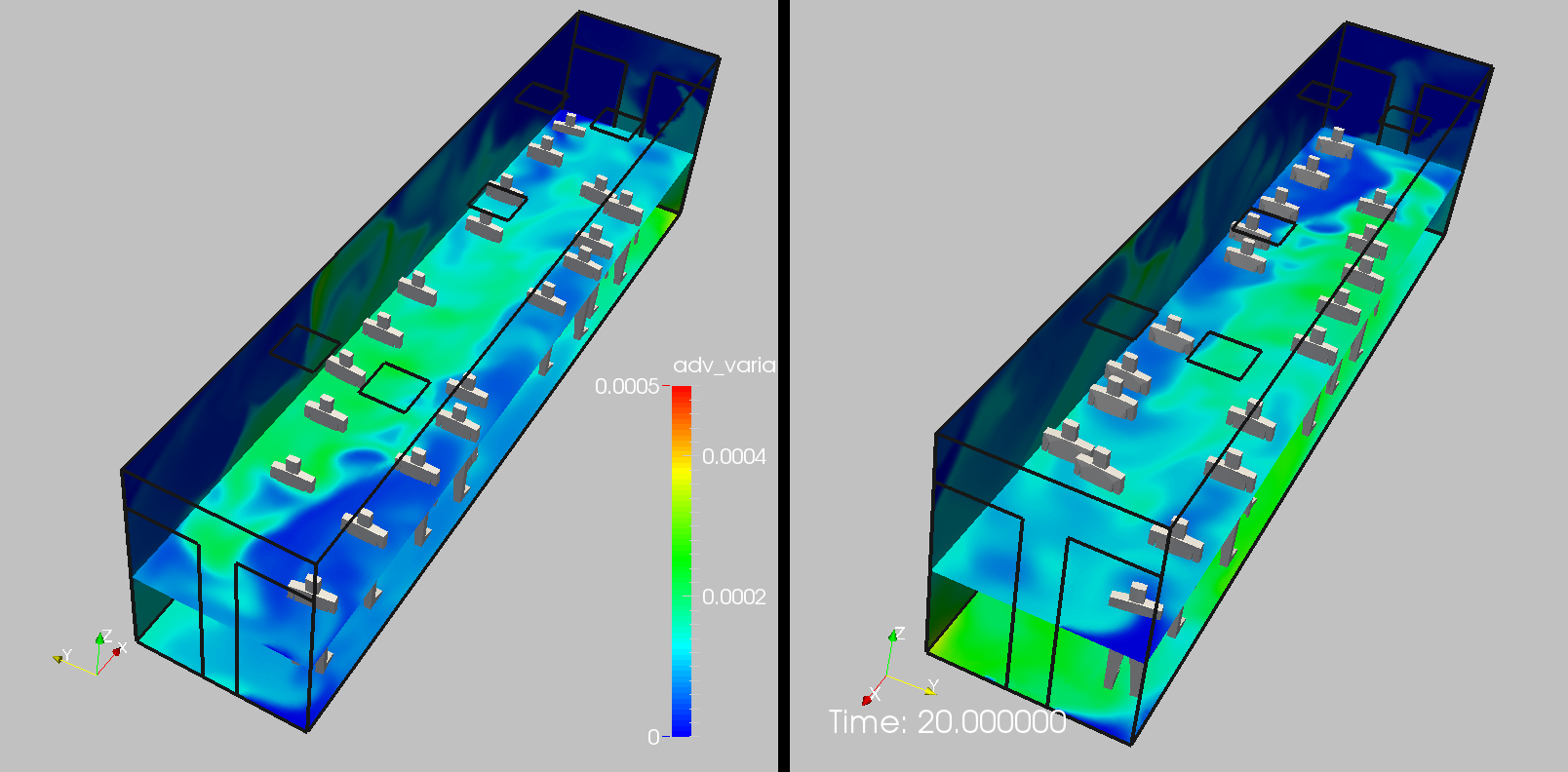}
	\caption{Counterflow Movement: Solution at $t=20.00~sec$}
\end{figure}

\begin{figure}
	\centering
	\includegraphics[width=10.0cm]{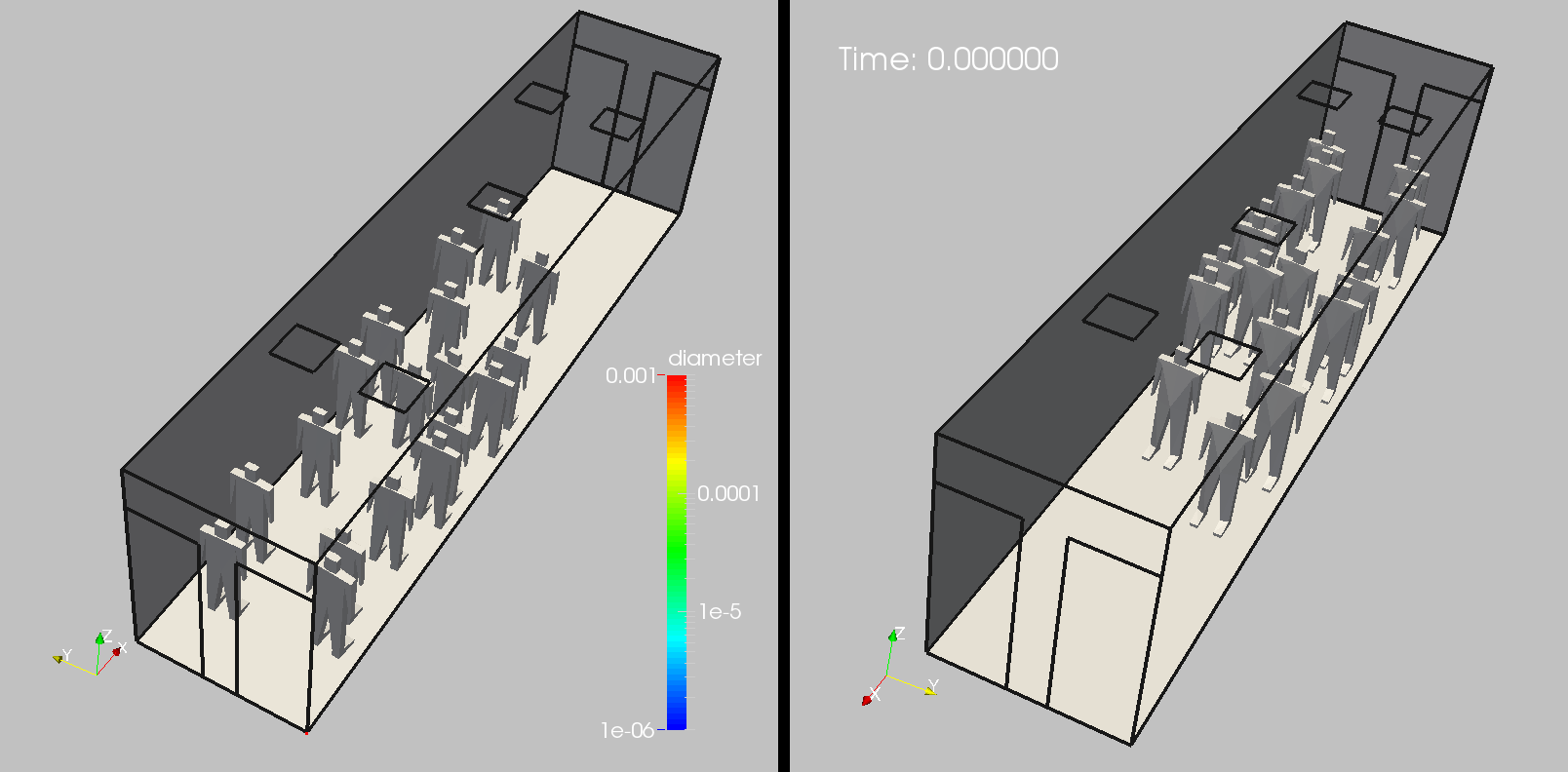}
	\vskip 10pt	
	\includegraphics[width=10.0cm]{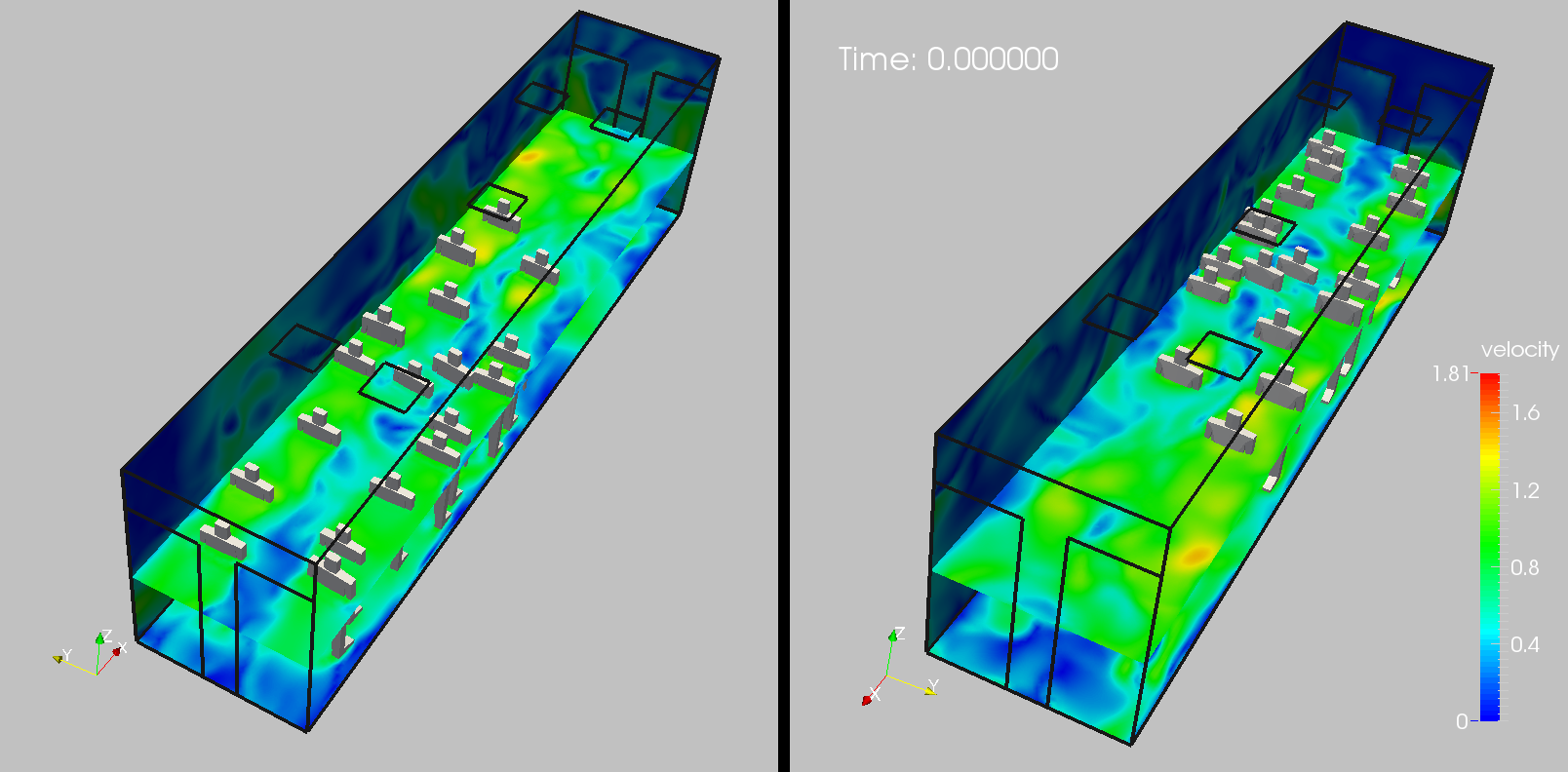}
	\vskip 10pt	
	\includegraphics[width=10.0cm]{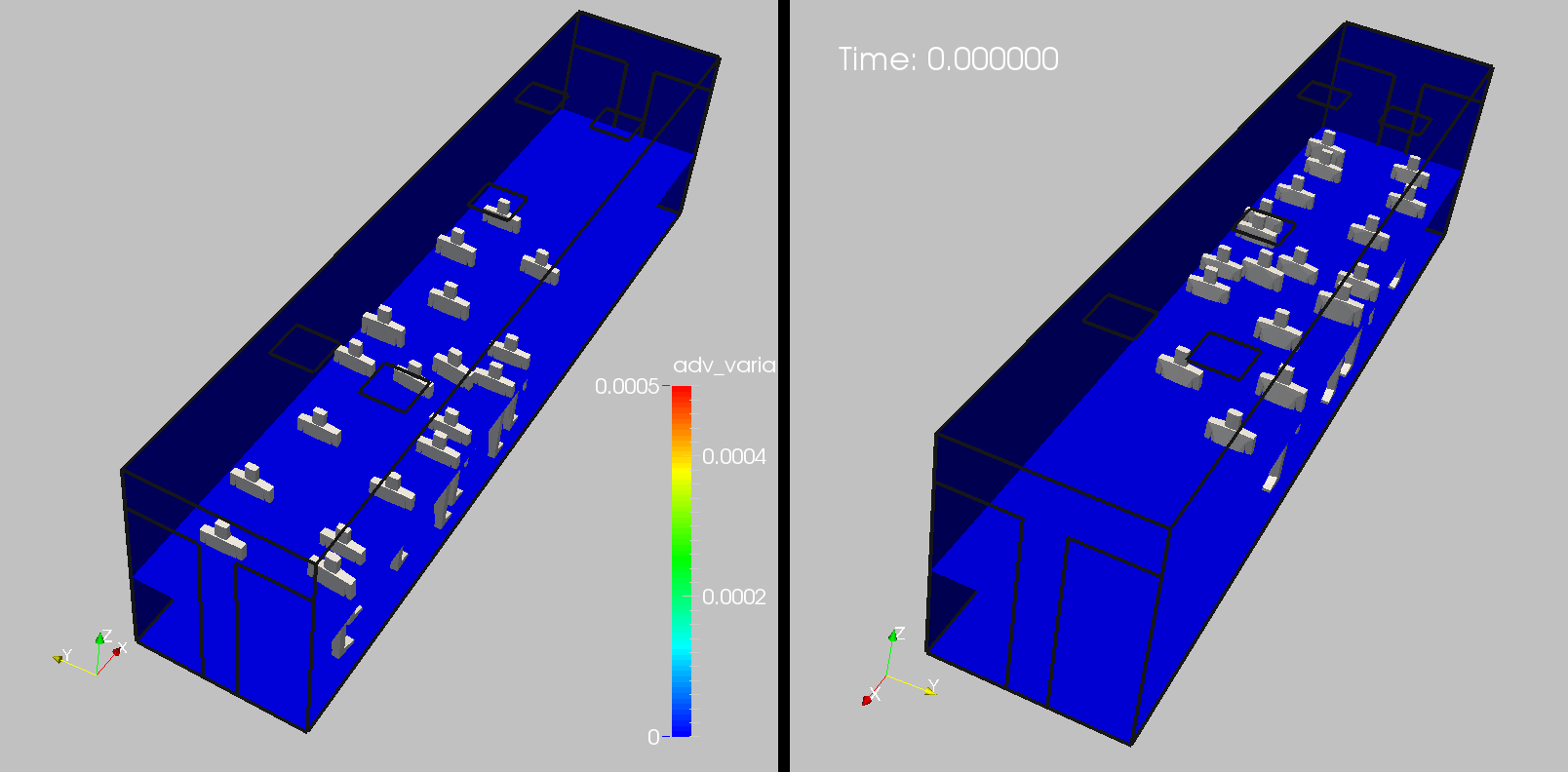}
	\caption{Parallel Movement: Solution at $t=0.00~sec$}
\end{figure}

\begin{figure}
	\centering
	\includegraphics[width=10.0cm]{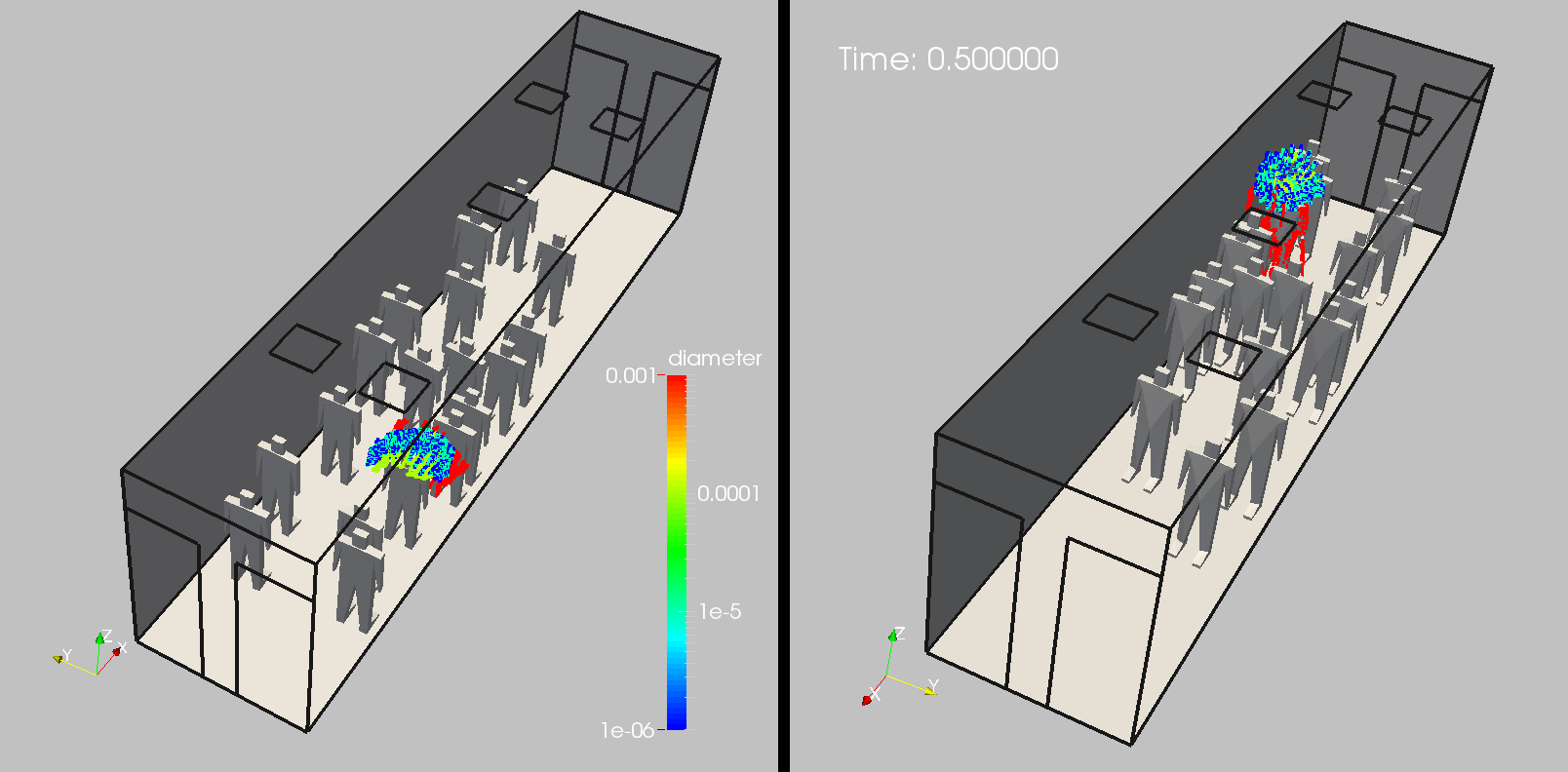}
	\vskip 10pt	
	\includegraphics[width=10.0cm]{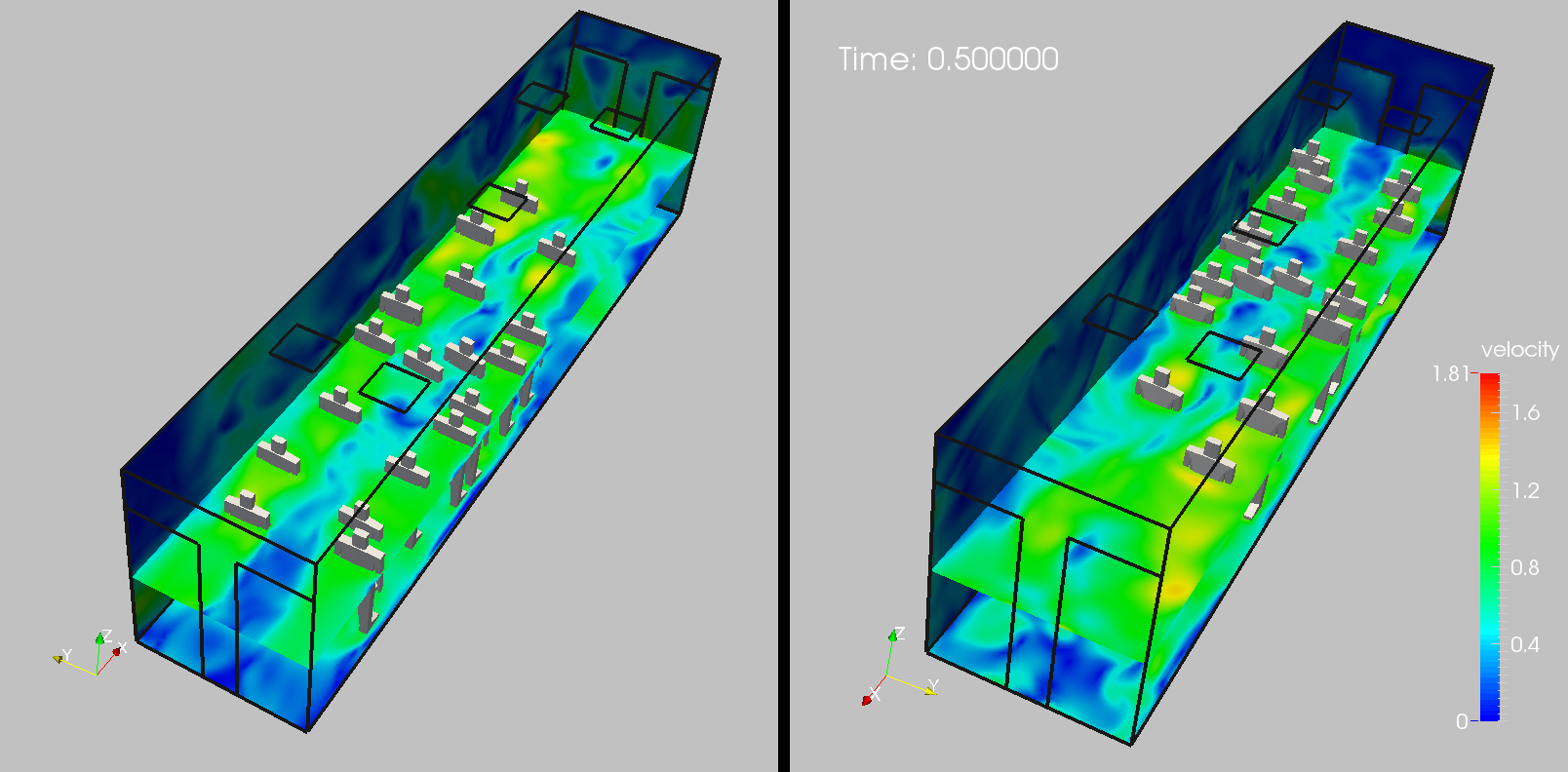}
	\vskip 10pt	
	\includegraphics[width=10.0cm]{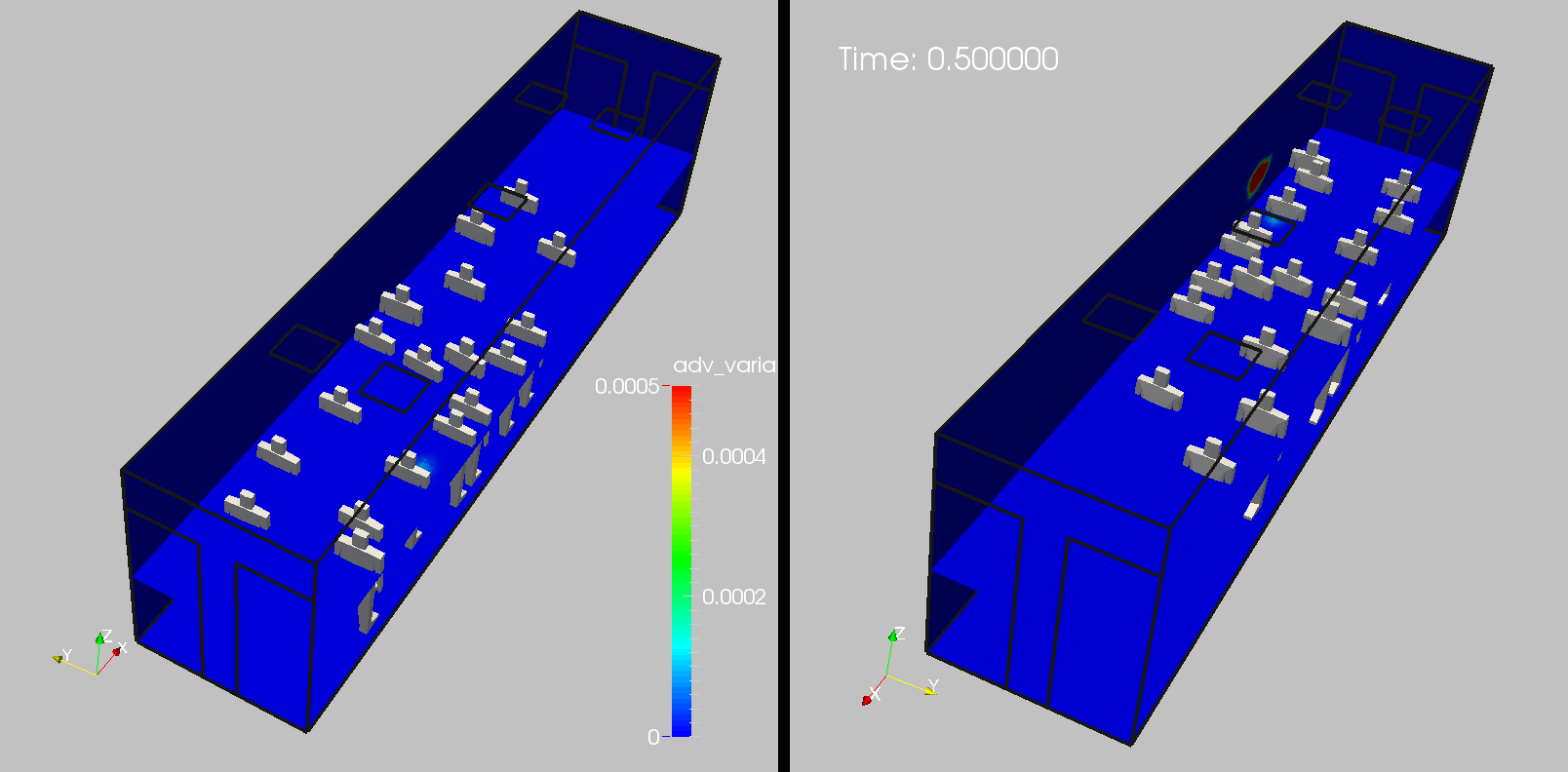}
	\caption{Parallel Movement: Solution at $t=0.50~sec$}
\end{figure}

\begin{figure}
	\centering
	\includegraphics[width=10.0cm]{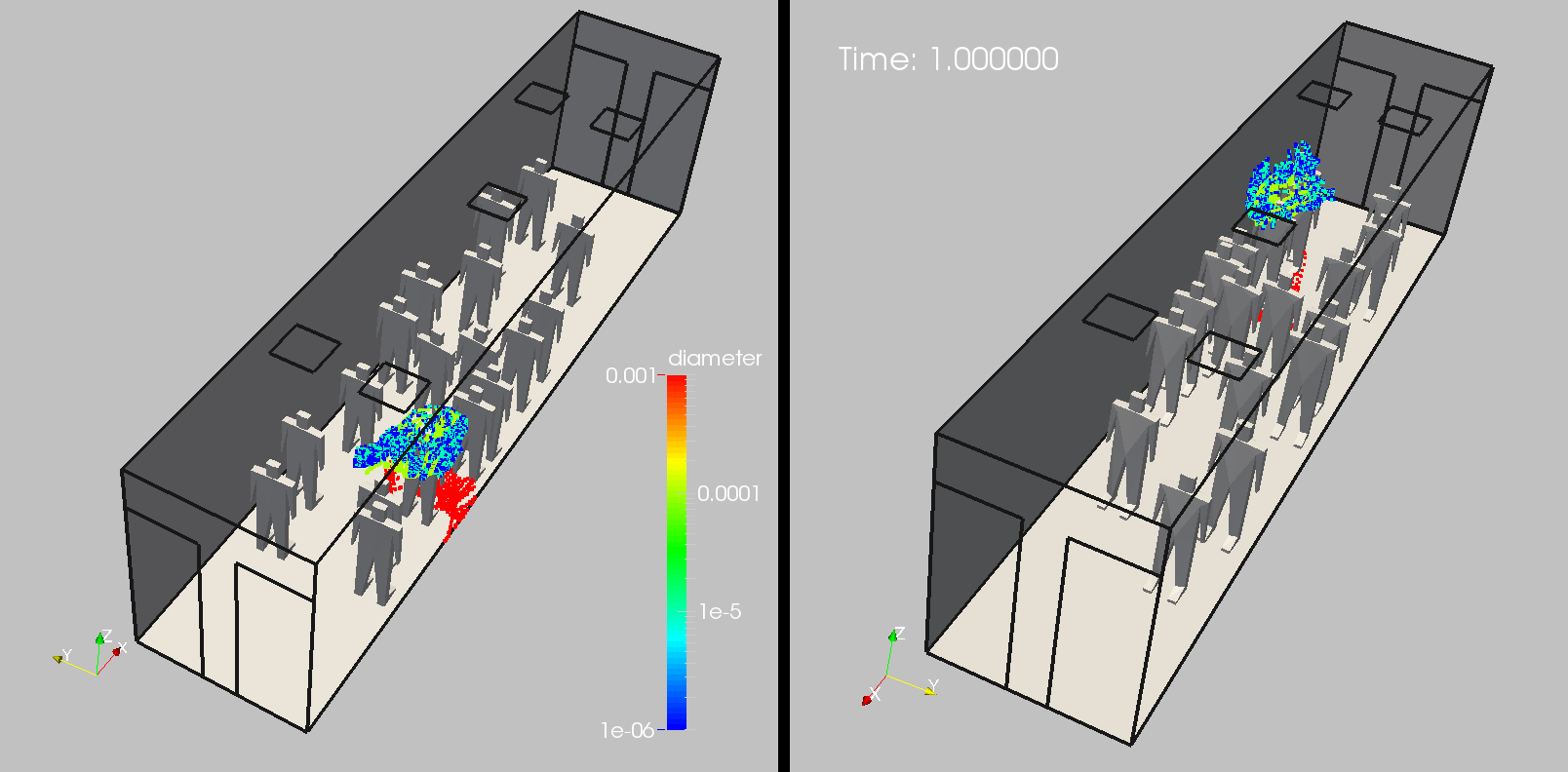}
	\vskip 10pt	
	\includegraphics[width=10.0cm]{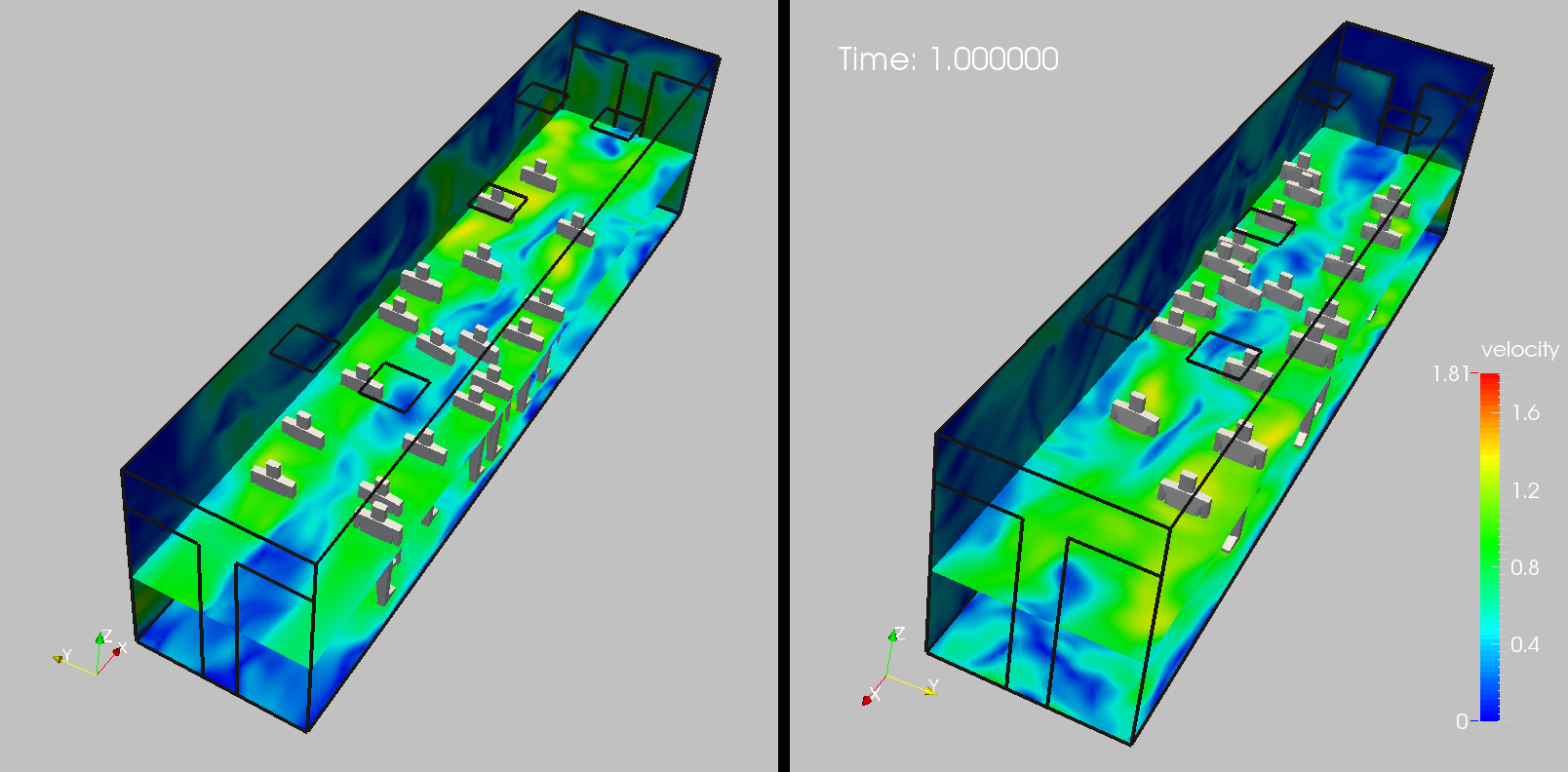}
	\vskip 10pt	
	\includegraphics[width=10.0cm]{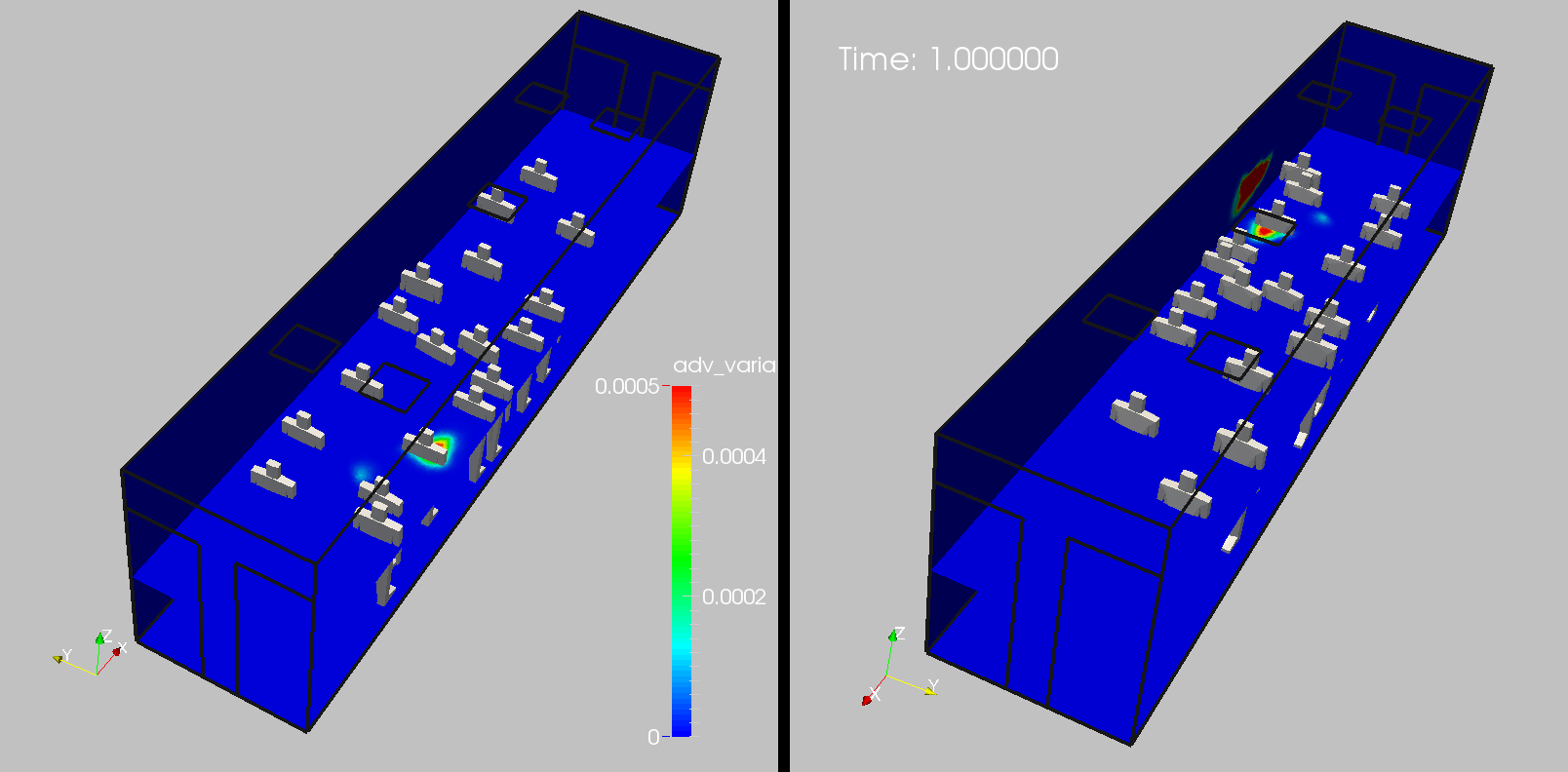}
	\caption{Parallel Movement: Solution at $t=1.00~sec$}
\end{figure}

\begin{figure}
	\centering
	\includegraphics[width=10.0cm]{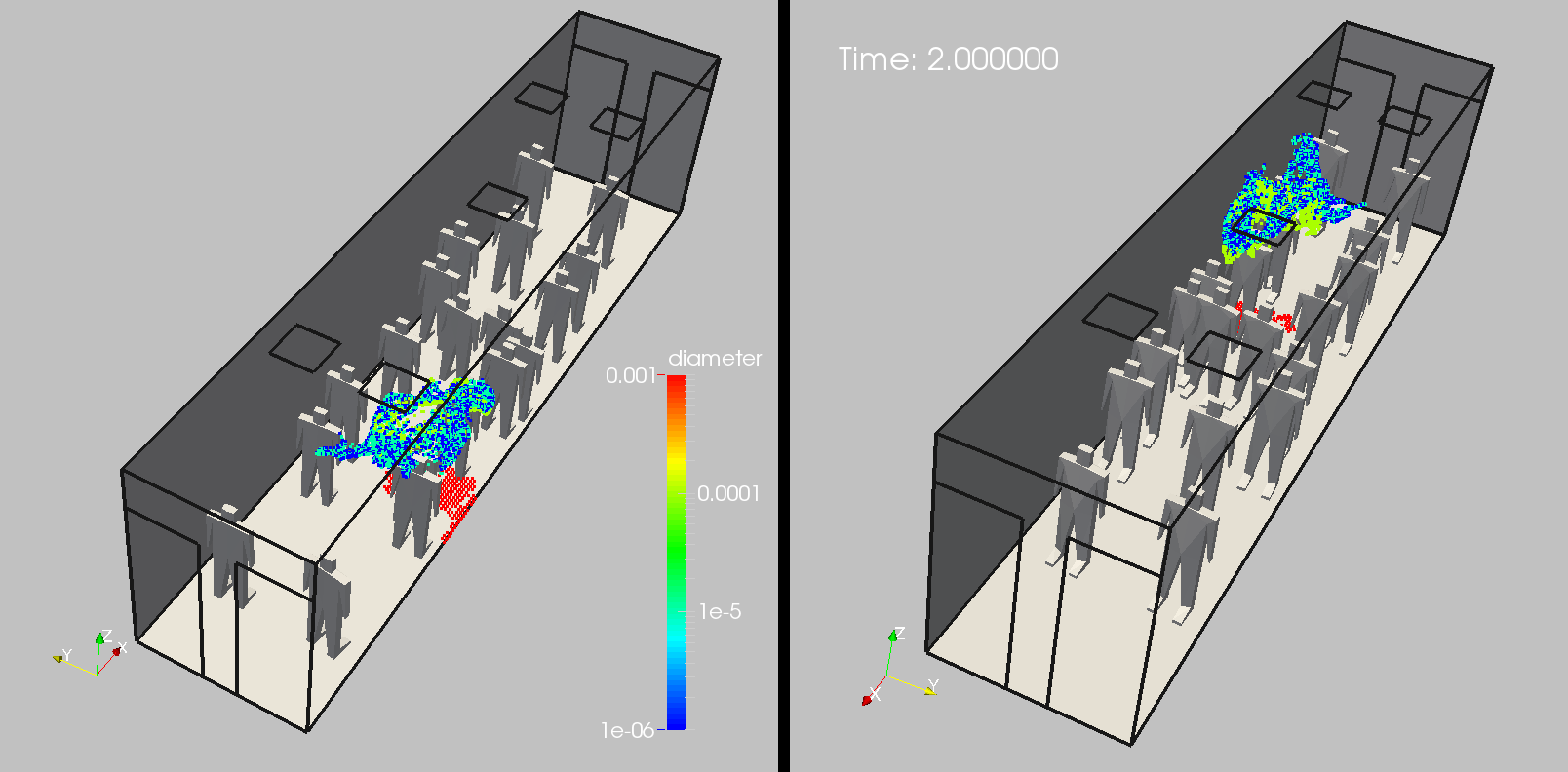}
	\vskip 10pt	
	\includegraphics[width=10.0cm]{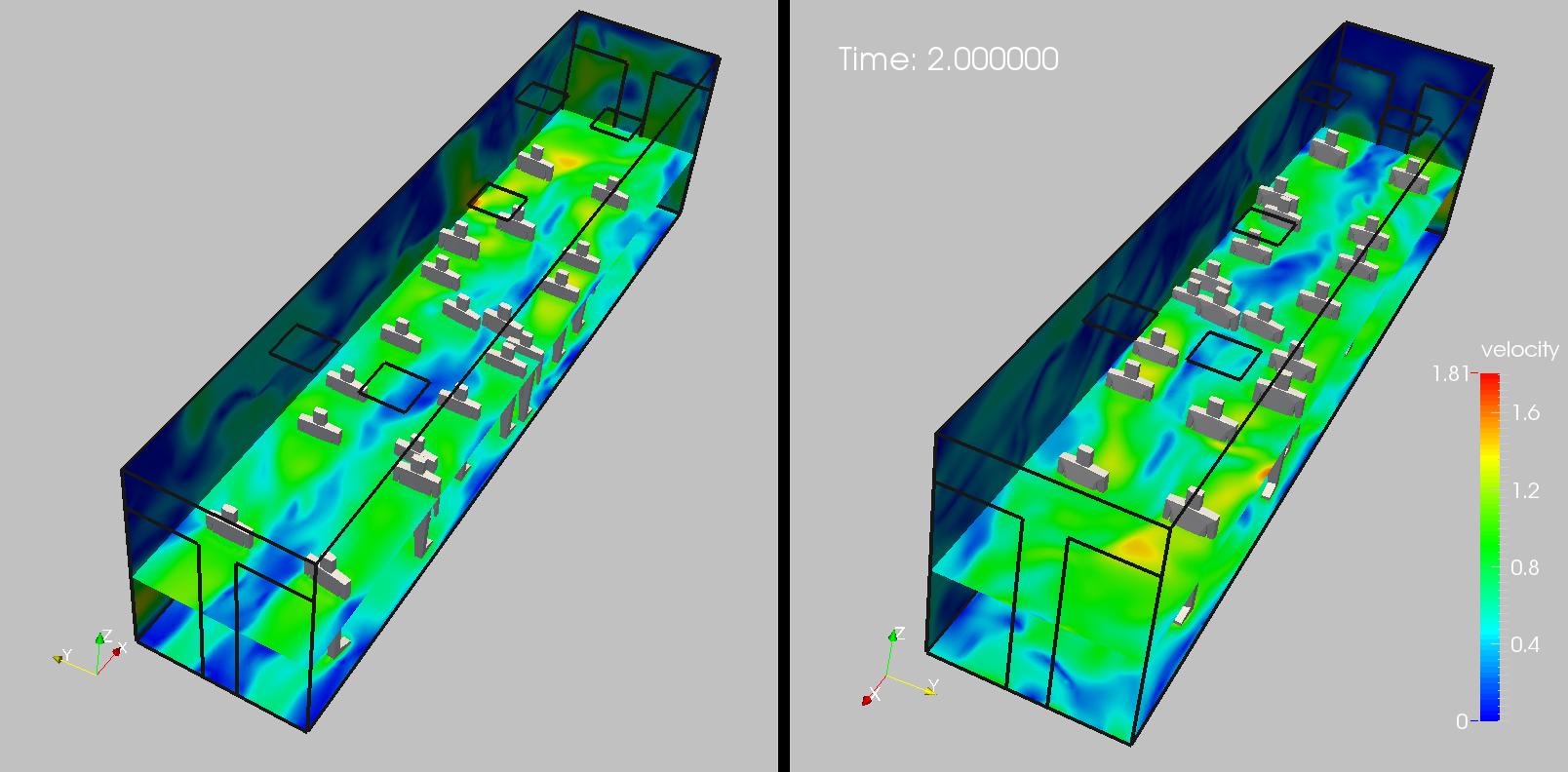}
	\vskip 10pt	
	\includegraphics[width=10.0cm]{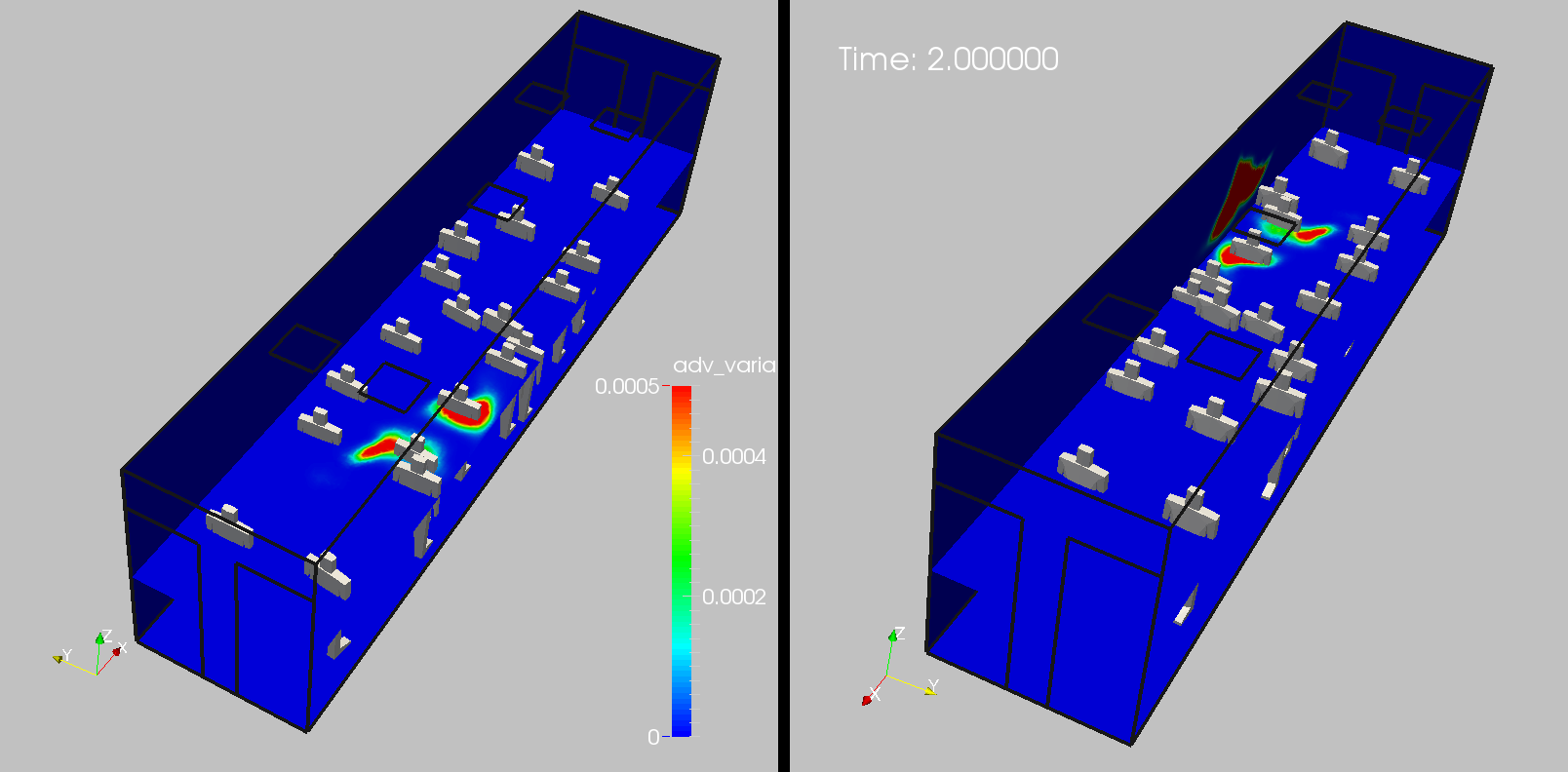}
	\caption{Parallel Movement: Solution at $t=2.00~sec$}
\end{figure}

\begin{figure}
	\centering
	\includegraphics[width=10.0cm]{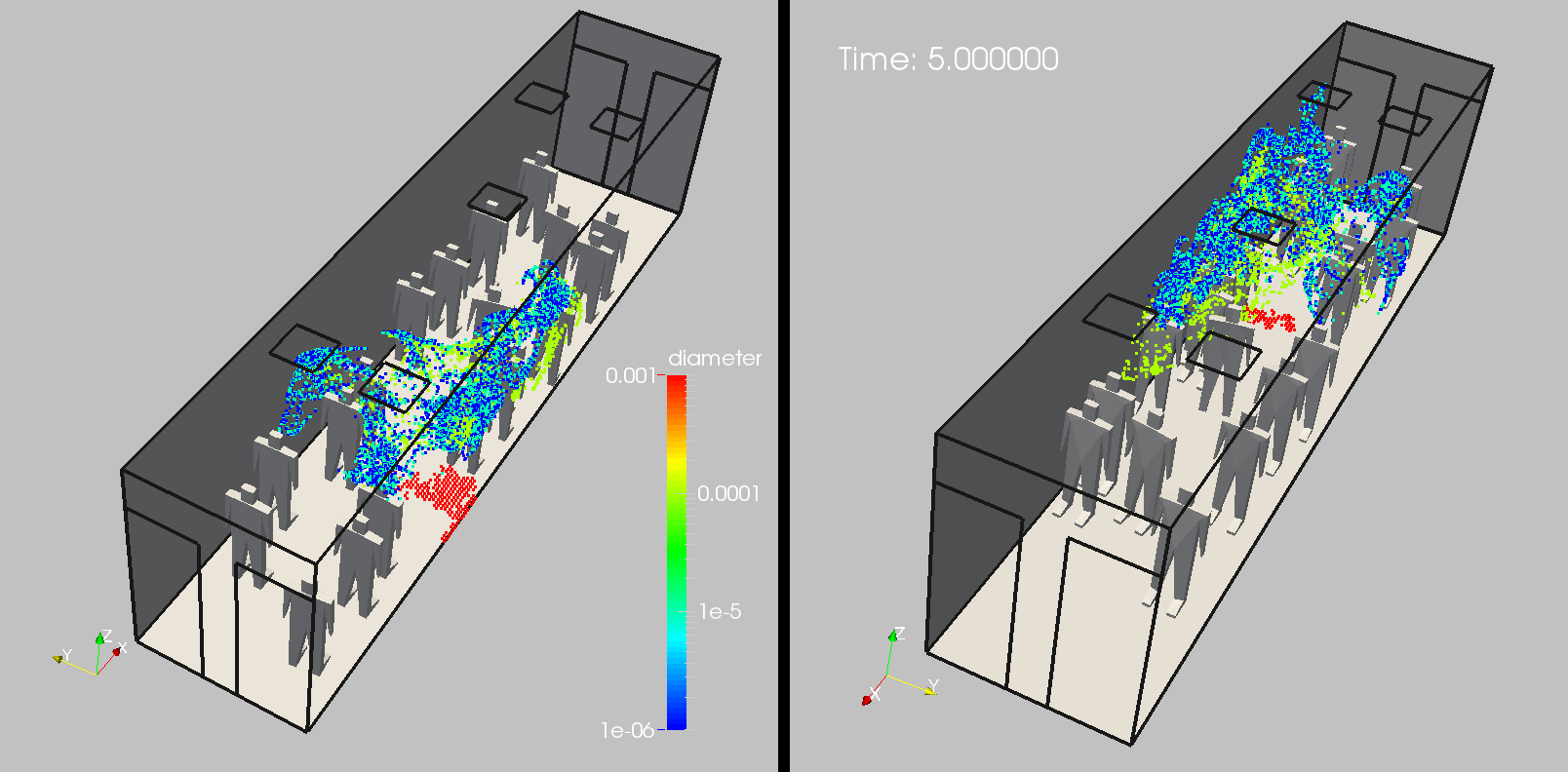}
	\vskip 10pt	
	\includegraphics[width=10.0cm]{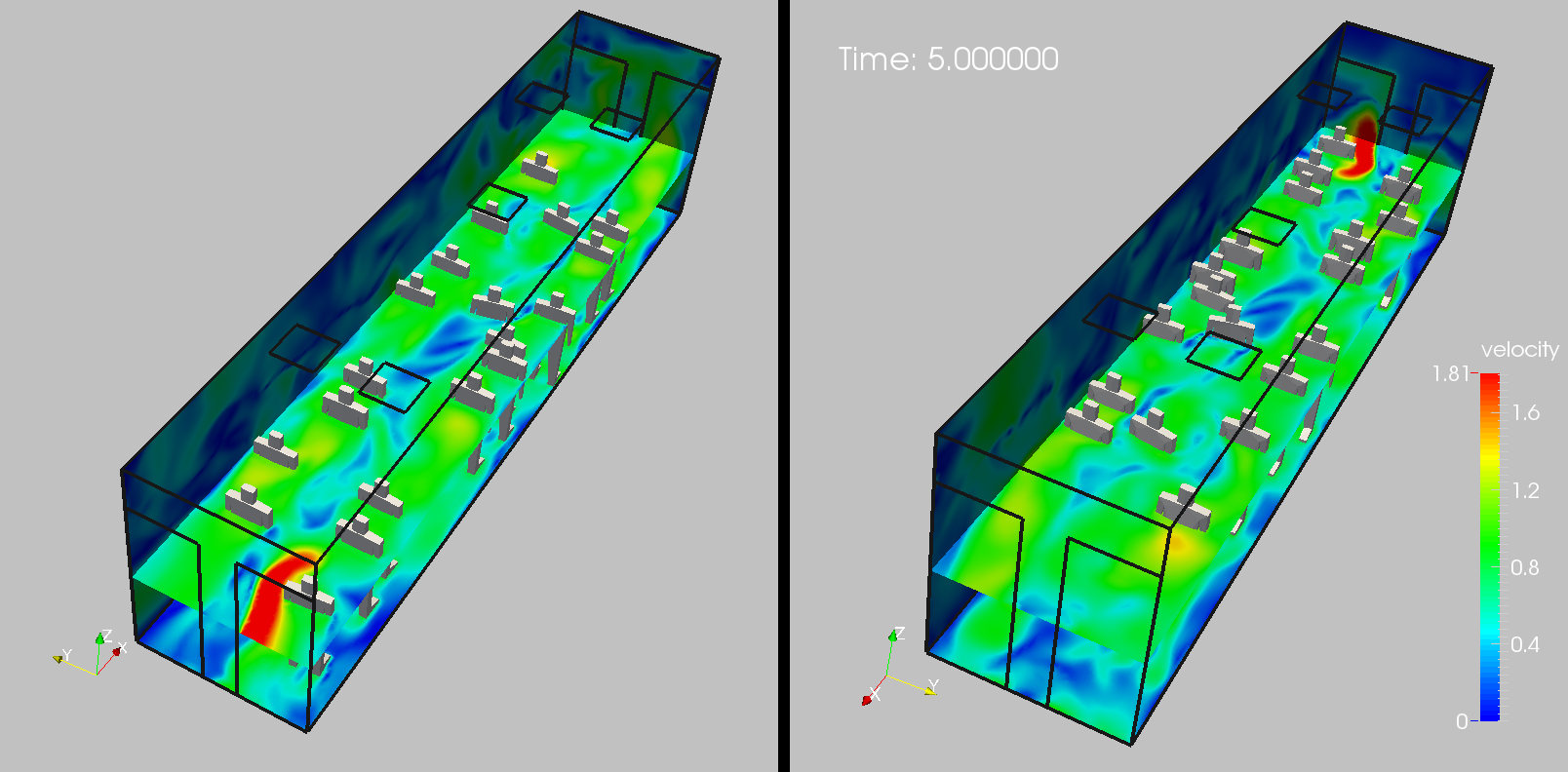}
	\vskip 10pt	
	\includegraphics[width=10.0cm]{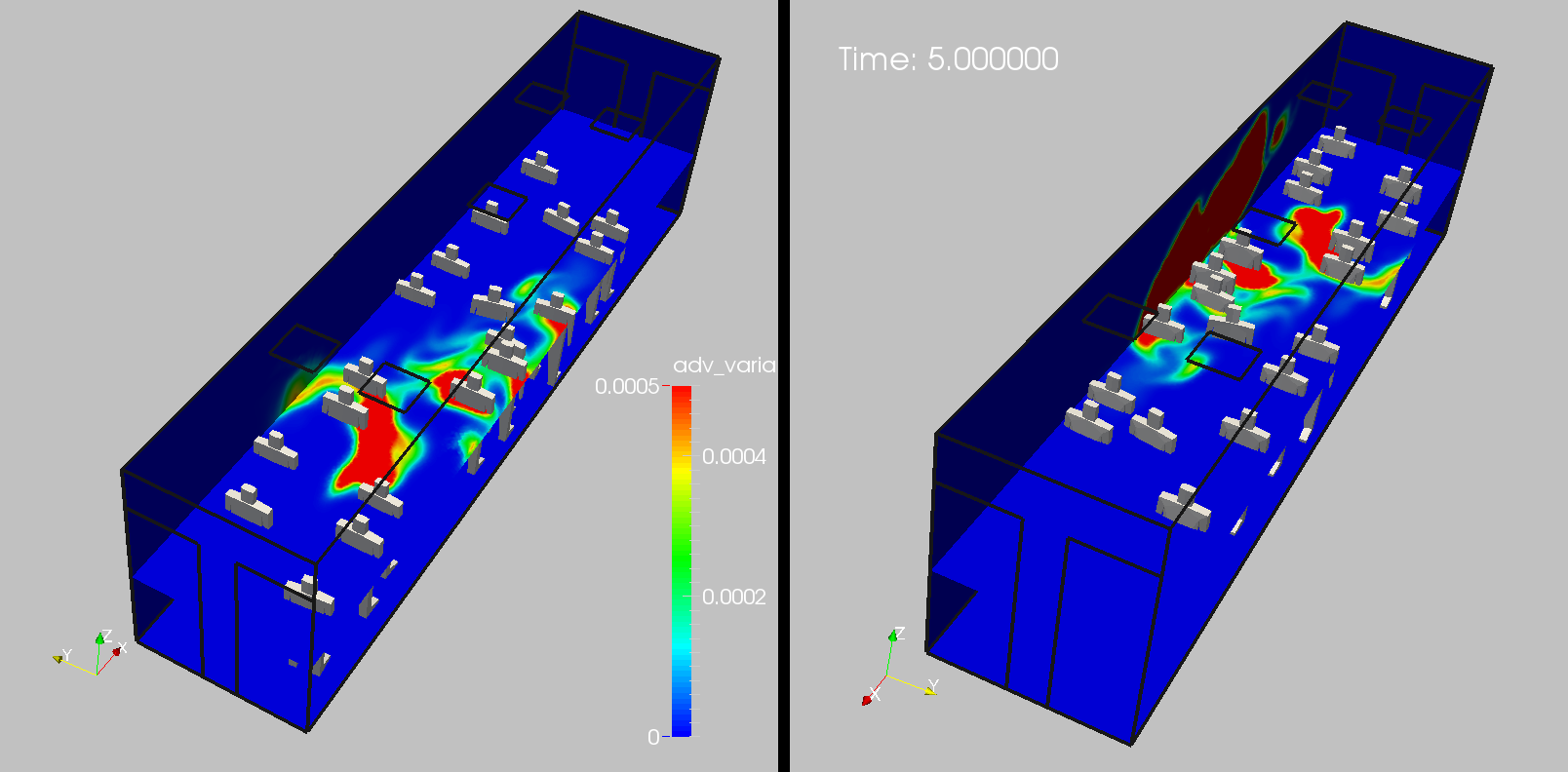}
	\caption{Parallel Movement: Solution at $t=5.00~sec$}
\end{figure}

\begin{figure}
	\centering
	\includegraphics[width=10.0cm]{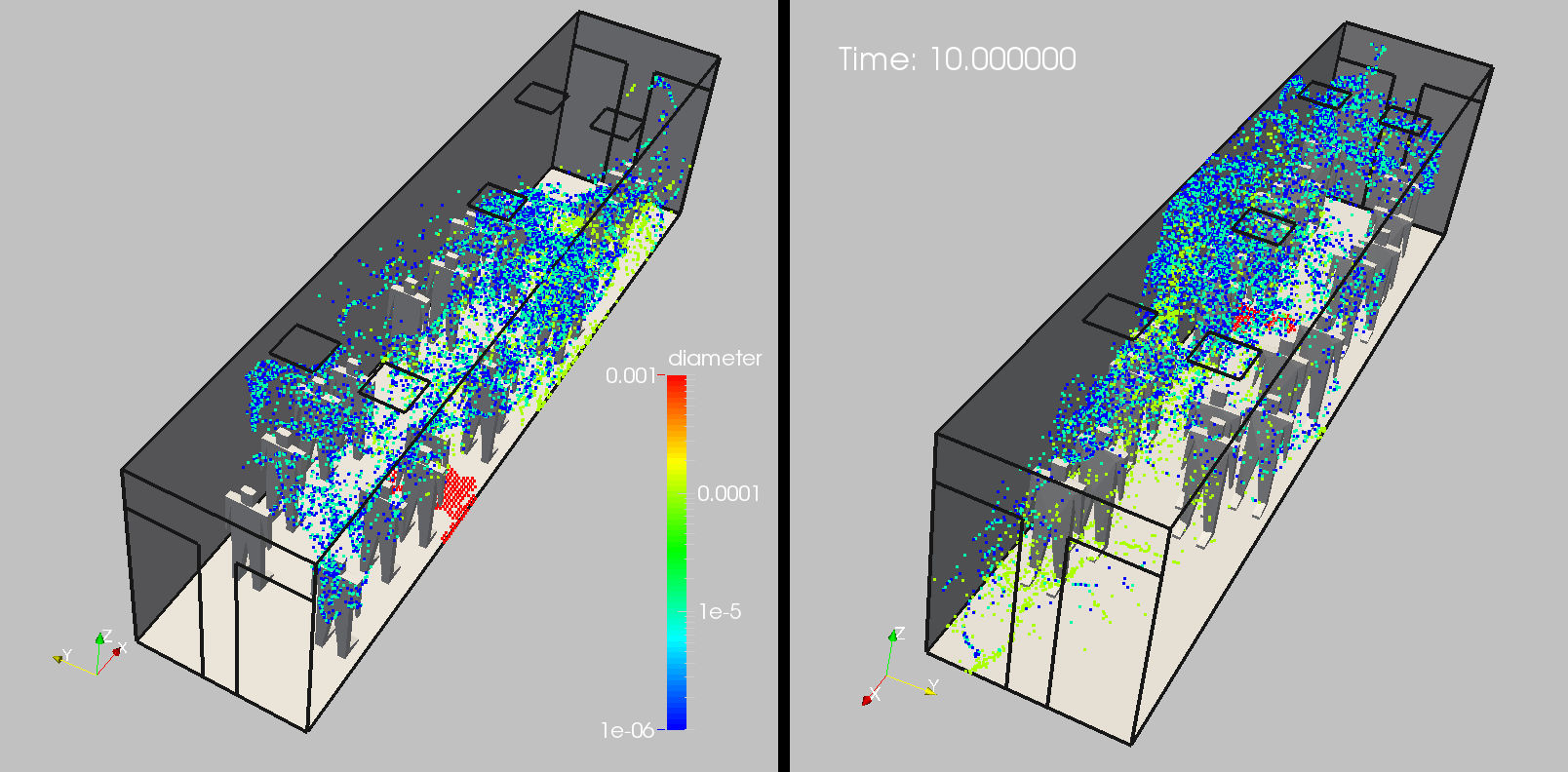}
	\vskip 10pt	
	\includegraphics[width=10.0cm]{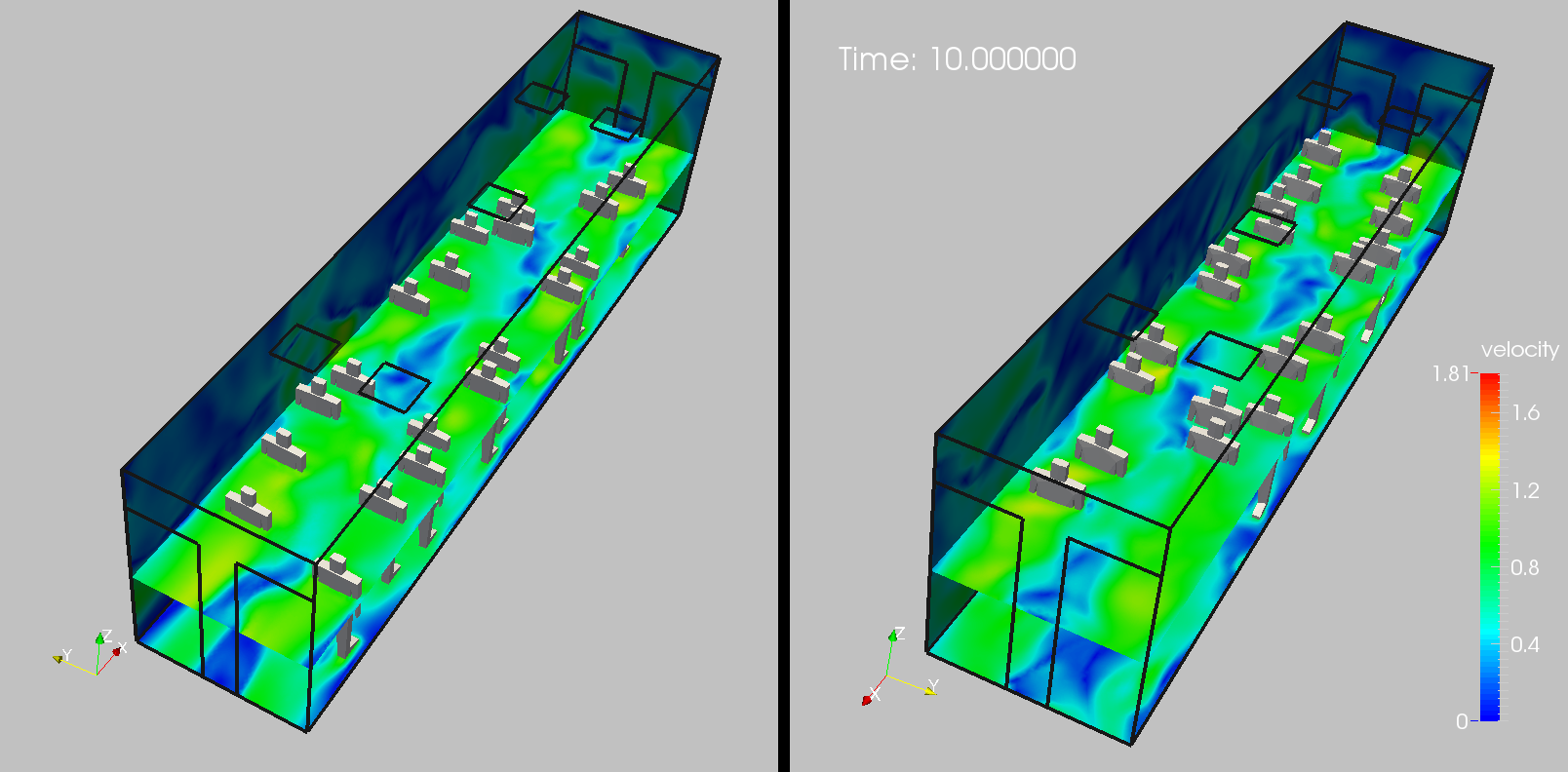}
	\vskip 10pt	
	\includegraphics[width=10.0cm]{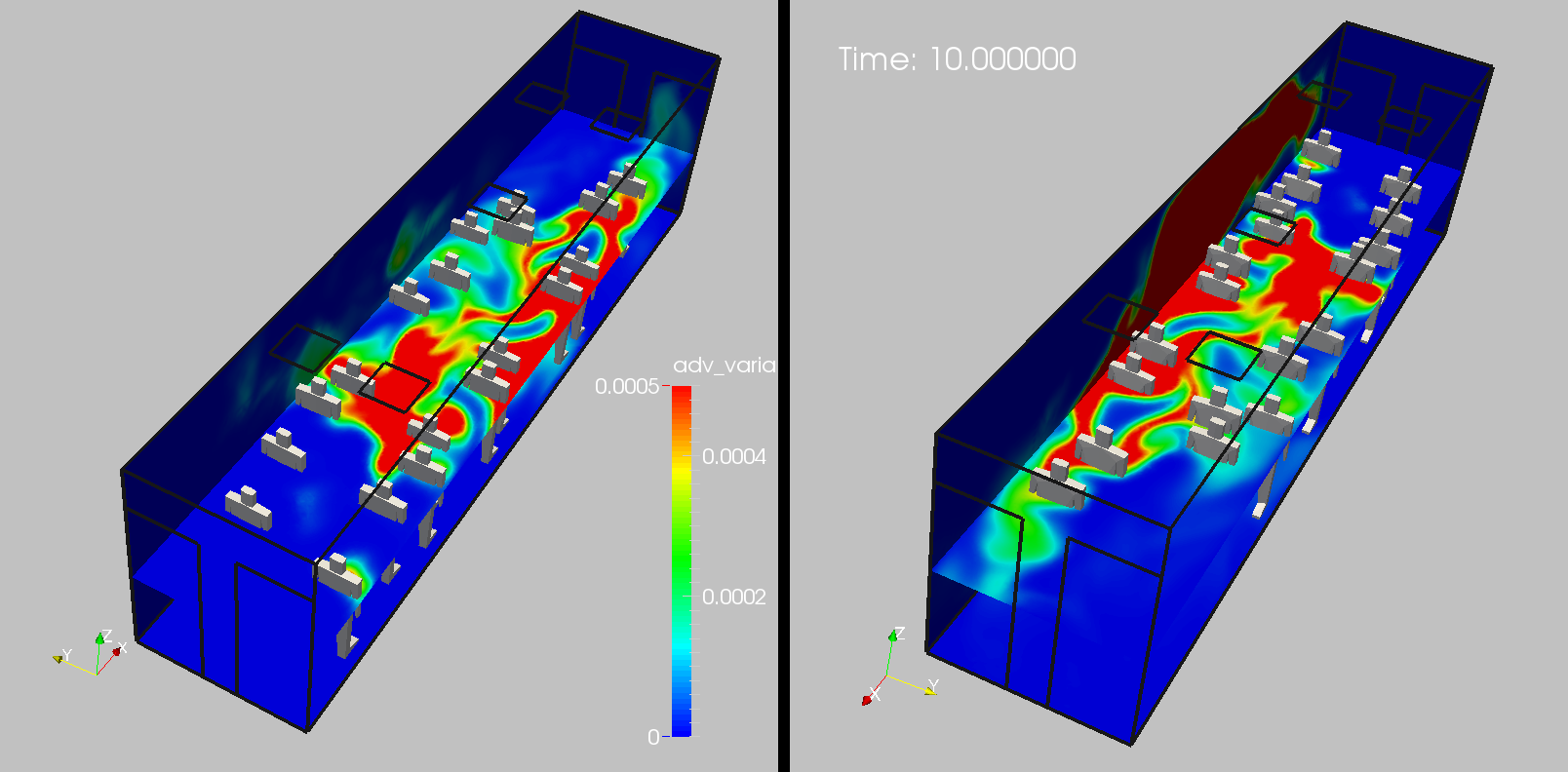}
	\caption{Parallel Movement: Solution at $t=10.00~sec$}
\end{figure}

\begin{figure}
	\centering
	\includegraphics[width=10.0cm]{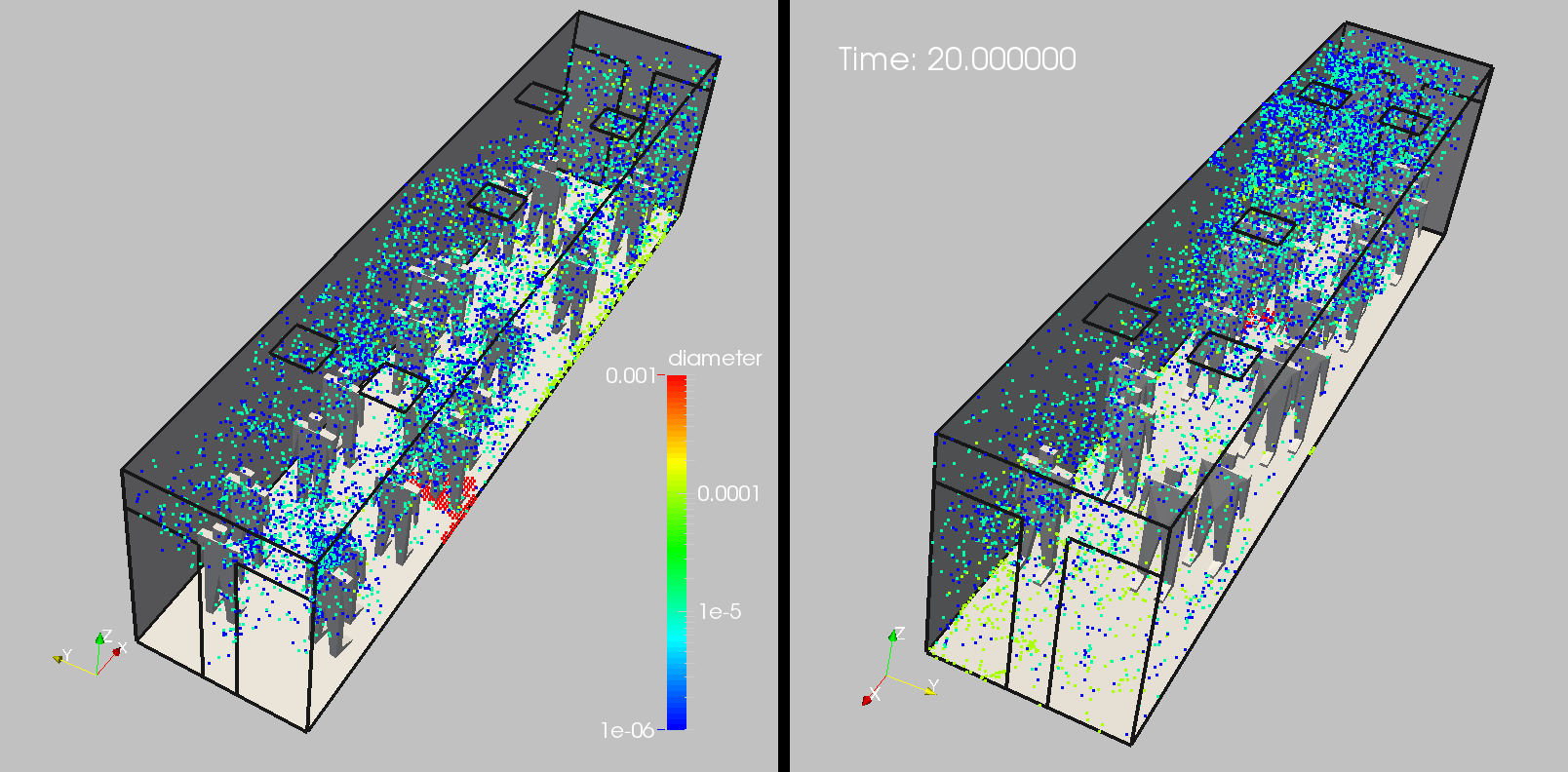}
	\vskip 10pt	
	\includegraphics[width=10.0cm]{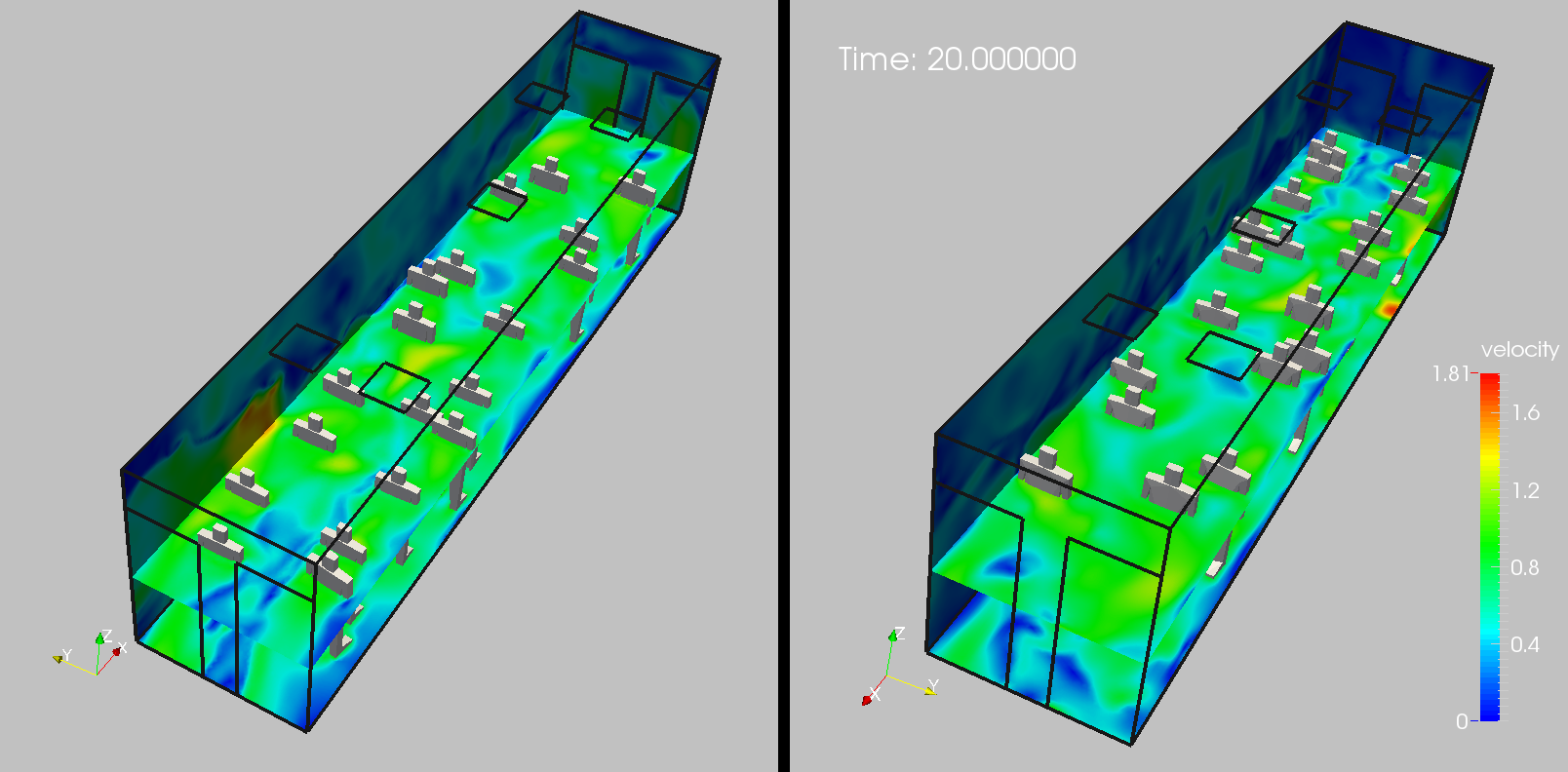}
	\vskip 10pt	
	\includegraphics[width=10.0cm]{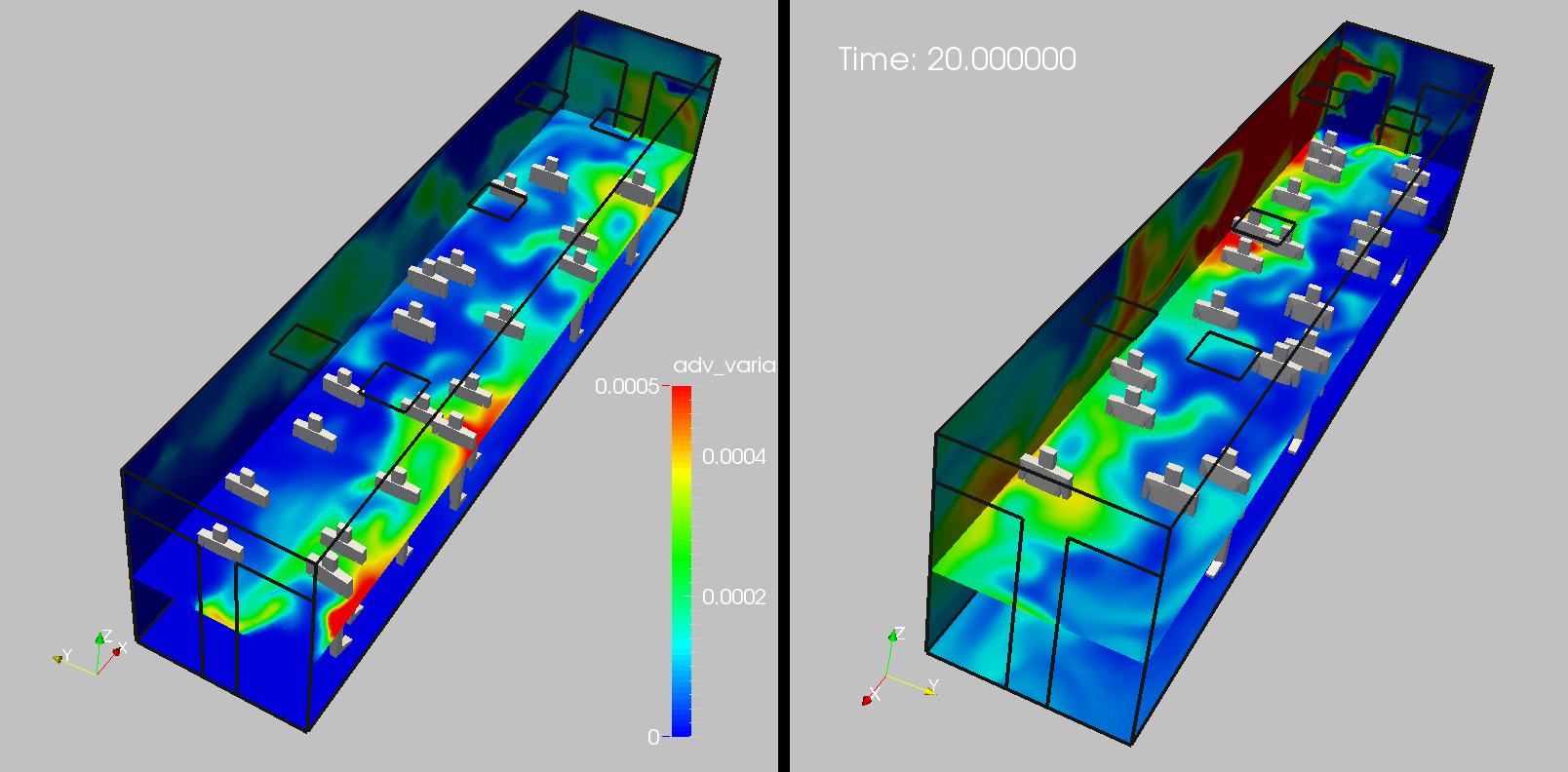}
	\caption{Parallel Movement: Solution at $t=20.00~sec$}
\end{figure}

\begin{figure}
	\centering
	\includegraphics[width=10.0cm]{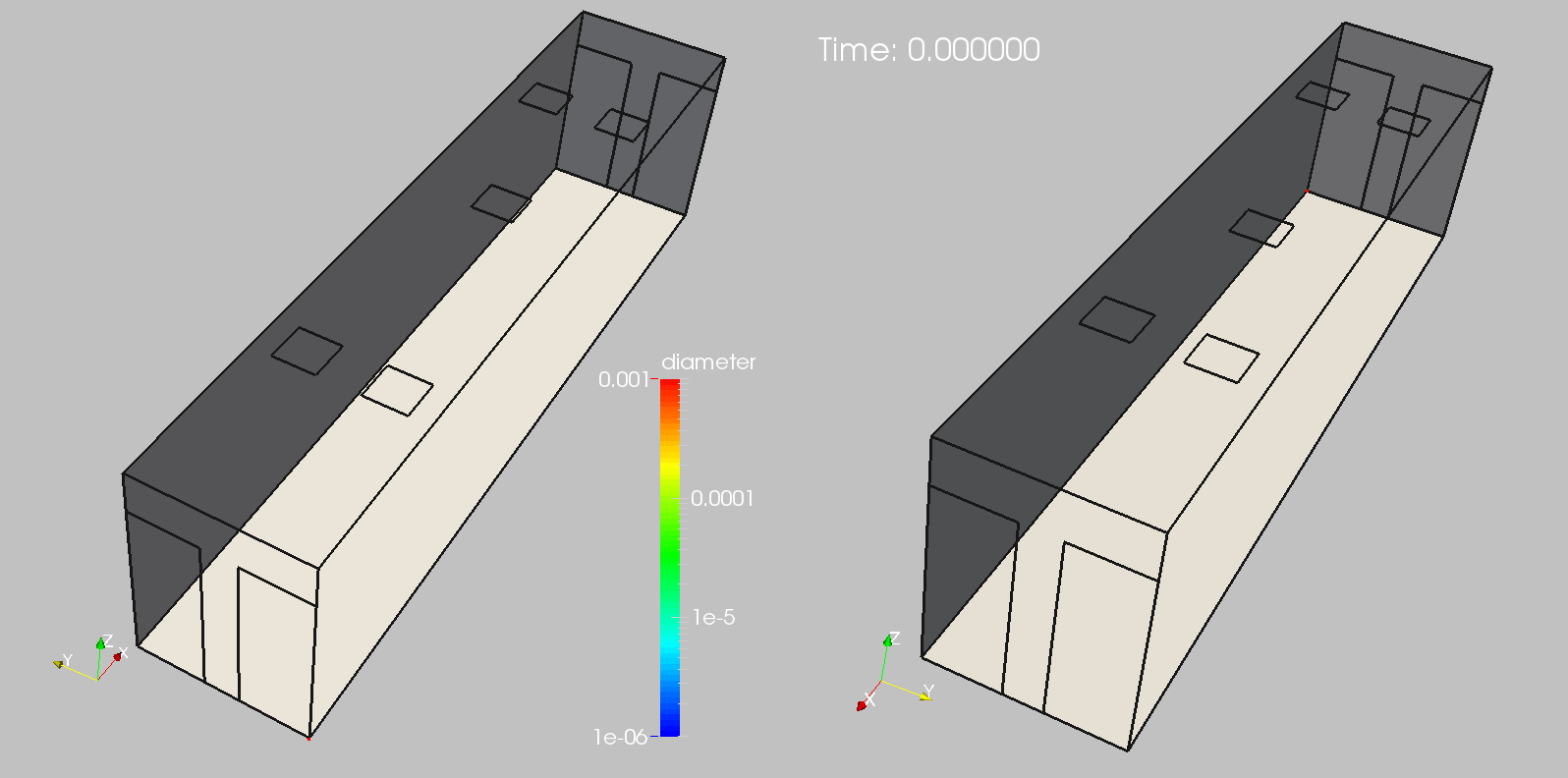}
	\vskip 10pt	
	\includegraphics[width=10.0cm]{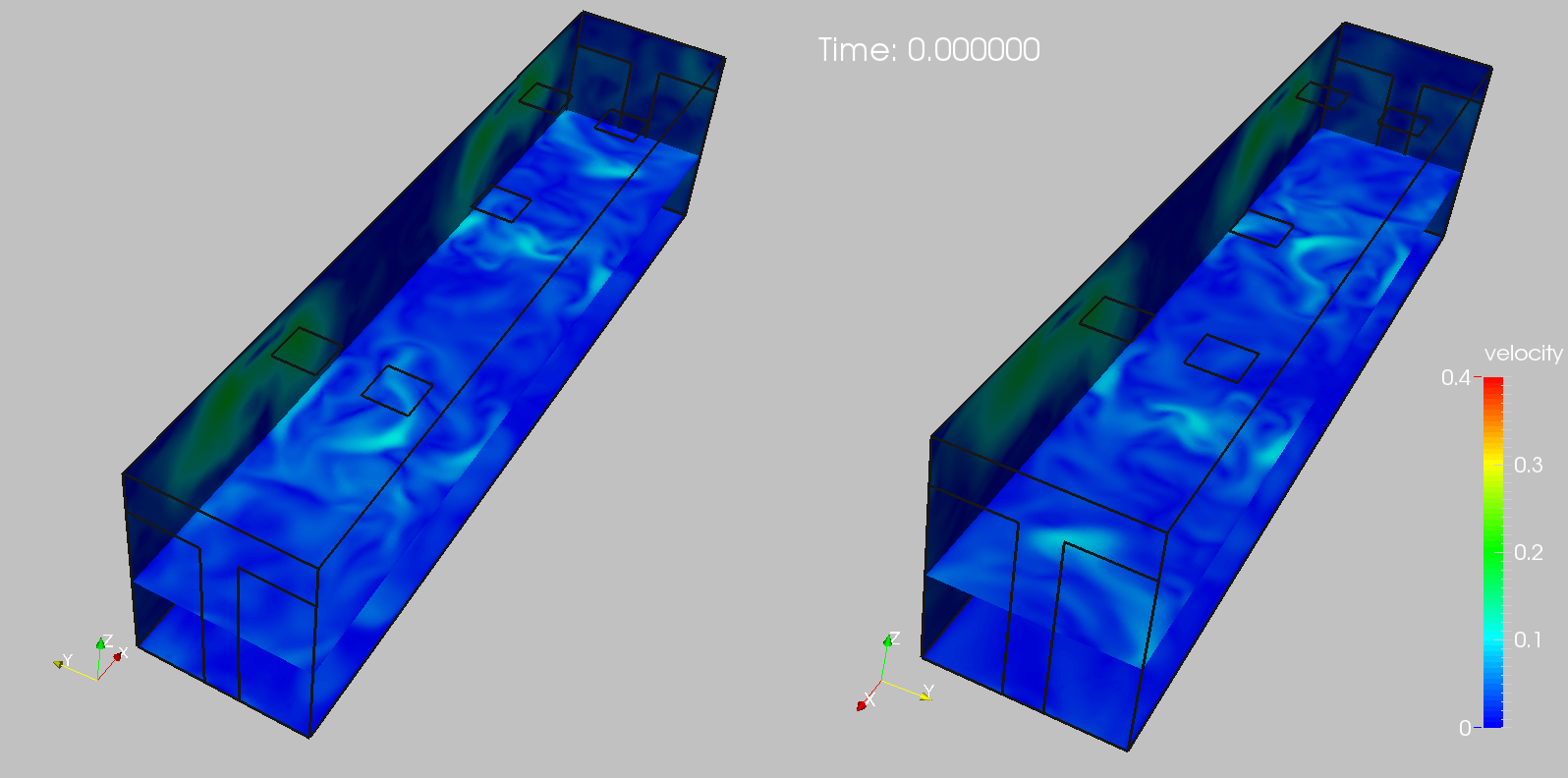}
	\vskip 10pt	
	\includegraphics[width=10.0cm]{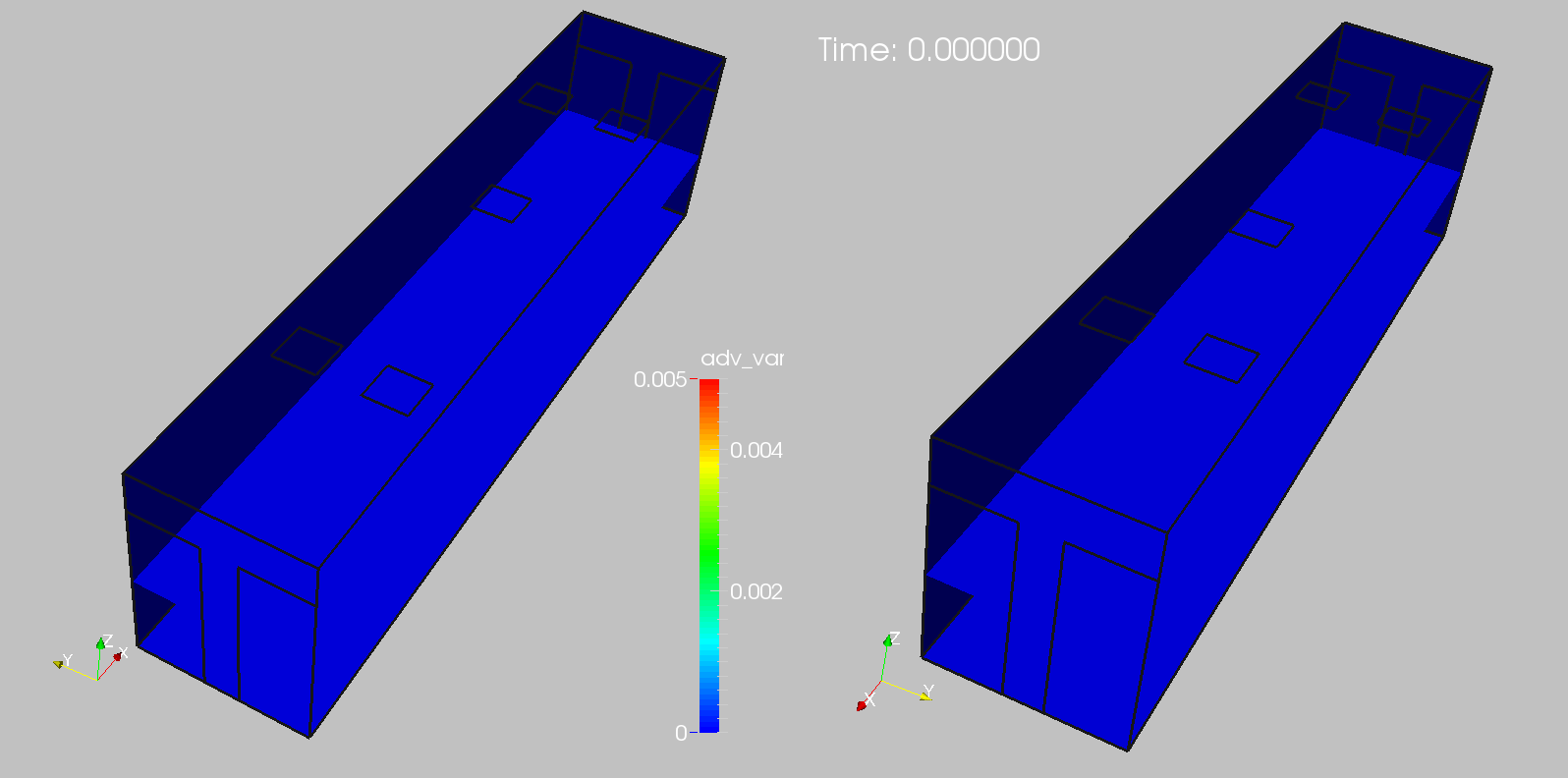}
	\caption{No Pedestrians: Solution at $t=0.00~sec$}
\end{figure}

\begin{figure}
	\centering
	\includegraphics[width=10.0cm]{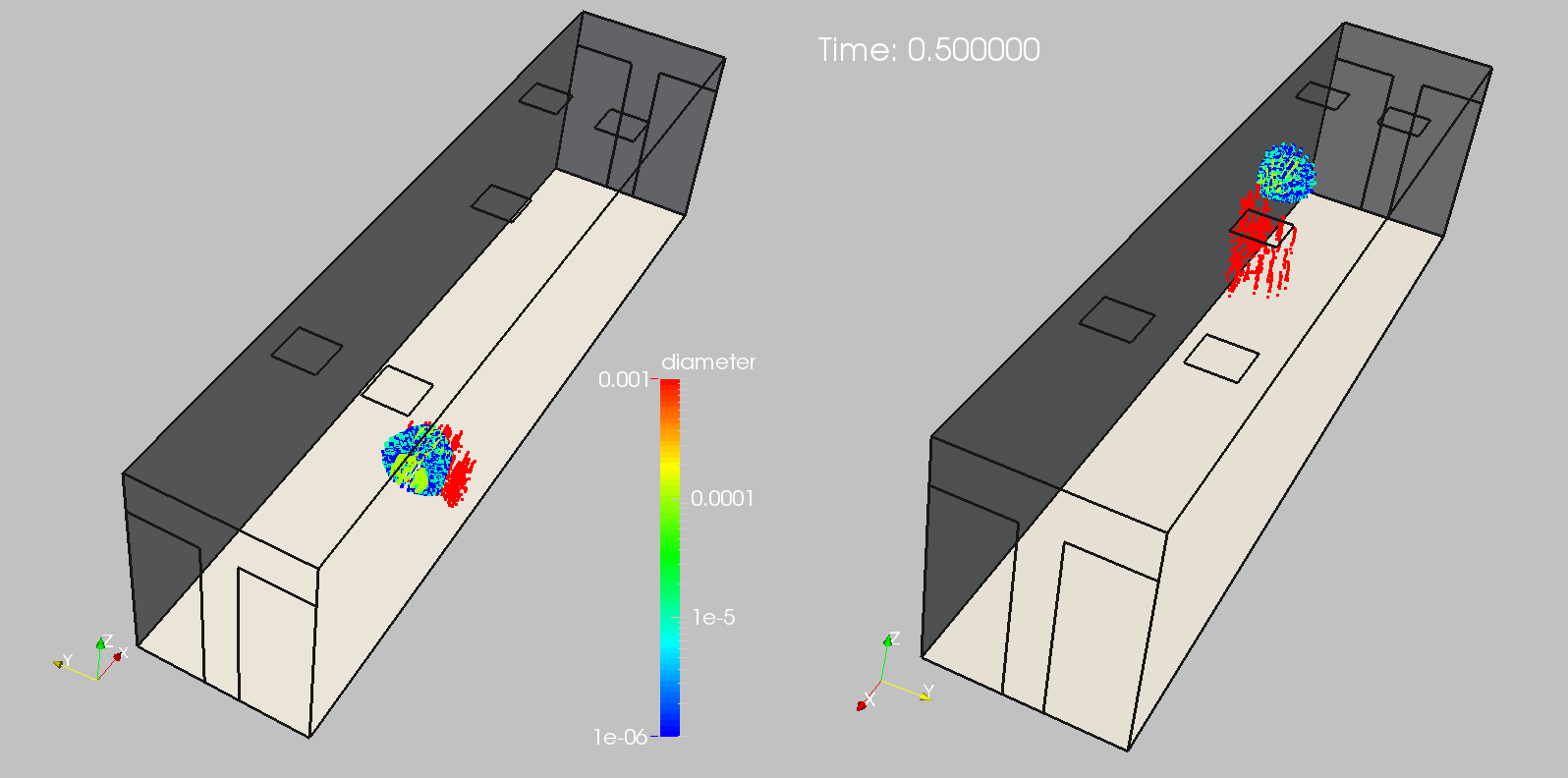}
	\vskip 10pt	
	\includegraphics[width=10.0cm]{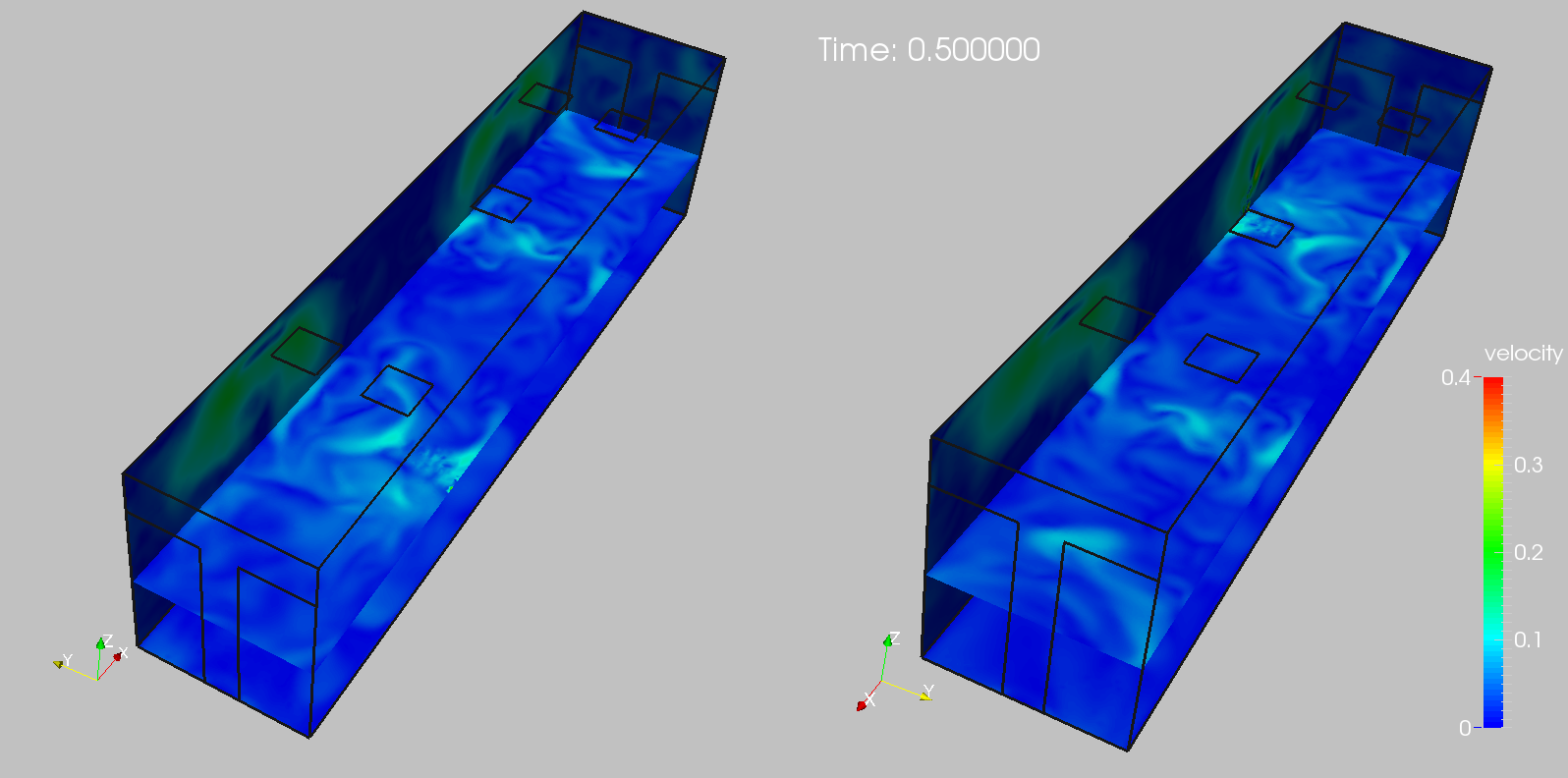}
	\vskip 10pt	
	\includegraphics[width=10.0cm]{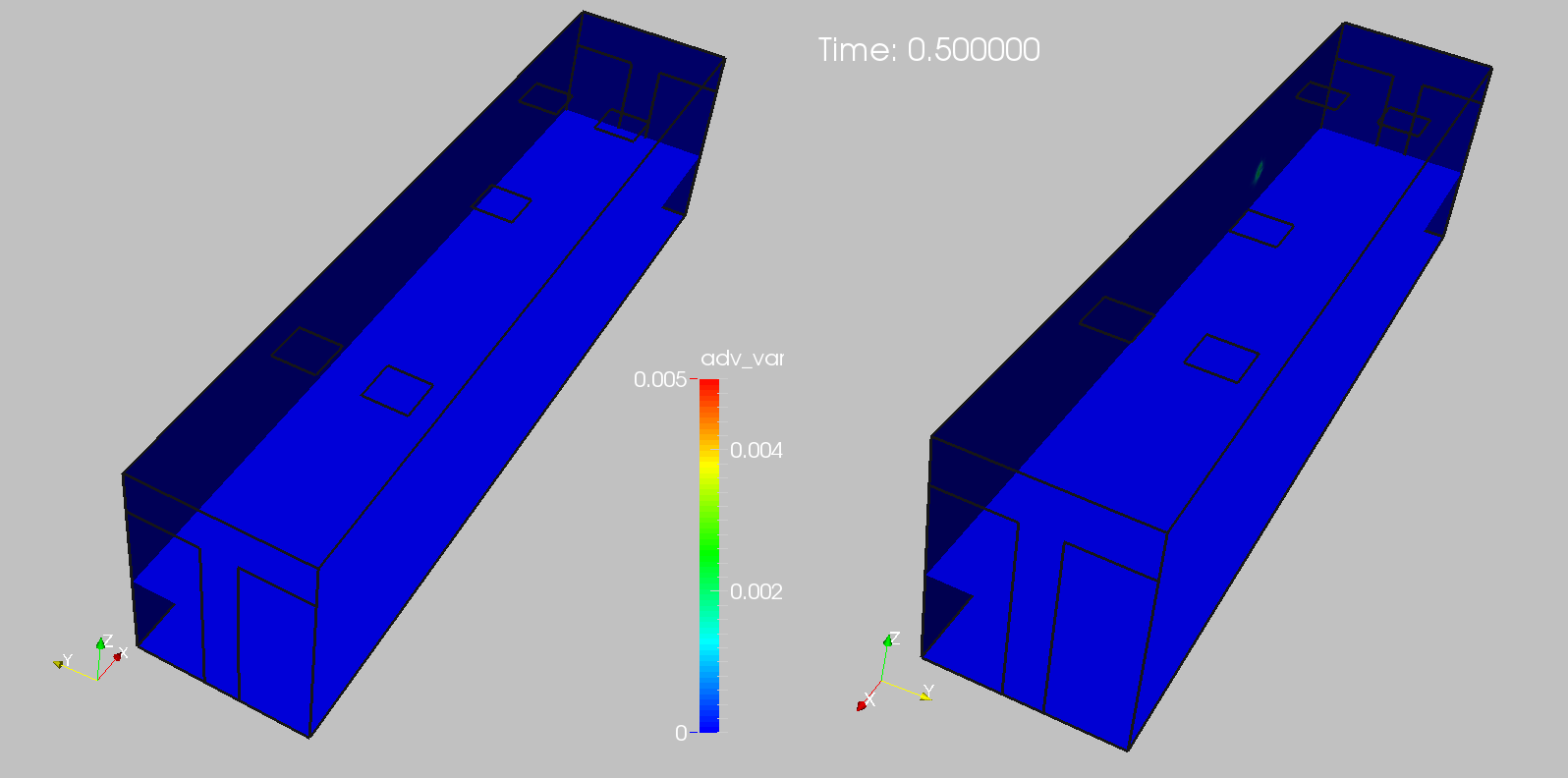}
	\caption{No Pedestrians: Solution at $t=0.50~sec$}
\end{figure}

\begin{figure}
	\centering
	\includegraphics[width=10.0cm]{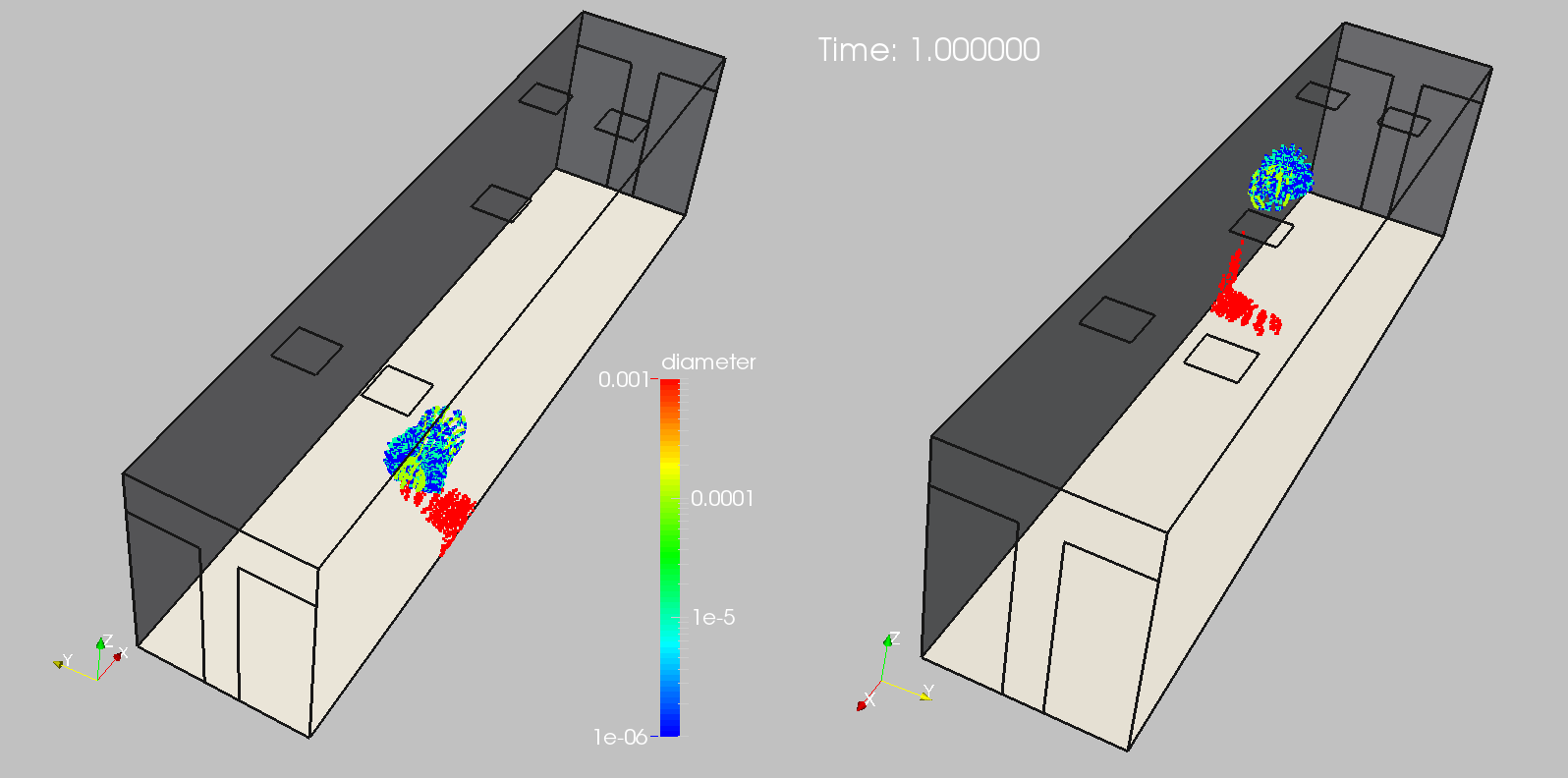}
	\vskip 10pt	
	\includegraphics[width=10.0cm]{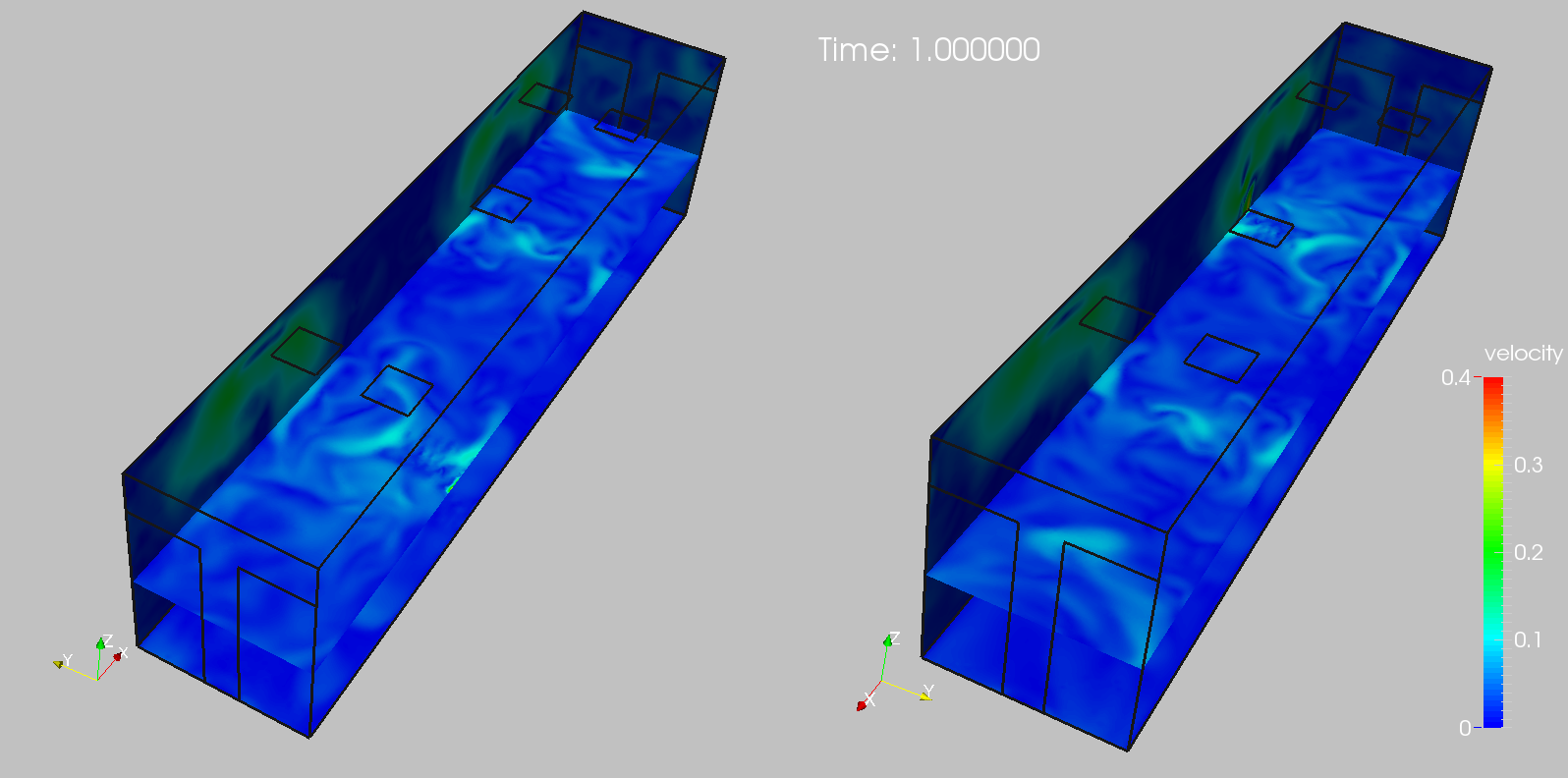}
	\vskip 10pt	
	\includegraphics[width=10.0cm]{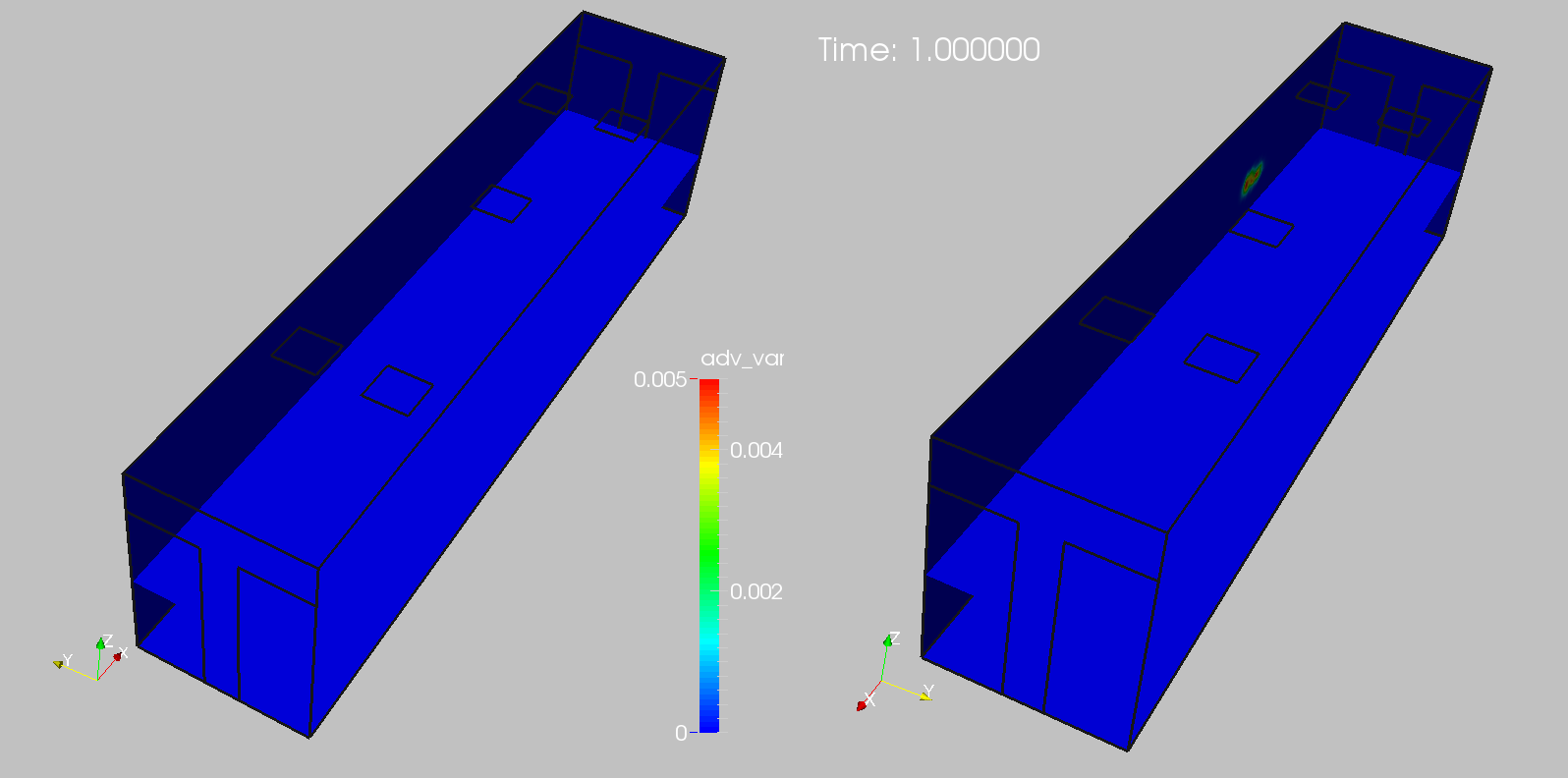}
	\caption{No Pedestrians: Solution at $t=1.00~sec$}
\end{figure}

\begin{figure}
	\centering
	\includegraphics[width=10.0cm]{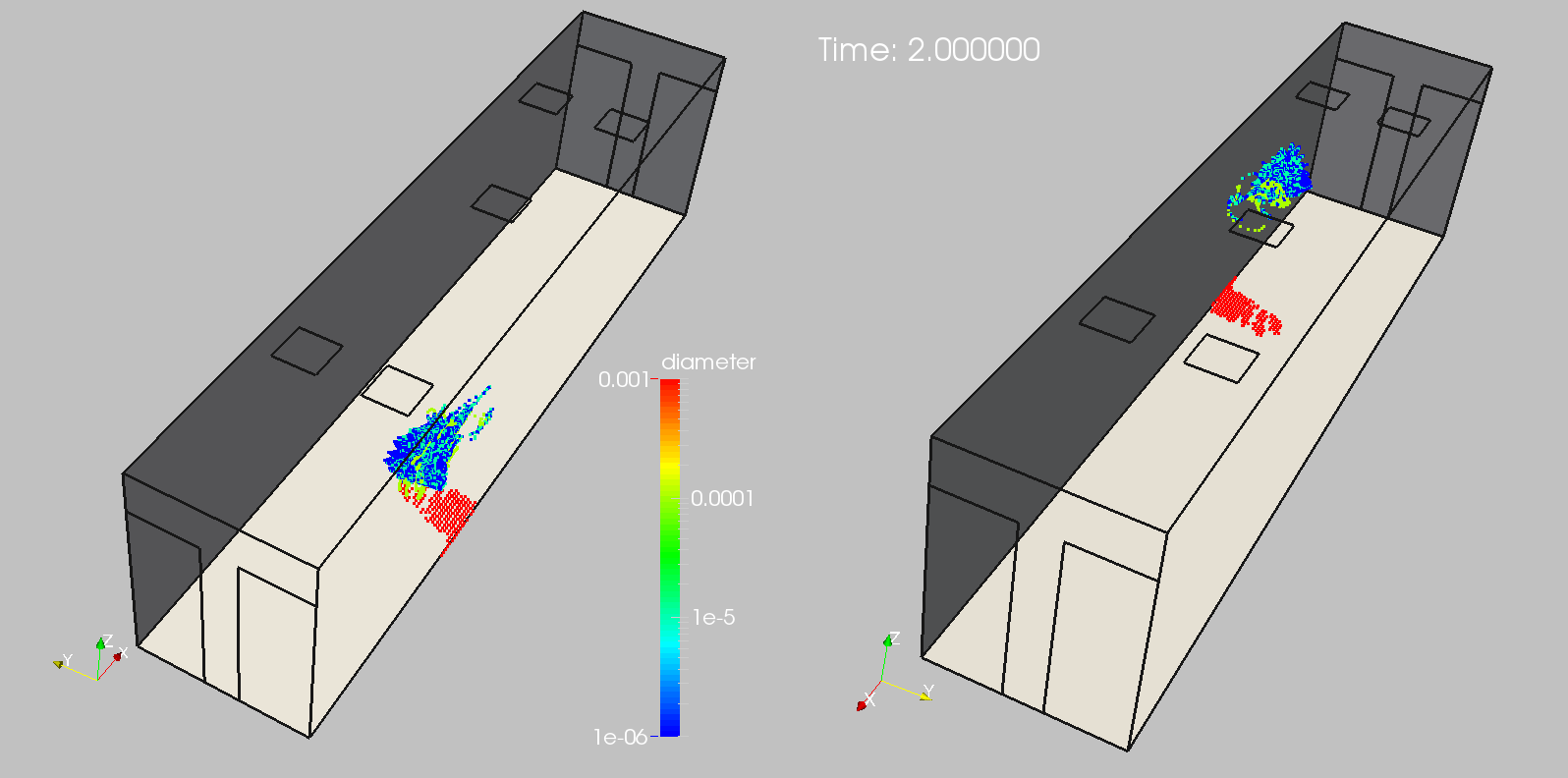}
	\vskip 10pt	
	\includegraphics[width=10.0cm]{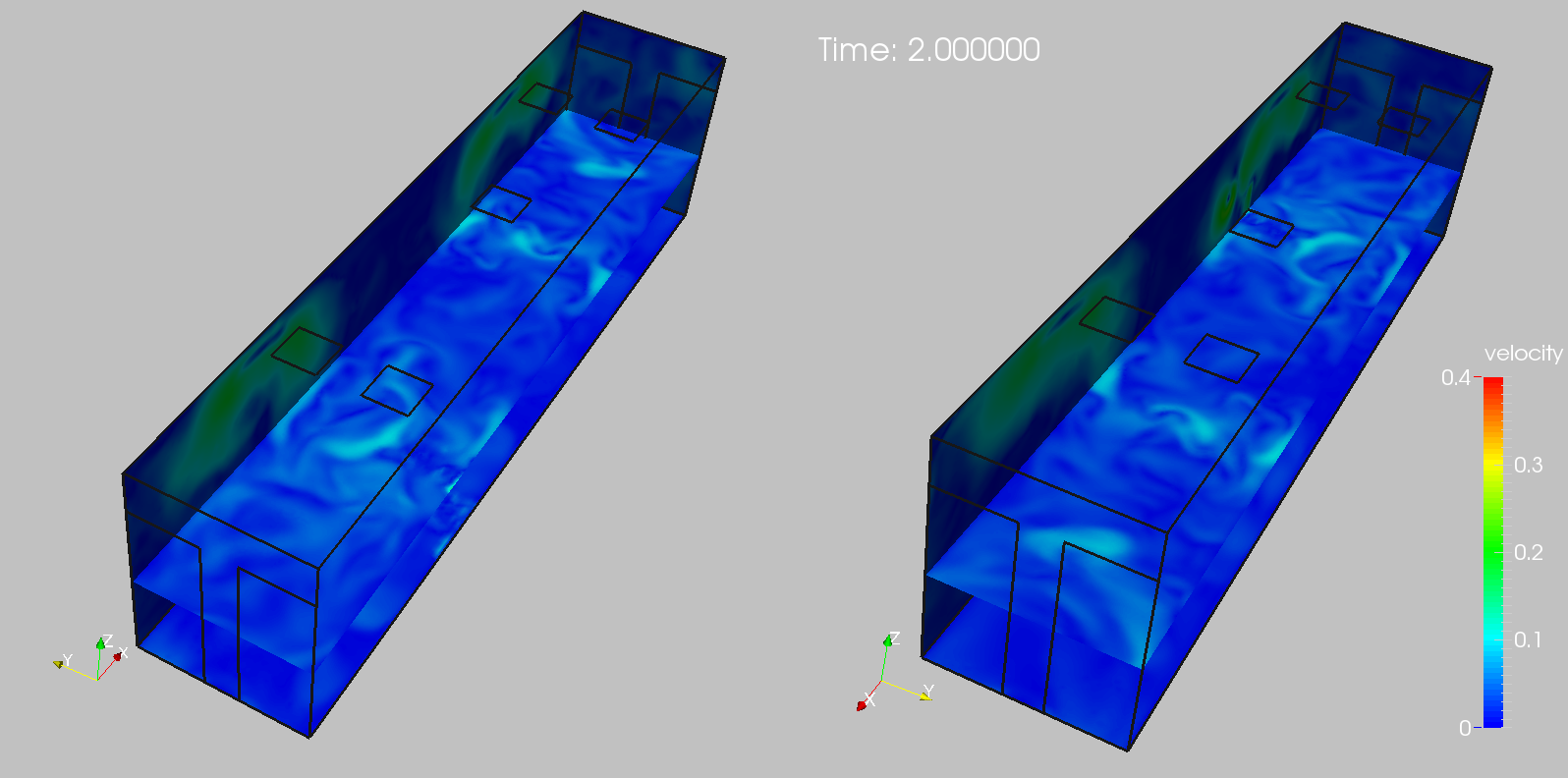}
	\vskip 10pt	
	\includegraphics[width=10.0cm]{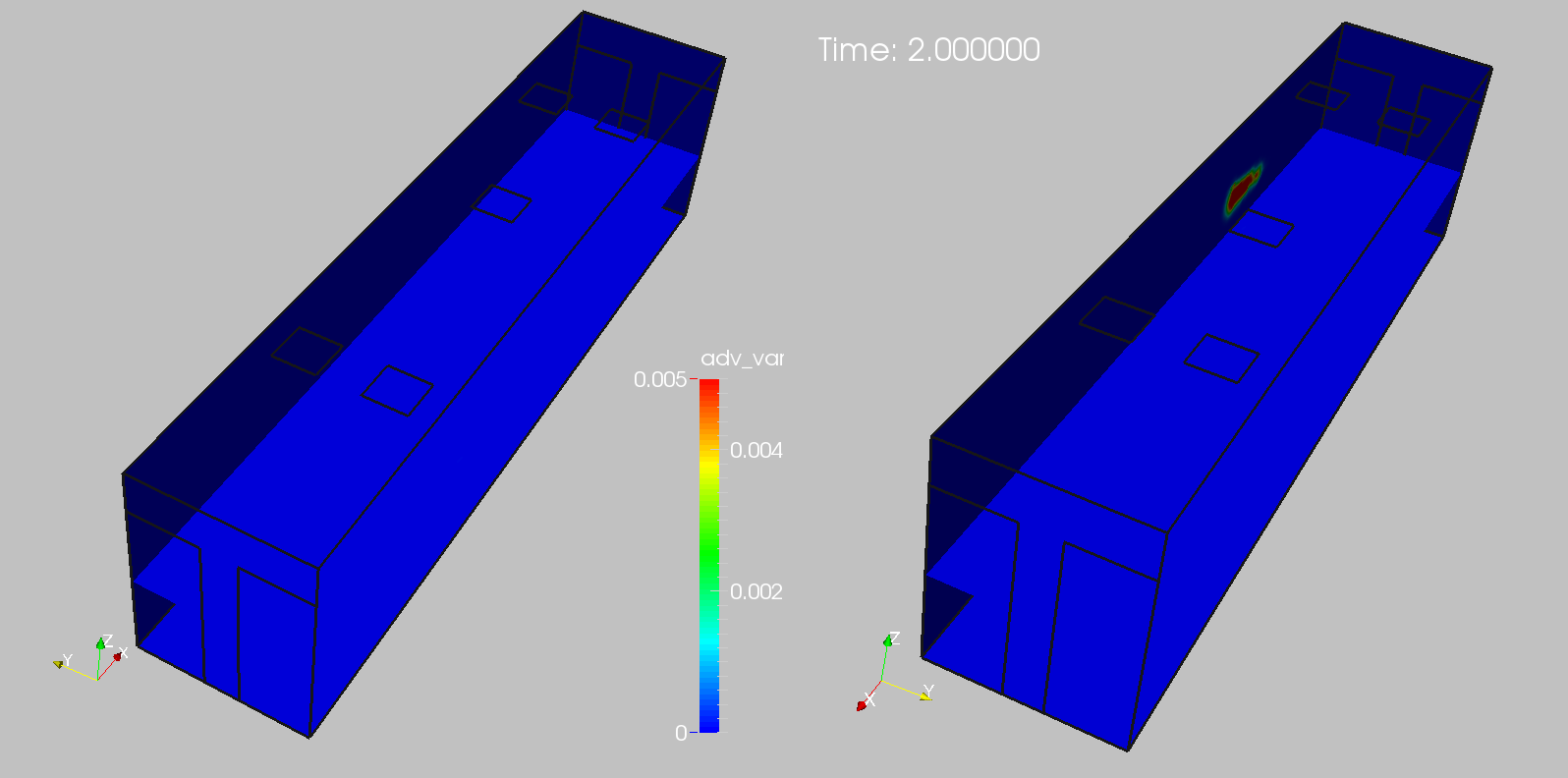}
	\caption{No Pedestrians: Solution at $t=2.00~sec$}
\end{figure}

\begin{figure}
	\centering
	\includegraphics[width=10.0cm]{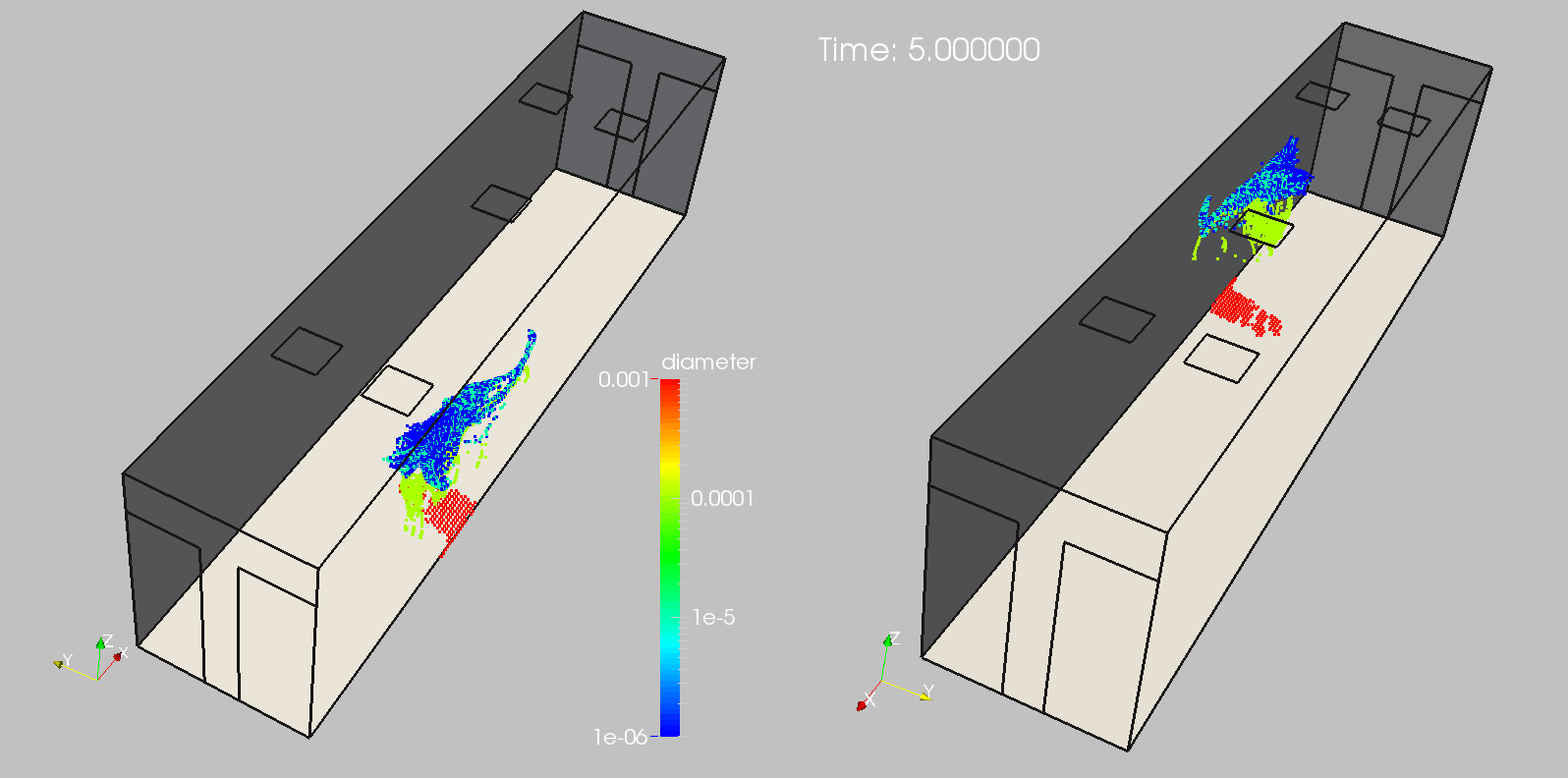}
	\vskip 10pt	
	\includegraphics[width=10.0cm]{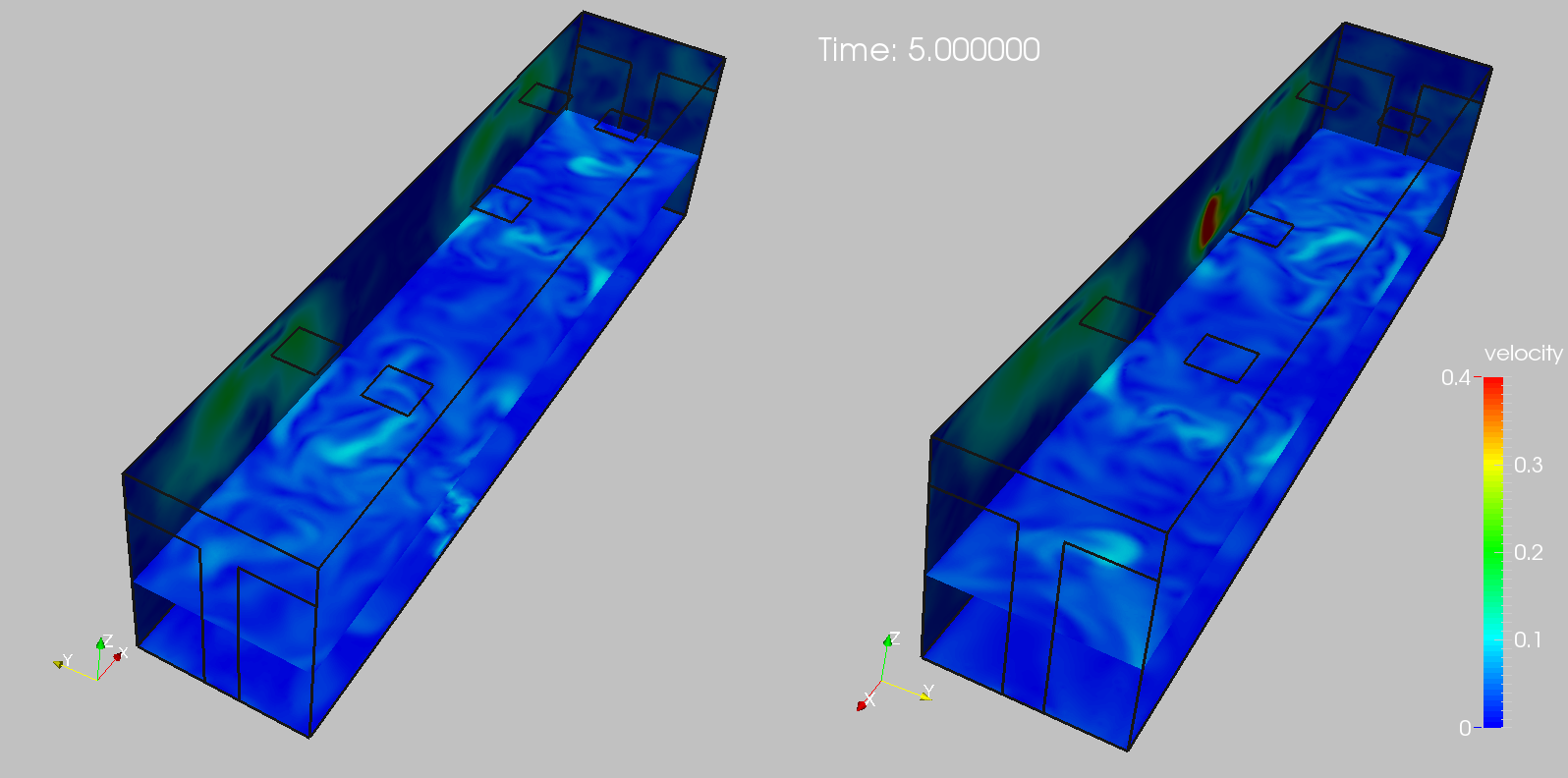}
	\vskip 10pt	
	\includegraphics[width=10.0cm]{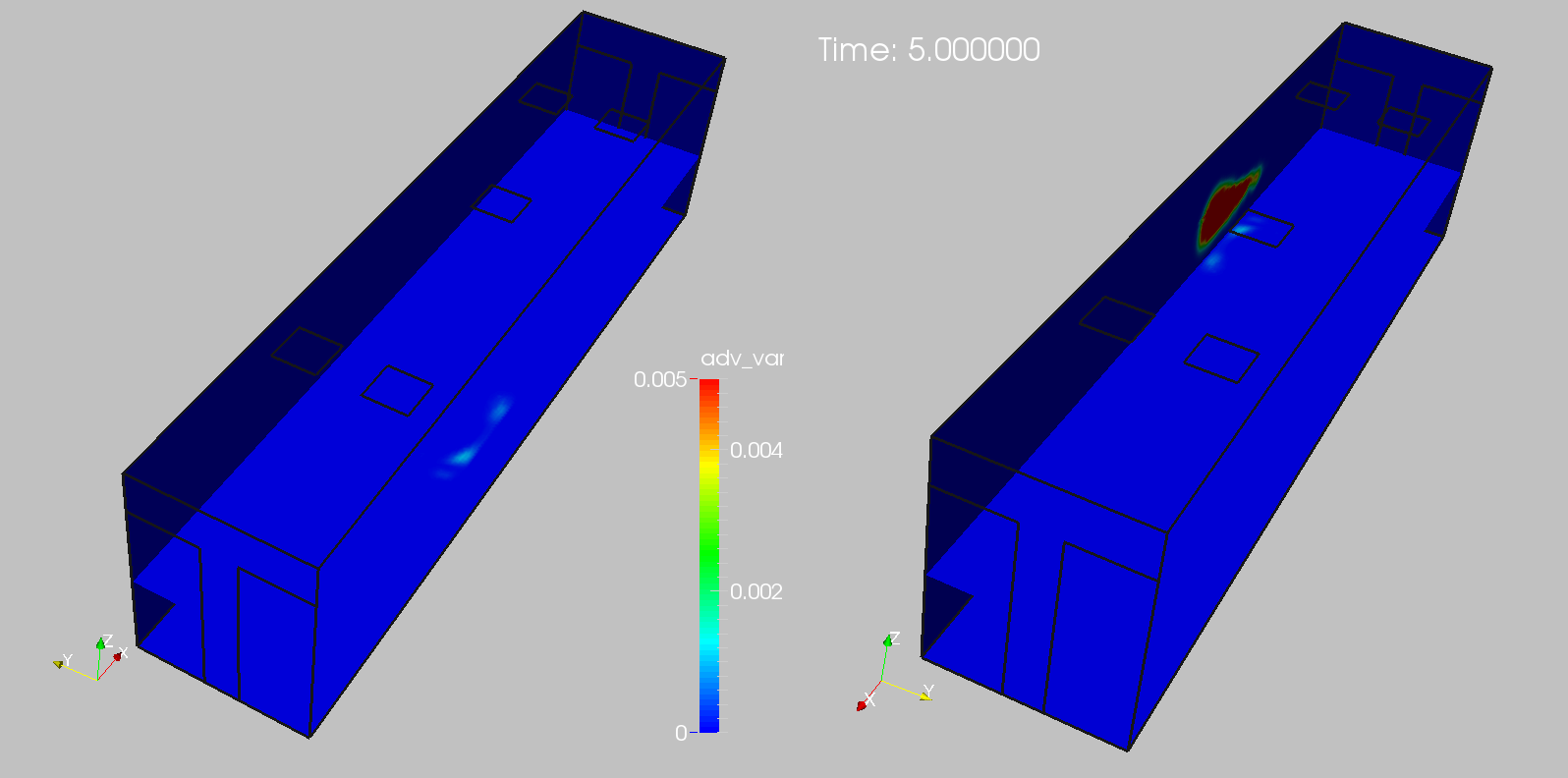}
	\caption{No Pedestrians: Solution at $t=5.00~sec$}
\end{figure}

\begin{figure}
	\centering
	\includegraphics[width=10.0cm]{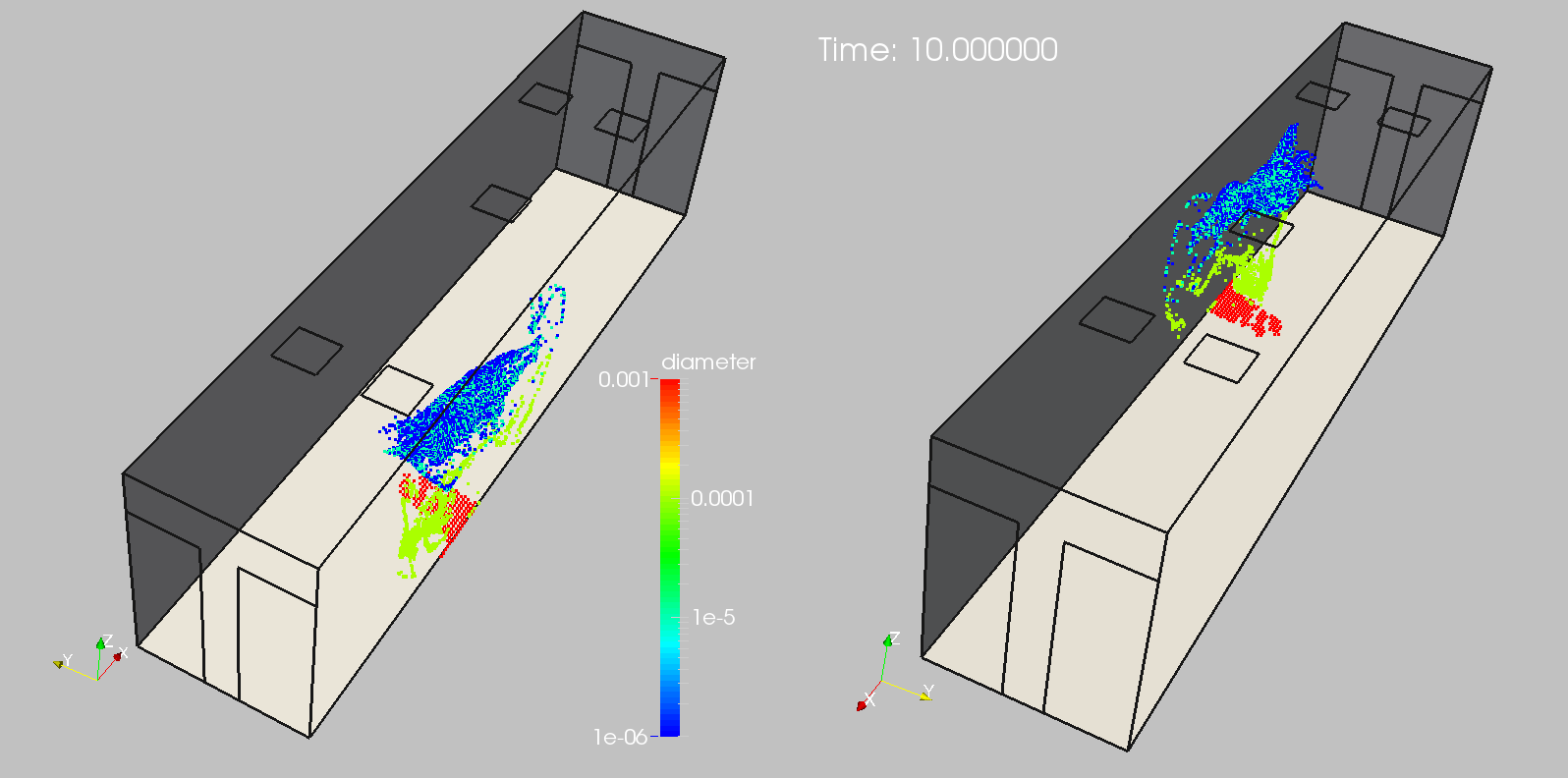}
	\vskip 10pt	
	\includegraphics[width=10.0cm]{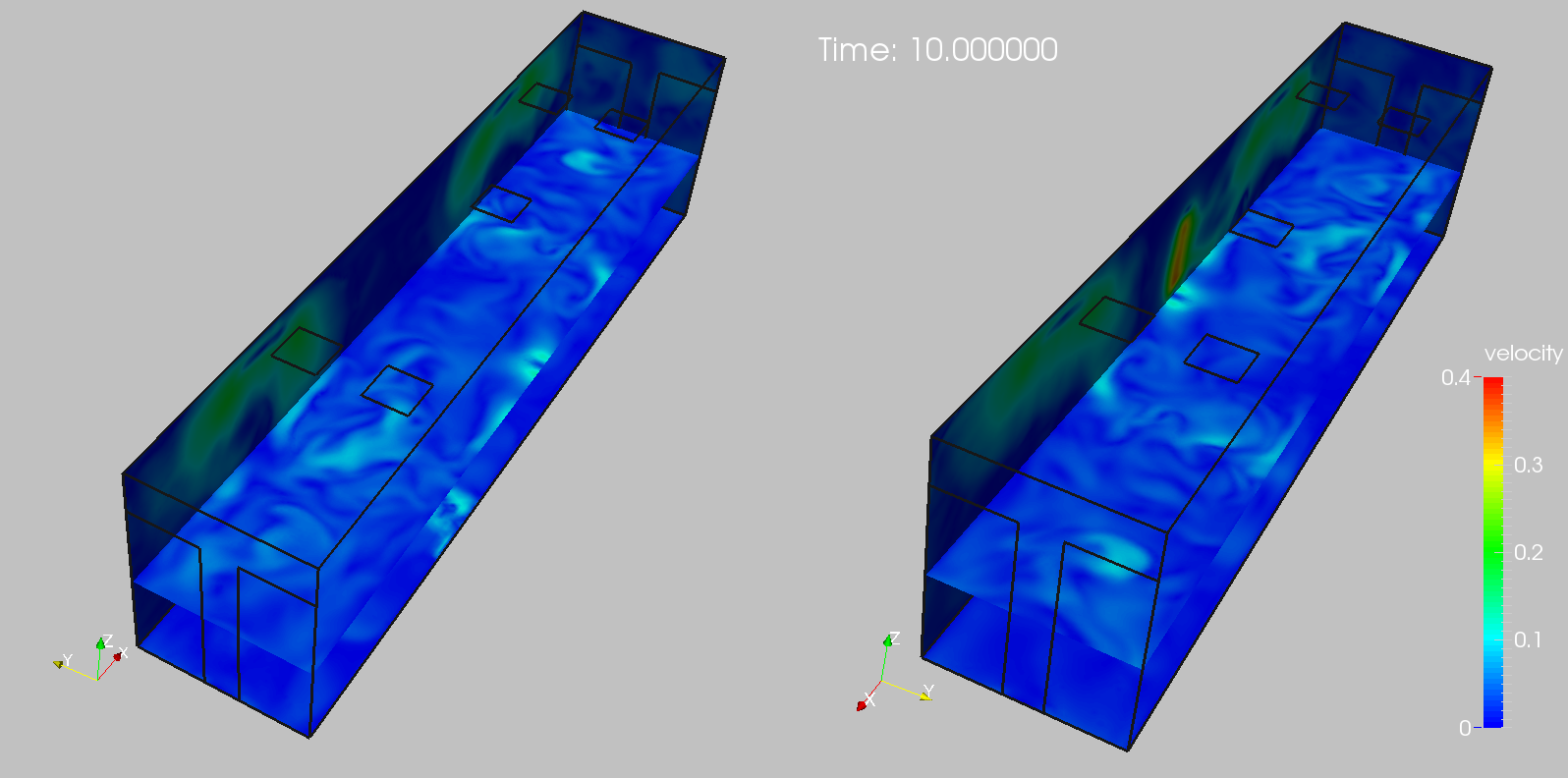}
	\vskip 10pt	
	\includegraphics[width=10.0cm]{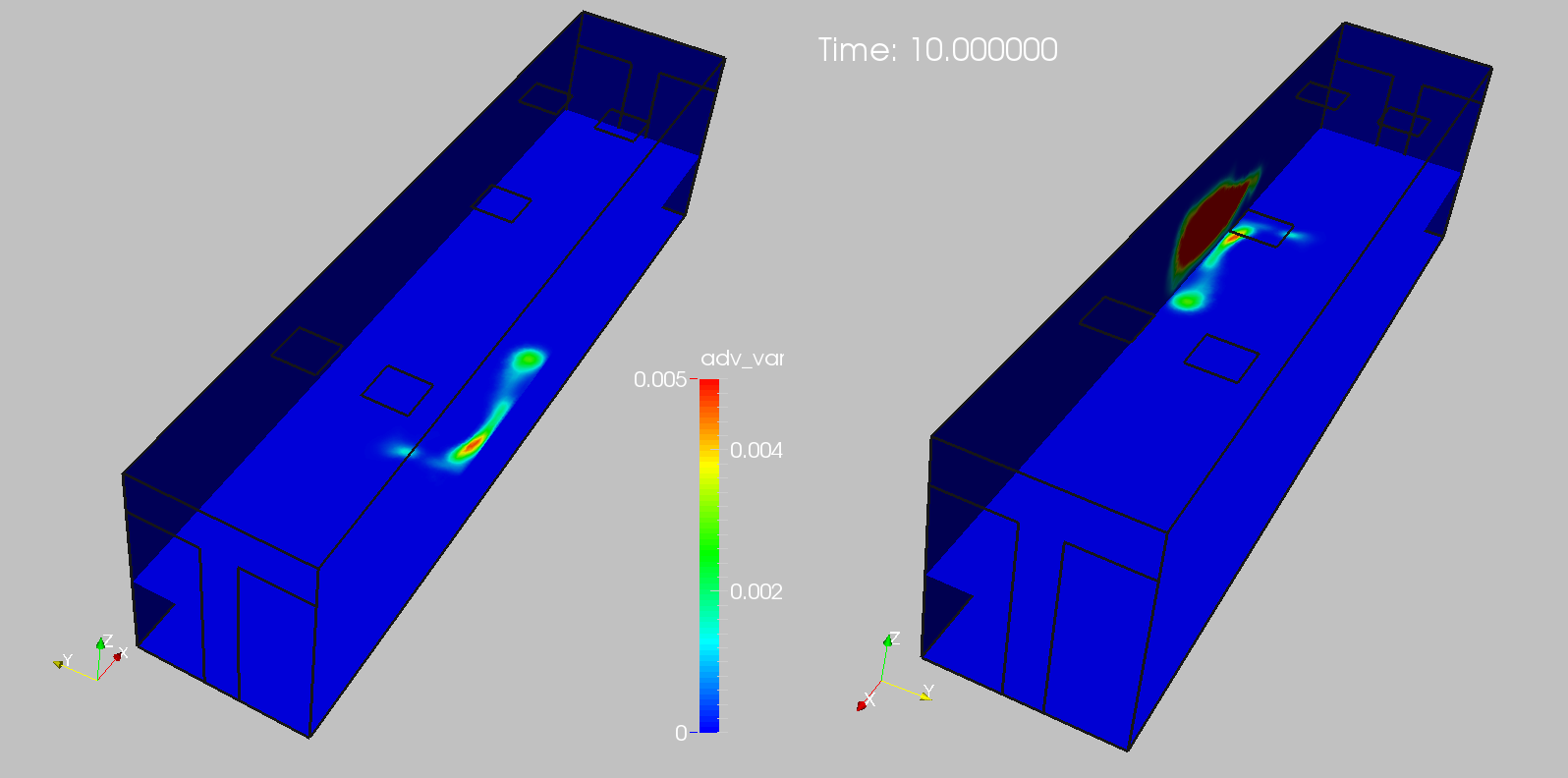}
	\caption{No Pedestrians: Solution at $t=10.00~sec$}
\end{figure}

\begin{figure}
	\centering
	\includegraphics[width=10.0cm]{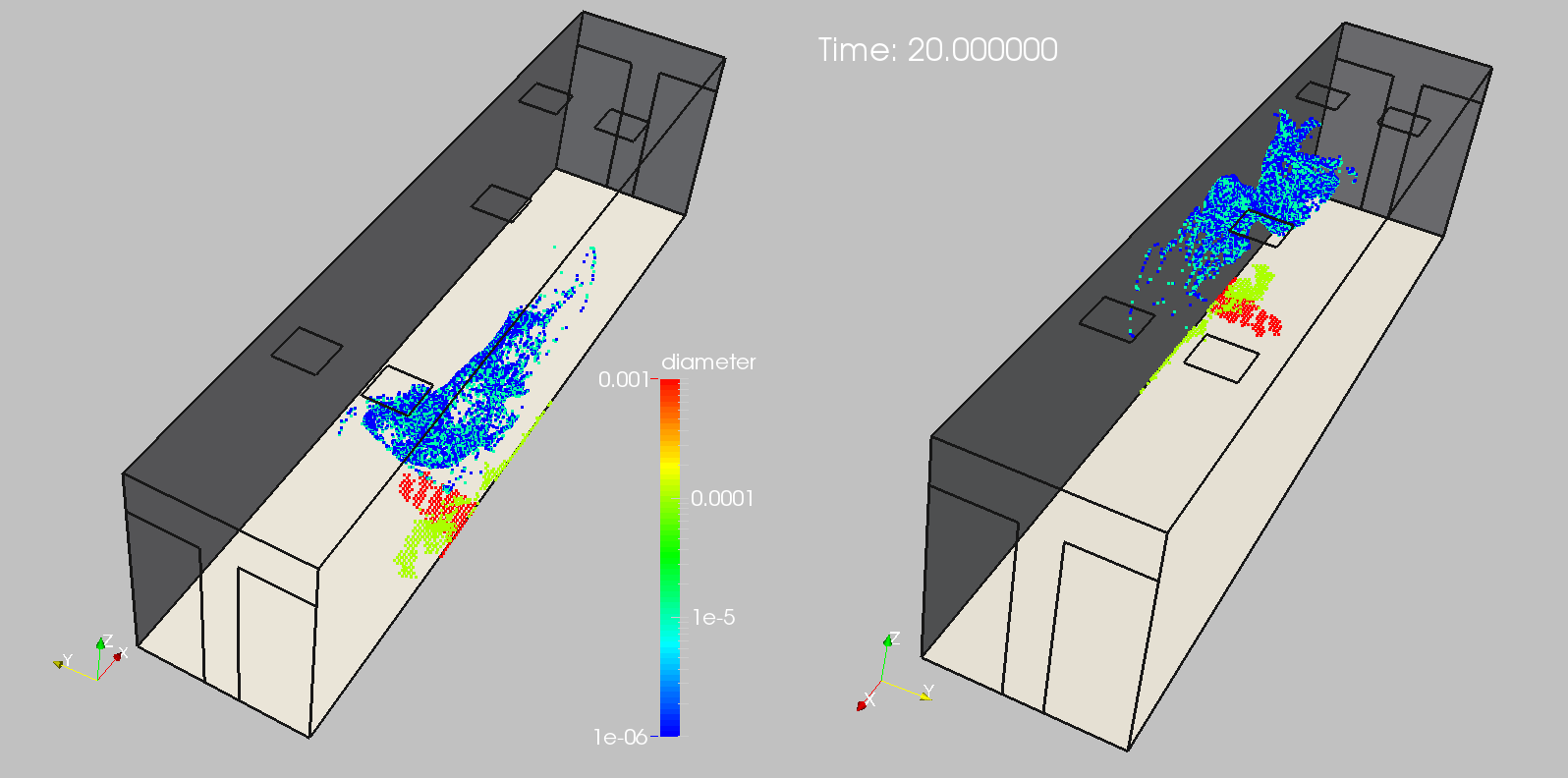}
	\vskip 10pt	
	\includegraphics[width=10.0cm]{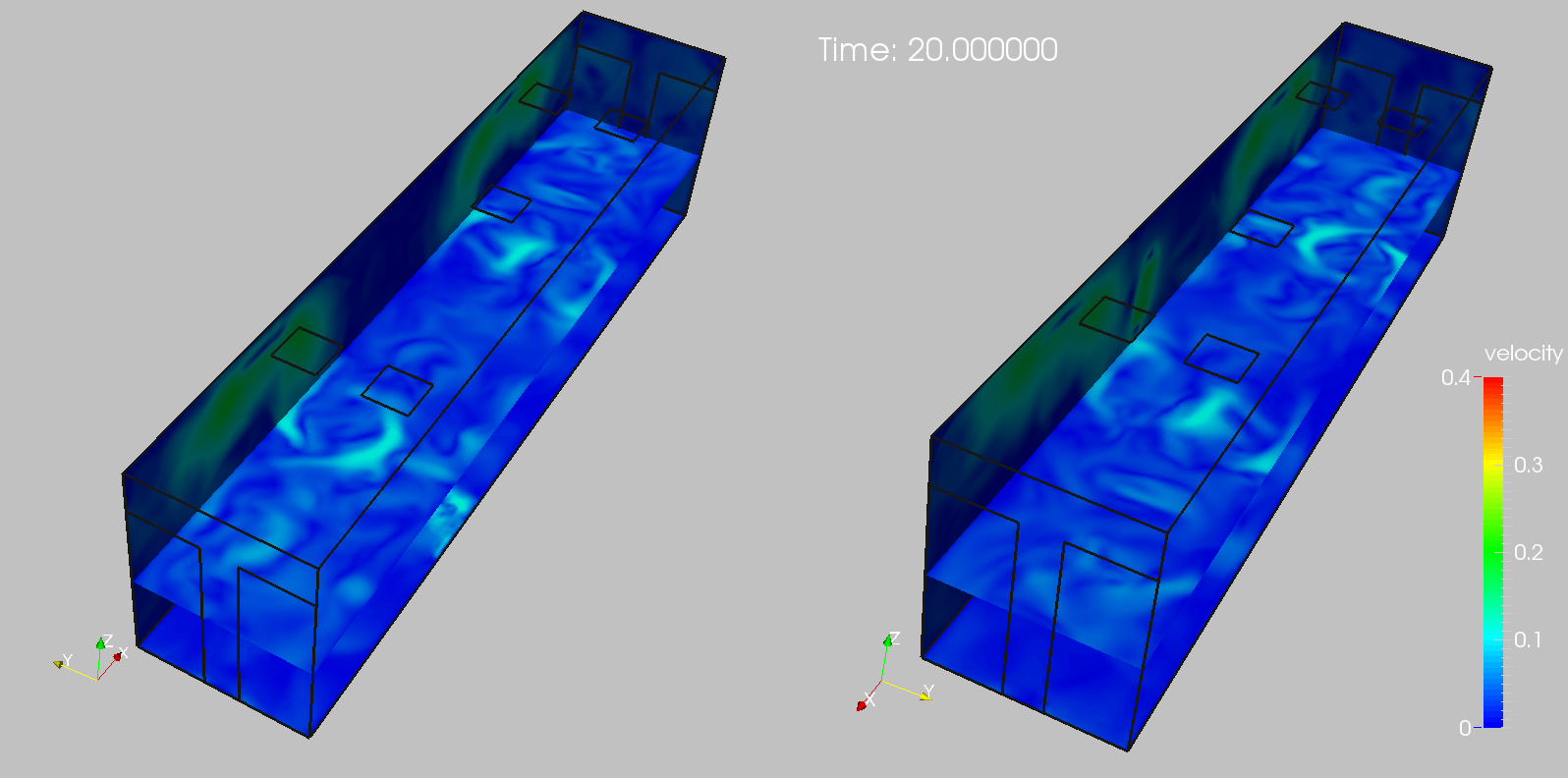}
	\vskip 10pt	
	\includegraphics[width=10.0cm]{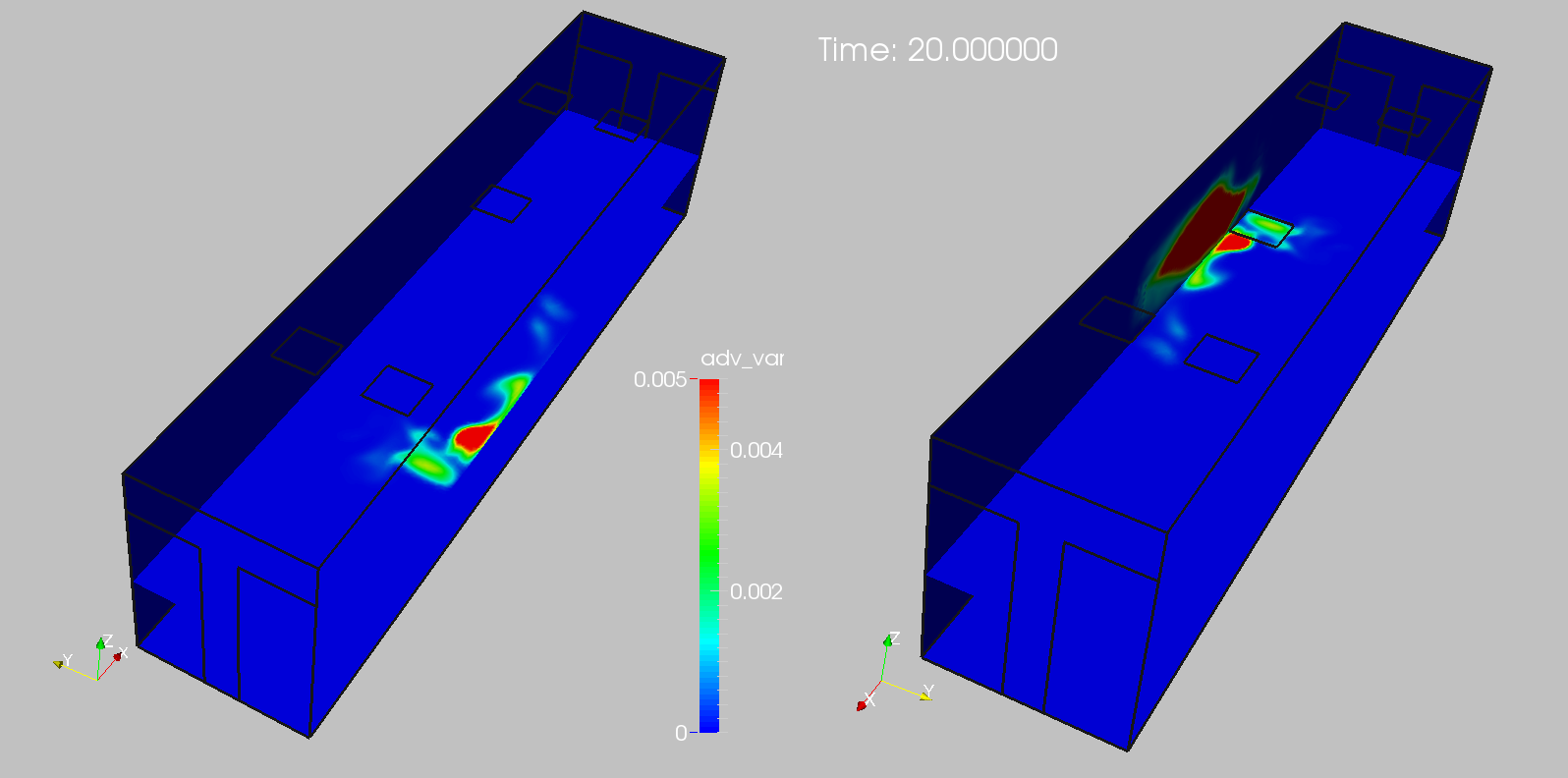}
	\caption{No Pedestrians: Solution at $t=20.00~sec$}
\end{figure}

\ms \noindent
Note the very large difference in mixing and viral transmission
due to the presence of pedestrians.

\section{Conclusions and Outlook}
\label{sec:conclusions}
A high fidelity model for the propagation of pathogens via
aerosols in the presence of moving pedestrians has been 
implemented.
The key idea is the tight coupling of computational fluid
dynamics and computational crowd dynamics in order to
capture the emission, transport and inhalation of pathogen
loads in space and time in the presence of moving pedestrians. \\
The example of a narrow corridor with moving pedestrians clearly
shows the considerable effect that pedestrian motion has on airflow,
and hence on pathogen propagation and potential infectivity. \\
At present, the `pedestrians' appear in the flow code as rigid
bodies. The incorporation of leg and arm movement while walking
would be a possible improvement to the model.


\bibliographystyle{aiaa}

\end{document}